\documentclass[12pt]{article}
\usepackage{amsmath}
\usepackage{amssymb}
\usepackage{graphicx}
\numberwithin{equation}{section} \numberwithin{equation}{section}

\begin{document}

\setcounter{page}{0}

\thispagestyle{empty}

\begin{center}
{\large \bf MAGNETARS }
\end{center}

\renewcommand{\thefootnote}{\fnsymbol{footnote}}

\begin{center}
{
Paolo Cea$^{1,2}$\protect\footnote{Electronic address: {\tt
Paolo.Cea@ba.infn.it}}} \\[0.5cm]
$^1${\em Dipartimento Interateneo di Fisica, Universit\`a di Bari,  Bari, Italy }\\[0.3cm]
$^2${\em INFN - Sezione di Bari, Bari,
Italy} \\[0.3cm]
\end{center}

\vspace*{0.5cm}

\vspace*{1.0cm}

\renewcommand{\abstractname}{\normalsize Abstract}
\begin{abstract}
P-stars are compact stars made of up and down quarks in $\beta$-equilibrium with electrons in a chromomagnetic condensate.
P-stars are able to account for  compact stars like {\it {RXJ 1856.5-3754}} and {\it {RXJ 0720.4-3125}},  stars with
radius comparable with canonical neutron stars, as well as super massive compact objects like {\it {SgrA}}$^*$. We discuss
p-stars endowed with super strong dipolar magnetic field which, following  consolidated tradition in literature, are
referred to as magnetars. We show that soft gamma-ray repeaters and anomalous $X$-ray pulsars can be understood within our
theory. We find a well defined criterion to distinguish rotation powered pulsars from magnetic powered pulsars. We show
that glitches, that in our magnetars are triggered by magnetic dissipative effects in the inner core, explain both the
quiescent emission and bursts in soft gamma-ray repeaters and anomalous $X$-ray pulsars. We are able to account for the
braking glitch from {\it {SGR 1900+14}} and the normal glitch from {\it {AXP 1E 2259+586}} following a giant burst. We
discuss and explain the observed anti correlation between hardness ratio and intensity. Within our magnetar theory we are
able to account quantitatively for  light curves for both gamma-ray repeaters and anomalous $X$-ray pulsars. In particular
we explain the puzzling light curve after the June 18, 2002 giant burst from {\it {AXP 1E 2259+586}}, so that  we feel
this last event as the Rosetta Stone for our theory. Finally, in Appendix we discuss the origin of the soft emission not
only for soft gamma-ray repeaters and anomalous $X$-ray pulsars, but also for isolated $X$-ray pulsars. We also offer an
explanation for the puzzling spectral features in {\it {1E 1207.4-5209}}.
\end{abstract}
%
%

\newpage

\section{\normalsize{INTRODUCTION}}
\label{Introduction}
In few years since their discovery~\cite{hewish:1968}, pulsars have been identified with rotating neutron stars, first
predicted theoretically by W.~Baade and F.~Zwicky~\cite{baade:1934}, endowed with a strong magnetic
field~\cite{pacini:1968,gold:1968}. The exact mechanism by which a pulsar radiates the energy observed as radio pulses is
still a subject of vigorous debate~\cite{michel:1982,michel:1991}, nevertheless the accepted standard model based on the
picture of a rotating magnetic dipole has been developed since long
time~\cite{goldreich:1969,sturrock:1971}. \\
Nowadays, no one doubts that pulsars are  neutron stars, even though it should be remembered that there may be other
alternative explanations for pulsars.  Up to present time it seems that there are no alternative models able to provide as
satisfactory an explanation for the wide variety of pulsar phenomena as those built around the rotating neutron star
model. However, quite recently we have proposed~\cite{cea:2003} a new class of compact stars, named p-stars, which is
challenging the two pillars of modern astrophysics, namely neutron stars and black holes. We are, however, aware that such
a dramatic change in the standard paradigm of relativistic astrophysics which is based on neutron stars and black holes
needs a careful comparison with the huge amount of observations collected so far for pulsar and black hole candidates. In
our opinion we feel that there are already clear observational evidences pointing towards the need of a drastic revision
of the standard paradigm. So that, before addressing the main subject of the present paper, it is worthwhile to briefly
discuss some observational evidences for p-stars and against neutron stars and black hole candidates.\\
As concern black holes, we point out that the most interesting and intriguing aspect of our theory is that p-stars are
able to overcome the gravitational collapse even for mass much greater $10^6 M_{\bigodot}$. Indeed, from the equation of
of state of degenerate up and down quarks in a chromomagnetic condensate described for the first time in
Ref.~\cite{cea:2003}, we have on dimensional ground:
\begin{equation}
\label{dimensional}
 M \; = \; \frac{1}{G^{3/2} gH} \; \;
 f(\overline{\varepsilon}_c) \;
\; \; \; \; \; R \; = \; \frac{1}{G^{1/2} gH} \; \; g(\overline{\varepsilon}_c) \; ,
\end{equation}
where $G$ is the gravitational constant, $\overline{\varepsilon}_c \, = \, \varepsilon_c/(gH)^2 $, and $\varepsilon_c$ is
the central density. As a consequence we get:
\begin{equation}
\label{ratio}
\frac{2 \; G \; M}{R} \; = \; 2 \; \frac{f(\overline{\varepsilon}_c)}{g(\overline{\varepsilon}_c)} \; \; \equiv \;
h(\overline{\varepsilon}_c).
\end{equation}
From Equation~(\ref{dimensional}) we see that by decreasing the strength of the chromomagnetic condensate we increase the
mass and radius of the star. However, the ratio $\frac{2 \; G \; M}{R}$ depends on  $\overline{\varepsilon}_c $ only. It
turns out that the function $h(x)$ defined in Eq.~(\ref{ratio}) is less than 1 for any allowed values of
$\overline{\varepsilon}_c$~\cite{cea:2003}. Thus, we infer that our p-stars do not admit the existence of an upper limit
to the mass of a completely degenerate configuration. In other words, our peculiar equation of state of degenerate up and
down quarks in a chromomagnetic condensate allows the existence of finite equilibrium states for stars of arbitrary mass. \\
The accepted arguments for the evidence of black holes is based on the fact that there is spectroscopic evidence of
compact objects with mass exceeding $3 \, M_{\bigodot}$. Indeed, a white dwarf cannot have a mass exceeding the
Chandrasekhar limit, about $1.4 \, M_{\bigodot}$, while even for neutron stars there is a maximum mass which probably is
about $3 \, M_{\bigodot}$. Then, compact objects with mass exceeding  $3 \, M_{\bigodot}$ are classified as black holes.
However, as we argued before, this argument cannot distinguish  between a massive p-star and a black hole. Indeed, the
fundamental difference between massive p-stars and black holes resides in the lack of stellar surface in black holes. As
it is well known, from general relativity it follows that the black hole boundary is a geometric surface called the event
horizon. Recently, there are claims in literature for evidence of  event horizons. For instance, in
Ref.~\cite{Garcia:2000} it is claimed that the $X$-ray luminosities of black hole candidates in quiescence are much less
than the corresponding $X$-ray luminosities of compact solar mass stellar systems. However, the authors of
Ref.~\cite{Campana:2000} argued that it is not correct to compare only the $X$-ray luminosities.  If one compares the
bolometric luminosities, then it turns out that there are no observable differences between compact solar mass stellar
systems and black holes candidates. So that, up to now there is no compelling evidence in favour of event horizons. On the
other hand, interestingly enough the authors of Ref.~\cite{Robertson:2002}, by using the standard analysis of  magnetic
propeller effect~\cite{Frank:2002} for pulsar in low mass $X$-ray binaries, found that the spectral properties of galactic
black hole candidates could be accounted for by compact objects with an intrinsic magnetic moment. Subsequently, in
Ref.~\cite{Robertson:2004} these authors,  extending their analysis to active galactic nuclei, showed how a standard
accretion disk can interact with the intrinsically magnetized central compact object to drive low state jets. Even though
these authors believe that massive intrinsically magnetized central objects can be accounted for within general relativity
as highly red shifted, extremely long lived, collapsing, radiating objects~\cite{Robertson:2003,Leiter:2003}, it is
evident that massive p-stars endowed with magnetic fields are indeed  the natural candidates for massive compact objects
with an
intrinsic magnetic moment. \\
As we will discuss in a future paper, in p-stars there is a natural mechanism to generate a dipolar magnetic field. As a
matter of fact, it turns out that  the generation of the dipolar magnetic field is enforced by the formation of a dense
inner core composed mainly by down quarks. As a consequence the surface dipolar magnetic field $B_S$ is proportional to
the strength of the chromomagnetic condensate $gH$. More precisely we have (here and in the following we shall adopt
natural units $\hbar \; = \; c \; = \; k_B \; = \; 1$):
\begin{equation}
\label{magn-condensate}
B_S\; \; \simeq \; \; \frac{e}{96 \, \pi }  \; \; gH\; \; \left (\frac{R_c}{R } \right )^3 \; ,
\end{equation}
where $e$ is the electric charge, $R$ and $R_c$ are the stellar and inner core radii respectively. It is interesting to
observe that for  p-stars with canonical mass $ M \, \simeq \, 1.4 \, M_{\bigodot}$ we get $\sqrt{gH} \; \simeq \; 0.55 \,
GeV$. On the other hand, massive p-stars with $ M \, \simeq \, 10 \, M_{\bigodot}$ require a chromomagnetic condensate
$gH$ smaller by about a factor $4 - 5$ with respect to canonical p-stars. So that, from Eq.~(\ref{magn-condensate}) it
follows that massive p-stars are characterized by surface magnetic fields reduced by less than one order of magnitude
with respect to pulsar magnetic fields. \\
In addition, as we said, there are finite equilibrium states for stars of arbitrary mass. For instance, {\it{SgrA}}$^*$,
the super massive compact object at the galactic center~\cite{Melia:2001}, could be interpreted as a p-star  with mass $M
\, \simeq 3.2 \, 10^6 \, M_{\bigodot}$ and radius  $R \, \simeq 1.4 \, 10^7 \, Km$. The
corresponding strength of the chromomagnetic condensate turns out to be $\sqrt{gH} \, \simeq \, 0.4 \, MeV$. \\
Recent {\it{CHANDRA}} observations~\cite{Baganoff:2003,Muno:2004} of diffuse emission around the galactic center have
confirmed that {\it{SgrA}}$^*$ is underluminous in $X$-ray by a factor of about $10^7 - 10^8$ compared to the standard
thin accretion disk model. The low luminosity of {\it{SgrA}}$^*$ may be explained within the standard paradigm either by
accretion at a rate far below the estimate Bondi rate, or accretion at the Bondi rate of gas that is radiating very
inefficiently. Since the gas from the winds of the surrounding young massive stars should be able to maintain the
accretion rate at a sizeable fraction of the Bondi accretion rate, which has been estimate~\cite{Baganoff:2003} to be
about $10^{-6} \, M_{\bigodot} \, years^{-1}$, it is believed that only the latter possibility is  viable. However, in
Ref.~\cite{Nayakshin:2004}, using close stars as probe of accretion flow in {\it{SgrA}}$^*$, it has been pointed out that
non radiative accretion flows are constrained to accretion rates no larger than $10^{-7} \,  M_{\bigodot} \, years^{-1}$.
It is not yet clear if this constrain could be reconciled with the accretion rate estimate in Ref.~\cite{Baganoff:2003}.
Furthermore, Zhao {\it et al.}~\cite{Zhao:2001} reported the presence of 106 days cycle variability at centimeter
wavelengths in the radio flux density of {\it{SgrA}}$^*$. This peculiar periodicity has been confirmed by observations at
millimeter wavelengths~\cite{Miyazaki:2003}. The very low $X$-ray luminosity and the periodicity in the  flux density of
{\it{SgrA}}$^*$ look  puzzling within the standard interpretation based on accreting black holes, while these can be
accounted for if we assume that {\it{SgrA}}$^*$ is a super massive p-star. Indeed, the periodicity is naturally explained
assuming a rotation period $P \, \simeq 106 \,
 days \, \simeq 9.2 \, 10^6 \, sec$. Moreover, from the strength of the chromomagnetic condensate and from
Eq.~(\ref{magn-condensate}) we estimate that the  dipolar surface magnetic field of  {\it{SgrA}}$^*$ is reduced by about
$10^{-6}$ with respect to  pulsar magnetic fields. So that  $B_{SgrA^*}$ should lie in the range $10^6 - 10^9 \, Gauss$.
From $B_{SgrA^*} \, \lesssim \, 10^9 \, Gauss$, we infer for the age of {\it{SgrA}}$^*$  $\tau \, \sim \, 10^{10} years$,
i.e. {\it{SgrA}}$^*$ is a primordial p-star. Finally, the low $X$-ray quiescent luminosity could be interpreted as thermal
emission from
the stellar surface. \\
As discussed in Refs.~\cite{cea:2003,cea:2004} p-stars are compact stars made of up and down quarks in $\beta$-equilibrium
with electrons in an abelian chromomagnetic condensate. It turns out that these compact stars are more stable than both
neutron stars and strange stars whatever the value of the  chromomagnetic condensate $\sqrt{gH}$. In other words, p-stars,
once formed, are absolutely stable. The logical consequence is that now  we must admit that  supernova explosions give
rise to p-stars. In other words, we are lead to identify pulsars with p-stars instead of neutron or strange stars. Such a
dramatic change in the standard paradigm of relativistic astrophysics has been already advanced in our previous
paper~\cite{cea:2003} where we suggested that, if we assume that pulsars are p-stars, then we could  solve the supernova
explosion problem. As is well known, the binding energy is the energy released when the core of an evolved massive star
collapses. Actually, only about one percent of the energy appears as kinetic energy in the supernova
explosion~\cite{bethe:1990}. Now, in Ref.~\cite{cea:2003} we showed that there is an extra gain in kinetic energy of about
$ \, 1 \, - \, 10 \, foe$ ($1 \, foe \, = \, 10^{51} \, erg$), which is enough to fire the supernova explosions. Further
support to our theory comes from cooling properties of p-stars. In fact, we found that
cooling curves of p-stars compare rather well with available observational data. \\
In our previous papers~\cite{cea:2003,cea:2004} we showed that p-stars are also able to account for compact stars with $R
\, \lesssim 6 \, Km$. In particular, we  convincingly  argued that the nearest isolated pulsars {\it {RXJ 1856.5-3754}}
and {\it {RXJ 0720.4-3125}} (for a recent review see~\cite{haberl:2003,pavlov:2003}) are  p-stars. From the $X$-ray
emission spectrum we argued that the most realistic interpretation is that these objects are compact p-stars with $M \,
\simeq 0.8 \, M_{\bigodot}$ and $R \, \simeq 5 \, Km$. However, it should be stressed that in the observed spectrum there
is also an optical emission  in excess over the extrapolated $X$-ray blackbody. By interpreting the optical emission as a
Rayleigh-Jeans tail of a thermal blackbody emission, one finds that the optical data are also fitted by the blackbody
model yielding an effective radius $R^\infty \, > \, 16 \, Km \, \frac{d}{120 \, pc}$~\cite{burwitz:2002}. However,
interestingly enough, quite recently the distance measurement of {\it {RXJ 1856.5-3754}} has been reassessed and it is now
estimated to be at $180 \; pc$ instead of $120 \; pc$~\cite{kaplan:2003}. This new determination of the distance of {\it
{RXJ 1856.5-3754}} rules out the two blackbody interpretation of the spectrum, for this model leads now to an effective
radius $R^\infty \, > \, 24 \, Km$, which is too large for a neutron star. Thus, the new determination of the distance of
{\it {RXJ 1856.5-3754}} strongly supports our p-star theory, and indicates clearly that the faint optical emission
originates in the magnetosphere.  In general, it remarkable that isolated $X$-ray pulsars do display a faint soft
emission, in excess over the extrapolated $X$-ray thermal emission, which is best fitted by a non thermal power law. The
origin of this faint emission is puzzling. Nevertheless, our previous considerations point toward a general mechanism in
the magnetosphere responsible for the faint emission. This problem, which to the best of our knowledge has never addressed
before, is thoroughly analyzed in Appendix, where we show that a subtle quantum mechanical effect related to strong enough
magnetic fields in the polar cap regions leads to a faint non thermal power law soft emission. \\
Quite recently it has been proposed that the compact accreting object in the famous $X$-ray binary  {\it { Herculses
$X$-1}} is a strange star~\cite{Li:1995}. This proposal was based on the comparison of a phenomenological  mass-radius
relation for {\it { Herculses $X$-1}}~\cite{shapiro:1983} with theoretical $M-R$ curves for neutron and strange stars. The
analysis in Ref.~\cite{Li:1995} has, however, been criticized by the authors of Ref.~\cite{Reynolds:1997}.  These authors,
using a new mass estimate together with a revised distance, which leads to a somewhat higher $X$-ray luminosity, argued
that the hypothesis that  {\it { Herculses $X$-1}} is a neutron star is not disproved. As a matter of fact, the authors of
Ref.~\cite{Reynolds:1997} found that there is marginal consistency with observations if one adopts for neutron stars a
very soft equation of state. At the same time, these authors pointed out that the hypothesis of a strange star can be
ruled out since the theoretical  curves no longer intercept the observational relations within the permitted mass range.
Recent observations of millisecond pulsars in  the globular cluster {\it { Terzan~5}} using the Green Bank
Telescope~\cite{Ransom:2005} indicated that al least one of the pulsar is more massive than $M \simeq 1.7 \,
M_{\bigodot}$. Even more, there is emerging observational evidences for pulsars with mass in excess of $1.6 \,
M_{\bigodot}$. For instance, the pulsar in {\it {Vela $X$-1}} has mass $M = 1.86 \pm 0.16 \,
M_{\bigodot}$~\cite{Barziv:2001,Quaintrell:2003}. The very existence of such massive pulsars constrains the equation of
state of matter in neutron stars. In fact it seems that soft equations of state are almost certainly ruled out. As a
consequence we infer that the compact accreting pulsar in {\it { Herculses $X$-1}} cannot be a strange star nor a neutron
star. On the other hand, theoretical $M-R$ curves for p-stars are compatible with the phenomenological mass-radius
relation for {\it { Herculses $X$-1}}. Indeed, we find that the pulsar in {\it { Herculses $X$-1}} could be described by a
p-star with $M \simeq 1.5 \,
M_{\bigodot}$, $R \, \simeq  6.5 \, Km$, and  $\sqrt{gH} \, \simeq \, 0.62 \, GeV$. \\
In the present paper we  investigate the properties of p-stars with super strong surface magnetic field. As we shall show,
these p-stars allow us reach a complete understanding of several puzzling observational aspects of anomalous $X$-ray
pulsars (AXPs) and soft gamma-ray repeaters (SGRs). Anomalous $X$-ray pulsars and soft gamma-ray repeaters are two class
of intriguing objects that in our opinion are challenging the standard paradigm based on neutron stars. For a recent
review on the observational properties of anomalous $X$-ray pulsars see
Refs.~\cite{Mereghetti:1999,Mereghetti:2002,Kaspi:2004a}, for soft gamma-ray repeaters see
Refs.~\cite{Hurley:1999,Woods:2003}. Recently, these two groups have been linked by the discovery of persistent emission
from soft gamma ray repeaters that is very similar to anomalous $X$-ray pulsars, and bursting activity in anomalous
$X$-ray pulsars quite similar to  soft gamma ray repeaters (see, for instance Refs.~\cite{Kaspi:2004b,Woods:2004}). \\
Duncan and Thompson~\cite{Duncan:1992} and Paczy\'nski~\cite{Paczynski:1992} have proposed that soft gamma-ray repeaters
are pulsars whose surface magnetic fields exceed the $QED$ critical magnetic field:
\begin{equation}
\label{crit-magn}
B_{QED} \; \; = \; \;\frac{m_e^2}{e} \; \; \simeq \; \; 4.4 \; 10^{13} \; \; Gauss \; \; .
\end{equation}
Indeed, Duncan and Thompson in Ref.~\cite{Duncan:1992}  refer to these pulsar as {\it{magnetars}}. In particular Duncan
and Thompson~\cite{Duncan:1995,Duncan:1996} argued that the soft gamma-ray repeater bursts and quiescent emission were
powered by the decay of an ultra-high magnetic field. This interpretation is based on the observations that showed that
these peculiar pulsars are slowing down rapidly, with an inferred magnetic dipole field much greater than the quantum
critical field $B_{QED}$, while producing steady emission at a rate far in excess of the rotational kinetic energy loss.
The identification of anomalous $X$-ray pulsars with magnetars was more recently supported by the similarity of anomalous
$X$-ray pulsar emission to that of soft gamma-ray repeaters in quiescence. This was confirmed by the detection of SGR-like
bursts from two anomalous $X$-ray pulsars~\cite{Kaspi:2004b,Woods:2004}. \\
In the standard neutron star theory, magnetars ought to be born with millisecond initial period to ensure vigorous dynamo
process to occur~\cite{Duncan:1993}. This mechanism should generate huge surface dipolar magnetic field up to $10^{15} \;
Gauss$, and even stronger interior fields. However, strong magnetic fields in excess of the critical field $B_{QED}$ would
squeeze electrons into the lowest Landau levels. In this conditions, the electron gas pressure transverse to the magnetic
field may vanish leading to a transverse collapse of the star~\cite{Chaichian:1999}. On the other hand, p-stars do not
share this stability problem. In fact, Eq.~(\ref{magn-condensate}) shows that the dipolar magnetic field in p-star is a
tiny effect with respect to the chromomagnetic condensate. Moreover, the stability of p-stars is due to the quark
pressure, while the electron pressure is almost completely negligible. Indeed, from Eq.~(\ref{magn-condensate}) it follows
that canonical p-stars could support dipolar magnetic fields up to $10^{16} \; Gauss$. As we said before, in standard
magnetars appropriate conditions for true dynamo mechanism could exist if neutron star is born with a very short period.
However, up to now there is no direct observational evidences for such short initial periods  in radio pulsars or in young
supernova remnants. On the contrary, the peculiar mechanism to generate dipolar magnetic fields in p-stars indicates that
huge magnetic fields require rather large initial period.  It is worthwhile to stress that our mechanism for the
generation of dipolar magnetic fields in pulsars solves in a natural way the puzzling discrepancy between characteristic
ages and true ages
which is displayed by at least two anomalous $X$-ray pulsars. \\
As it is well known, the pulsar characteristic ages is defined as:
\begin{equation}
\label{characteristic}
\tau_c \; = \; \frac{P}{2 \, \dot{P}} \; ,
\end{equation}
while the true age is given by:
\begin{equation}
\label{age}
 \tau\; = \; \frac{P}{2 \, \dot{P}} \; \; \left [ 1 \, - \,
 \left ( \frac{P_0}{P}\right )^2 \right ]  \; \; \; .
\end{equation}
The true age can be significantly smaller than  $\tau_c$ if the initial period $P_0$ is close to $P$. \\
Both anomalous $X$-ray pulsars 1E 1841-045 and 1E 2259+586 have  $\tau_c$ greater than the ages estimated from their
respective supernova remnants (see Table~1 in Ref.~\cite{Mereghetti:2002}). In particular, for  1E 2259+586 the
characteristic age $\tau_c$ is more than one order of magnitude larger than the age of SNR G109.1-0.1. We may solve this
large discrepancy if we assume the initial period $P_0 \simeq 6.93 \; sec$. In the case of 1E 1841-045 we find that $P_0
\simeq 10.6 \; sec$ reconciles the age discrepancy. Note that this solution cannot be adopted within the standard neutron
star theory, for huge magnetic fields can be generated only for stars born with very small initial period. On the other
hand, the peculiar mechanism to generate dipolar magnetic fields in p-stars favors initial strong magnetic fields in
slowly rotating stars. In the next
Section we shall present further phenomenological evidences in support of our point of view. \\
The main purpose of this paper is to discuss in details p-stars endowed with super strong dipolar magnetic field which,
following well consolidated tradition in literature, will be referred to as magnetars. In particular we will show that,
indeed, soft gamma-ray repeaters and anomalous $X$-ray pulsars can be understood within our theory. Whenever possibly, we
shall critically compare our theory with the standard paradigm based on neutron stars. The plan of the paper is as
follows. In Section~\ref{rotation} we discuss the phenomenological evidence for the dependence of pulsar magnetic fields
on the rotational period. We argue that there is a well defined criterion which allows us to distingue between rotation
powered pulsars and magnetic powered pulsars. We explicitly explain why the recently discovered high magnetic field radio
pulsars are indeed rotation powered pulsars. In Section~\ref{death}  we introduce the radio death line, which in the
$\dot{P} - P$ plane separated radio pulsars from radio quiet magnetic powered pulsars, and compare with available
observational data. Section~\ref{glitches}  is devoted to the glitch mechanism in our magnetars and their observational
signatures. In Section~\ref{braking} we compare glitches in SGR 1900+14 and 1E 2259+586, our prototypes for soft gamma-ray
repeater and anomalous $X$-ray pulsar respectively. Sections~\ref{quiescent} and \ref{bursts} are devoted to explain the
origin of the quiescent luminosity, the bursts activity and the emission spectrum during bursts. In Section~\ref{hardness}
we discuss the puzzling anti correlation between hardness ratio and intensity. In Section~\ref{light} we develop a general
formalism to cope with light curves for both giant and intermediate bursts. In Sections~\ref{2259} through \ref{1806} we
careful compare our theory with the available light curves in literature. In particular, we are able  to account for the
peculiar light curve following the June 18, 2002 giant burst from the anomalous $X$-ray pulsar 1E 2259+586. Finally, we
draw our conclusions in Section~\ref{conclusion}. In Appendix we face with  the problem of the origin of soft emission in
isolated $X$-ray pulsars. We argue that the soft emission originates in the magnetosphere and it can be ascribed to a
subtle quantum mechanical effect related to strong magnetic fields in the polar cap regions.
\section{\normalsize{ ROTATION VERSUS MAGNETIC POWERED PULSARS}}
\label{rotation}
In Ref.~\cite{cea:2003} we introduced  p-stars, a new class of compact quark stars made of almost massless deconfined up
and down quarks immersed in a chromomagnetic field in $\beta$-equilibrium. The structure of p-stars is determined once the
equation of state appropriate for the description of deconfined quarks and gluons in a chromomagnetic condensate is
specified. In particular, the chemical potentials satisfy the constrains arising from $\beta$- equilibrium and charge
neutrality:
\begin{equation}
\label{constr1}
 \mu_e \, + \, \mu_u \; = \; \mu_d  \; \; \; \; \;
\; \;  \beta - equilibrium
\end{equation}
\begin{equation}
\label{constr2}
 \frac{2}{3 } n_u \, - \, \frac{1}{3 } n_d  \; = \;
n_e  \; \; \; \;  charge \; neutrality
\end{equation}
where:
\begin{equation}
\label{number}
  n_u \, = \, \frac{1}{2 \pi^2} \, gH \, \mu_u  \,
,  \;  n_d \, = \, \frac{1}{2 \pi^2} \, gH \, \mu_d  \, , \;  n_e \, = \, \frac{\mu_e^3}{3 \pi^2} \, .
\end{equation}
\begin{figure}[t]
\includegraphics[width=0.9\textwidth,clip]{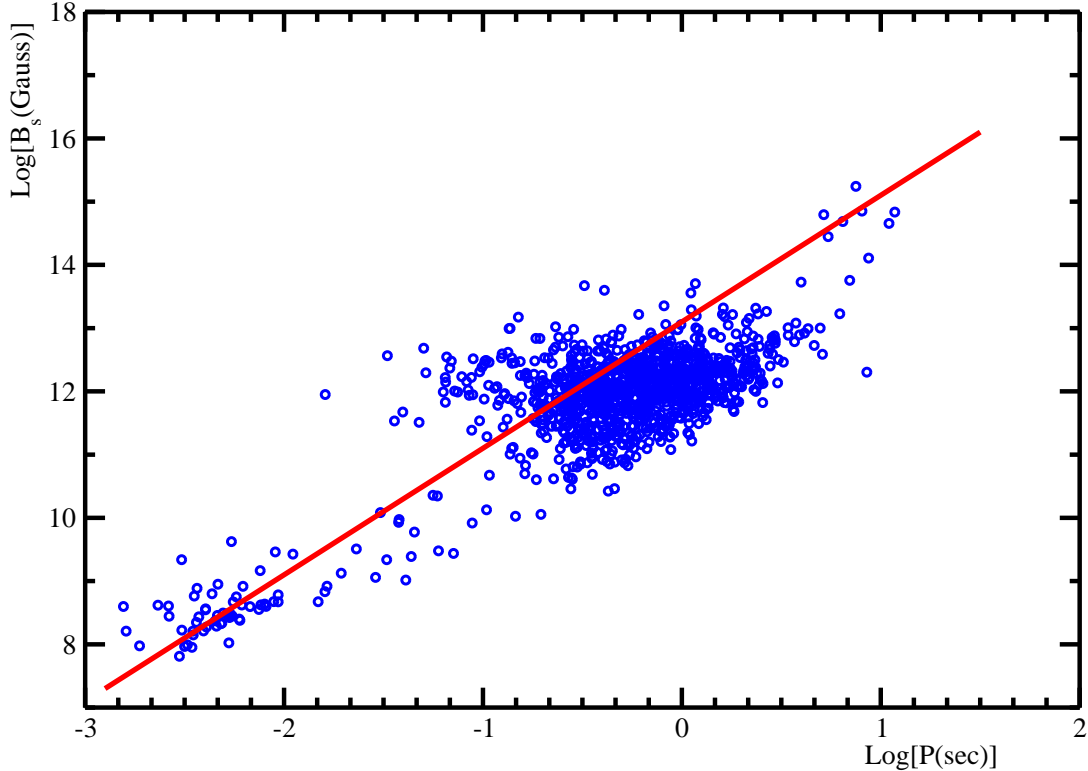}
\caption{\label{fig_1}
Inferred magnetic field $B_S$  plotted versus stellar period for 1194 pulsars taken from the ATNF Pulsar
Catalog~\cite{ATNF}. Full red line corresponds to Eq.~(\ref{magn-period}) with $B_1 \, \simeq \, 1.3 \, 10^{13} \;
Gauss$.}
\end{figure}
From previous equations it follows that $ \mu_d \, > \mu_u \, > \, \mu_e$. Moreover, it turns out that the chemical
potentials are monotonic decreasing smooth functions of the distance from the center of the star.  In general, the quark
chemical potentials $\mu_d$ and $\mu_u$ are smaller that the strength of the chromomagnetic condensate  $\sqrt{gH}$. So
that, up and down quarks occupy the lowest Landau levels.  However, for certain values of the central energy  density it
happens that $\frac{\mu_d}{\sqrt{gH}} \, \gtrsim \, 1$ in the stellar core. Thus, a fraction of down quarks must jump into
higher Landau levels leading to a central core with energy density $\varepsilon_c$ somewhat greater than the energy
density outside the core. Now, these quarks in the inner core produce a vector current in response to the chromomagnetic
condensate. Obviously, the quark current tends to screen the chromomagnetic condensate by a very tiny amount. However,
since the down quark has an electric charge $q_d  =  - \frac{1}{3} \, e$, the quark current generates in the core a
uniform magnetic field parallel to the chromomagnetic condensate with strength:
\begin{equation}
\label{magn-core}
B_c \; \; \simeq \; \; \frac{e}{96 \, \pi }  \; \; gH\; \;  .
\end{equation}
Outside the core the magnetic field is dipolar leading to surface magnetic field given by  Eq.~(\ref{magn-condensate}). In
general the formation of the inner core denser than the outer core is contrasted by the centrifugal force produced  by
stellar rotation. Since the centrifugal force is proportional to the square of the stellar rotation frequency, this leads
us to argue that the surface magnetic field strength is proportional to the square of the stellar period:
\begin{equation}
\label{magn-period}
B_S\; \; \simeq \; \; B_1 \; \left (\frac{P}{1 \, sec } \right )^2 \; \; \; \; ,
\end{equation}
where $B_1$ is the surface magnetic field for pulsars with nominal period $P \, = \, 1 \; sec$. Indeed, in
Fig.~\ref{fig_1} we we display the surface  magnetic field strength $B_S$ inferred from (for instance, see
Ref.~\cite{manchester:1977}):
\begin{equation}
\label{magn-surf}
B_S\; \; \simeq \; \; 3.1 \; 10^{19} \; \; \sqrt{P \;\dot{P} }  \; Gauss \; \; ,
\end{equation}
versus the period. Remarkably, assuming $B_1 \, \simeq \, 1.3 \, 10^{13} \; Gauss$, we find the Eq.~(\ref{magn-period})
accounts rather well the inferred magnetic field for pulsars ranging from millisecond pulsars up to anomalous $X$-ray
pulsars and soft-gamma repeaters.
As a consequence of Eq.~(\ref{magn-period}), we see that the dipolar magnetic field is time dependent. In fact, it is easy
to find:
\begin{equation}
\label{mag-time}
 B_S(t) \; \simeq \;  B_0 \; \; \left ( 1 \; + \; 2 \; \frac{\dot{P}}{P} \; t \;  \right )\;
 \;  \; \; ,
\end{equation}
where $B_0$ indicates the magnetic field at the initial observation time. Note that Eq.~(\ref{mag-time}) implies that the
magnetic field varies on a time scale given by the characteristic age. Equation~(\ref{mag-time}) leads to remarkable
consequences discussed in Ref.~\cite{cea:2004a}. Indeed, in Ref.~\cite{cea:2004a}, starting from
Equation~(\ref{mag-time}), we discussed a general mechanism which allows to explain naturally both radio and high energy
emission from pulsars. We also discuss the plasma distribution in the region surrounding the pulsar, the pulsar wind and
the formation of jet along the magnetic axis. We also suggested a plausible mechanism to generate pulsar proper motion
velocities. In particular, in our emission mechanism there is a well defined geometric  mapping between frequency and
distance from the star which seems to be consistent with observations. In particular, in the recently detected binary
radio pulsar system {\it{J0737-3039}}~\cite{Burgay:2003,Lyne:2004} it has been reported~\cite{McLaughlin:2004} the
detection of features similar to drifting subpulses with a fluctuation frequency which is exactly the beat frequency
between the period of the two pulsars. This direct influence of the electromagnetic radiation from one pulsar on the
electromagnetic emission from the other pulsar can be naturally accounted for within our emission mechanism due to the
geometric  mapping between emission frequencies and distances. On the other hand, that effect cannot easily reconciled
with the generally accepted model for pulsar radio emission which involves coherent radiation from very energetic
$e^+-e^-$ pairs. \\
It is widely accepted that pulsar radio emission is powered by the rotational energy:
\begin{equation}
\label{ener-rot}
E_{R} \; = \;  \frac{1}{2} \; I \; \; \omega^2 \; \; \; ,
\end{equation}
so that, the spin-down power output is given by:
\begin{equation}
\label{ener-rot-dot}
- \; \dot{E}_{R} \; = \; - \; I \; \; \omega \; \dot{\omega} \; \; = \; 4 \;
 \pi^2 \; I \; \frac{\dot{P}}{P^3} \; .
\end{equation}
On the other hand, an important source of energy is provided by the magnetic field. Indeed, the classical energy stored
into the magnetic field is:
\begin{equation}
\label{ener-mag}
 E_{B} \; = \; \frac{1}{8 \, \pi} \;  \int_{r \, \geq \, R} \; \; d^3r \; B^2(r) \; \;  \; \; ,
\end{equation}
Assuming a dipolar magnetic field:
\begin{equation}
\label{dip-mag}
 B(r) \; = \;  B_S \; \; \left ( \frac{R}{r} \right )^3 \; \; for \; \; \;
 r \, \geq \, R \; \;  \; \; ,
\end{equation}
Eq.~(\ref{ener-mag}) leads to:
\begin{equation}
\label{ener-mag-dip}
 E_{B} \; = \; \frac{1}{6 } \; B_S^2 \; \; R^3 \; \; .
\end{equation}
Now, from Eq.~(\ref{mag-time}) the surface magnetic field  is time dependent. So that, the magnetic power output is given
by:
\begin{equation}
\label{ener-mag-dot}
 \dot{E}_{B} \; = \; \frac{2}{3} \;  B_0^2 \; \; R^3 \; \;
  \; \frac{\dot{P}}{P} \; .
\end{equation}
For rotation-powered pulsars it turns out that $ \dot{E}_{B} \; \ll \; - \, \dot{E}_{R}$. However, if the dipolar magnetic
field is strong enough, then the magnetic power Eq.~(\ref{ener-mag-dot}) can be of the order, or even greater than the
spin-down power. Thus, we may formulate a well defined criterion to distinguish rotation  powered pulsars from  magnetic
powered pulsars. Indeed, until $ \dot{E}_{B} \; < \; - \, \dot{E}_{R}$ there is enough rotation power to sustain the
pulsar emission. On the other hand, when $ \dot{E}_{B} \; \geq \; - \, \dot{E}_{R}$ all the rotation energy is stored into
the increasing magnetic field and the pulsar emission is turned off. In fact, in the next Section we will derive the radio
death line, which is the line that in the $P-\dot{P}$ plane separates rotation-powered pulsars from magnetic-powered
pulsars. In the remainder of this Section, we discuss the recently detected radio pulsar with very strong surface magnetic
field. We focus on the two radio pulsars with the strongest surface magnetic field: {\it {PSR J1718-3718}} and {\it {PSR
J1847-0130}} . These pulsar have inferred surface magnetic fields well above the quantum critical field $B_{QED}$ above
which some models~\cite{Baring:1998} predict that radio emission should not occur. In particular, we have:
\begin{equation}
\label{1718-1847}
\begin{split}
  PSR J1718-3718~\cite{Hobbs:2004} &  \; \; \; \; \; \; \; P \; \simeq \; \; \; 3.4 \; \; sec  \; \; , \; \;
      B_S \, \simeq \, 7.4 \; 10^{13} \; Gauss \; \; , \\
  PSR J1847-0130~\cite{McLaughlin:2003}  &  \; \; \; \; \; \; \; P \; \simeq \; \; \; 6.7 \; \;  sec  \; \; , \; \;
    B_S \, \simeq \, 9.4 \; 10^{13} \; Gauss \; \; .
\end{split}
\end{equation}
Both pulsars have average radio luminosities and surface magnetic fields larger than that of {\it {AXP 1E 2259+586}}. Now,
using Eqs.~(\ref{ener-mag-dot}), (\ref{ener-rot-dot}) , together with $I = \frac{2}{5} \, M \, R^2$, and
Eq.~(\ref{1718-1847}) we get:
\begin{equation}
\label{1718-1847-power}
\begin{split}
  PSR J1718-3718 &  \; \; \; \;  - \; \dot{E}_{R} \;  \simeq \; 3.4 \; 10^{45} \;  erg \; \;  \;
  \frac{\dot{P}}{P} \; , \; \dot{E}_{B} \;  \simeq \; 4.7 \; 10^{44} \;  erg \; \;  \;
  \frac{\dot{P}}{P} \;  \; , \\
  PSR J1847-0130  &  \; \; \; \;  - \; \dot{E}_{R} \;  \simeq \; 8.8 \; 10^{44} \;  erg \; \;  \;
  \frac{\dot{P}}{P} \; , \; \dot{E}_{B} \;  \simeq \; 7.5 \; 10^{44} \;  erg \; \;  \;
  \frac{\dot{P}}{P}  \; \; .
\end{split}
\end{equation}
We see that in any case: $ \dot{E}_{B} \;  < \;  - \dot{E}_{R}$, so that there is enough rotational energy to power the
pulsar emission.

\section{\normalsize{RADIO DEATH LINE}}
\label{death}
As discussed in previous Section, until $ \dot{E}_{B} \; < \; - \, \dot{E}_{R}$ the  rotation power loss  sustains the
pulsar emission. We  have already shown that this explain the otherwise puzzling pulsar activity for pulsars with inferred
magnetic fields well above the critical field $B_{QED}$. Even more, the two pulsars {\it {PSR J1718-3718}} and {\it {PSR
J1847-0130}} have magnetic fields which are well above the magnetic field of the anomalous $X$-ray pulsar {\it {AXP 1E
2259+586}}. Up to now, it was unclear how these high-field pulsars and anomalous $X$-ray pulsars can have such similar
spin-down parameters but vastly different emission properties. We have offered a natural explanation for the pulsar
activity for high-field radio pulsars. In this Section we explain why anomalous $X$-ray pulsars and soft gamma repeaters
are radio quiet pulsars.
\begin{figure}[th]
\includegraphics[width=0.9\textwidth,clip]{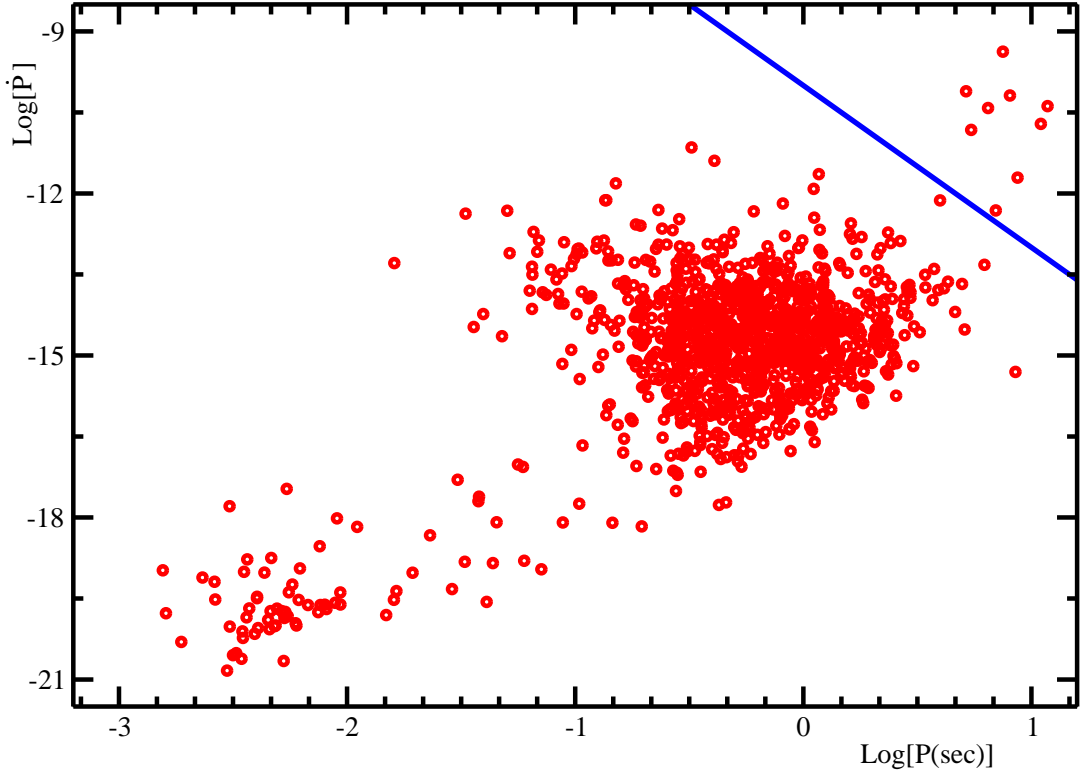}
\caption{\label{fig_2}
Period derivative  plotted versus stellar period for 1194 pulsars taken from the ATNF Pulsar Catalog~\cite{ATNF}. Full
blue line corresponds to Eq.~(\ref{death-log})}
\end{figure}
When $ \dot{E}_{B} \; \geq \; - \, \dot{E}_{R}$ all the rotation energy is stored into the increasing magnetic field and
the pulsar emission is turned off. As a consequence pulsars with strong enough magnetic fields are radio quiet.
Accordingly we see that the condition:
\begin{equation}
\label{radio-death}
 \dot{E}_{B} \; = \; - \; \dot{E}_{R} \;  \; .
\end{equation}
is able to distinguish rotation powered pulsars from magnetic powered pulsars. Now, using~\cite{manchester:1977}
\begin{equation}
\label{mag-brake}
 B_S \; \simeq \; \sqrt{\frac{3 \, I \, P \, \dot{P}}{8 \, \pi^2 \, R^6}}  \; \; \; ,
\end{equation}
we recast Eq.~(\ref{radio-death}) into:
\begin{equation}
\label{r-death}
 P^3 \, \dot{P} \; = \; 16 \, \pi^4 \, R^3   \; \; \; .
\end{equation}
Using the canonical radius $R \simeq 10 \, Km$, we get:
\begin{equation}
\label{death-log}
 3 \; \log (P) \, + \, \log (\dot{P})  \; \simeq  \; - \, 10  \; \; \; .
\end{equation}
Equation~(\ref{death-log}) is a straight line,  plotted in Fig.~\ref{fig_2}, in the $\log(P)-\log(\dot{P})$ plane. In
Figure.~\ref{fig_2} we have also displayed 1194 pulsars taken from the ATNF Pulsar Catalog~\cite{ATNF}. We see that
rotation powered pulsars, ranging from millisecond pulsars up to radio pulsars, do indeed lie below our
Eq.~(\ref{death-log}). Note that in Fig.~\ref{fig_2} the recently detected high magnetic field pulsars are not included.
However,  we have already argued in previous Section that these pulsars have spin parameters which indicate that these
pulsars are rotation powered. On the other hand, Fig.~\ref{fig_2} shows that all soft gamma-ray repeaters and anomalous
$X$-ray pulsars in the  ATNF Pulsar Catalog lie above our line Eq.~(\ref{death-log}). In particular, in Fig.~\ref{fig_2}
the pulsar above and nearest to the line Eq.~(\ref{death-log}) corresponds to {\it {AXP 1E 2259+586}}. So that, we see
that our radio dead line, Eq.~(\ref{death-log}), correctly predicts that {\it {AXP 1E 2259+586}} is not a radio pulsar
even though the magnetic field is lower than that in radio pulsars {\it {PSR J1718-3718}} and {\it {PSR J1847-0130}}. We
may conclude that pulsars above our dead line are magnetars, i.e. magnetic powered pulsars. The  emission properties of
magnetars are quite different from rotation powered pulsars. The emission from magnetars  consists in thermal blackbody
radiation form the surface. In addition, it could eventually also be present a faint power-law emission superimposed to
the thermal radiation. As discussed in Appendix, this  soft faint emission is caused by the thermal radiation reprocessed
in the magnetosphere. In Ref.~\cite{cea:2004} we suggested that {\it {RXJ 1856.5-3754}}  is exactly in this state. On the
other hand, the energy stored into the magnetic field can be released if the star undergoes a glitch. Indeed, as
thoroughly discussed in the next Section, glitches originate from dissipative effects in the inner core of the star
leading to a decrease of the strength of the core magnetic field. So that, soon after the glitch there is a release of
magnetic energy. We have already suggested in Ref.~\cite{cea:2004} that this picture is consistent with the long-term
variability in the $X$-ray emission of {\it {RXJ 0720.4-3125}}. Remarkably, a recent timing analysis of the isolated
pulsar {\it {RXJ 0720.4-3125}} performed in Ref.~\cite{Cropper:2004} suggested that, among different possibilities,
glitching may have occurred in this pulsar.
\section{\normalsize{GLITCHES IN MAGNETARS}}
\label{glitches}
\begin{figure}[th]
\includegraphics[width=0.9\textwidth,clip]{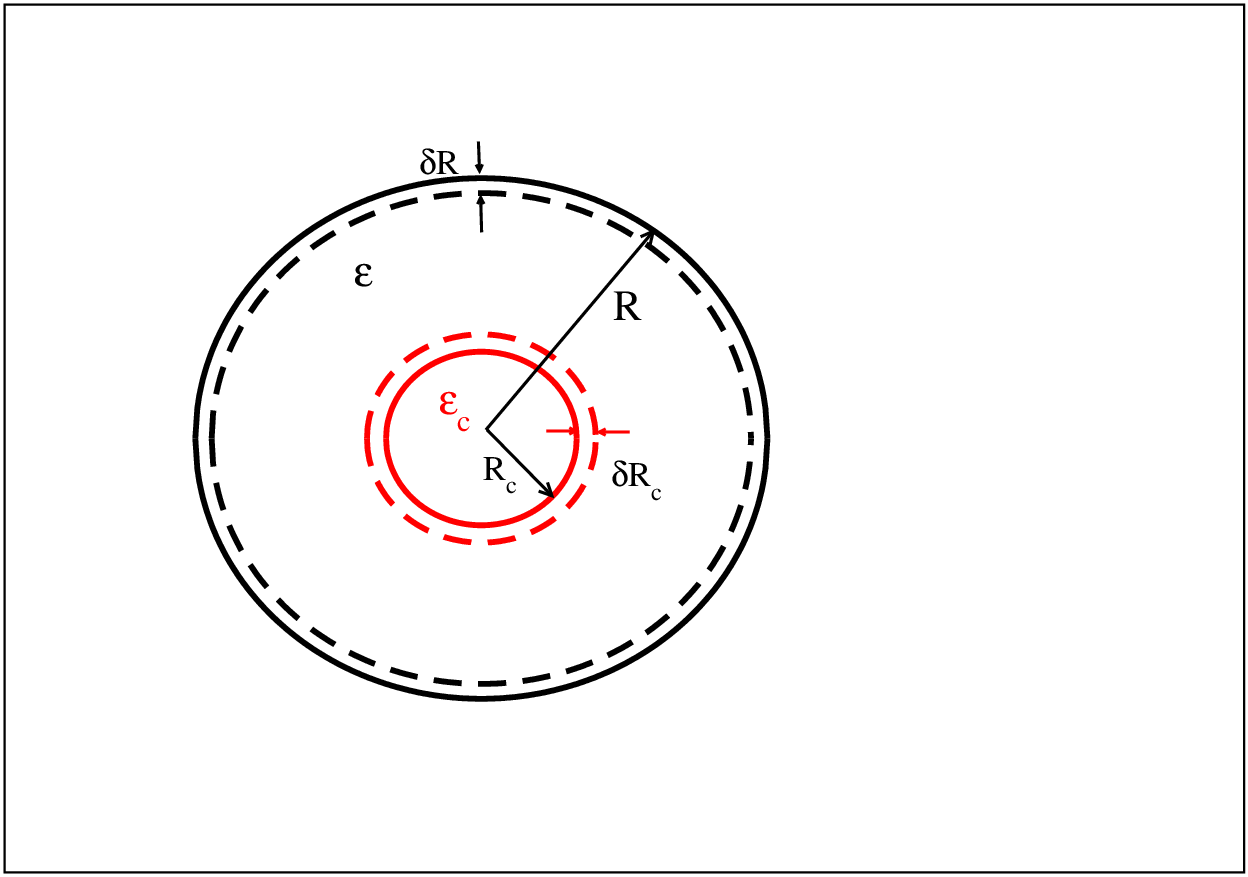}
\caption{\label{fig_3}
Schematic view of the interior of a p-star. $R_c$ and $R$ represent the inner core and stellar radii respectively;
$\varepsilon_c$ is the energy density of the inner core, $\varepsilon$ is the energy density outside the core.}
\end{figure}

In  p-stars there is a natural mechanism to generate a dipolar magnetic field, namely the generation of the dipolar
magnetic field is enforced by the formation of a dense inner core composed mainly by down quarks. A quite straightforward
calculation, which will be presented elsewhere, leads to the conclusion that down quarks in the inner core produce a
vector current in response to the chromomagnetic condensate. This quark current, in turn,  generates in the core a uniform
magnetic field parallel to the chromomagnetic condensate with strength given by Eq.~(\ref{magn-core}). Outside the core
the magnetic field is dipolar. Now, we note that the inner core is characterized by huge conductivity, while outer core
quarks are freezed into the lowest Landau levels. So that, due to the energy gap between the lowest Landau levels and the
higher ones, the quarks outside the core cannot support any electrical current. As a consequence, the magnetic field in
the region outside the core is not screened leading to our previous Eq.~(\ref{magn-condensate}). For later convenience,
after taking into account Eq.~(\ref{magn-core}), we rewrite Eq.~(\ref{magn-condensate}) as:
\begin{equation}
\label{magn-core-surf}
B_S\; \; \simeq \; \;B_c \; \; \left (\frac{R_c}{R } \right )^3 \;  \; ,  \;  \; B_c \; \simeq \; \; \frac{e}{96 \, \pi }
\; \; gH\; \; .
\end{equation}
In Figure~\ref{fig_3} we display a schematic view of the interior of a p-star. \\
The presence of the inner core uniformly magnetized leads to well defined glitch mechanism in p-stars. Indeed, dissipative
effects, which are more pronounced in young stars, tend to decrease the strength of the core magnetic field. On the other
hand, when $B_c$ decreases due to dissipation effects, then the magnetic flux locally decreases and, according to Lenz's
law, induces a current which resists to the reduction of the magnetic flux. This means that some quarks must flow into the
core by jumping onto higher Landau levels. In other words, the core radius must increase. Moreover, due to very high
conductivity of  quarks in the core, we have:
\begin{equation}
\label{magn-flux}
B_c \; \; R_c^2  \simeq \; constant \; ,
\end{equation}
which implies:
\begin{equation}
\label{magn-var-1}
\frac{\delta \, B_c}{B_c} \; + \; 2 \; \frac{\delta \, R_c}{R_c} \; \simeq \; 0 \; ,
\end{equation}
or
\begin{equation}
\label{magn-var-2}
\frac{\delta \, R_c}{R_c} \; \simeq \; - \; \frac{1}{2} \; \frac{\delta \, B_c}{B_c}  \; .
\end{equation}
Equation~(\ref{magn-var-2}) confirms that to the decrease of the core magnetic field, $\delta B_c < 0$, it corresponds an
increase of the inner core radius $\delta R_c > 0$. This sudden variation of the radius of the inner core leads to
glitches. Indeed, it is straightforward to show that  the magnetic moment:
\begin{equation}
\label{mag-mom}
 \emph{m} \; = \; B_S \; R^3  \; =  \; \; B_c \; R_c^3  \;
\end{equation}
where we used Eq.~(\ref{magn-core-surf}), must increase in the glitch.  Using Eq.~(\ref{magn-var-2}), we get:
\begin{equation}
\label{mom-var}
\frac{\delta \, \emph{m}}{\emph{m}} \; =  \; \frac{\delta \, B_c}{B_c} \; + \; 3 \; \frac{\delta \, R_c}{R_c} \; \simeq \;
 \frac{\delta \, R_c}{R_c} \; > \;0 \; ,
\end{equation}
Another interesting consequence of the glitch is that the stellar radius $R$ must decrease, i.e. the star contracts. This
is an inevitable consequence of the increase of the inner core, which is characterized by an energy  density
$\varepsilon_c$ higher then the outer core density  $\varepsilon$. As a consequence the  variation of radius is negative:
$\delta R < 0$ (see Fig.~\ref{fig_3} ). In radio pulsar, where the magnetic energy can be neglected, conservation of the
mass leads to:
\begin{equation}
\label{rad-var-1}
\frac{\delta \, R}{R} \; \simeq \; - \; \frac{\varepsilon_c - \varepsilon}{\overline{\varepsilon}}  \; \left  ( \frac{R_c
}{ R} \right )^3
 \;   \frac{\delta \, R_c}{R_c} \; ,
\end{equation}
where $\overline{\varepsilon}$ is the average energy density. In general, we may assume that $\frac{\varepsilon_c -
\varepsilon}{\overline{\varepsilon}}$ is a constant of order unity. So that Eq.~(\ref{rad-var-1}) becomes:
\begin{equation}
\label{rad-var-2}
\frac{\delta \, R}{R} \; \simeq \; - \;  \left ( \frac{R_c }{ R} \right )^3 \;  \frac{\delta \, R_c}{R_c} \; .
\end{equation}
Note that the ratio $ \left  ( \frac{R_c }{ R} \right )^3$ can be estimate from Eq.~(\ref{magn-core-surf}) once the
surface magnetic field is known. We find that, even for magnetars,  $\left  ( \frac{R_c }{ R} \right )^3$ is of order
$10^{-2}$ or less. So that our Eqs.~(\ref{magn-core-surf}) and~(\ref{rad-var-2}) show that:
\begin{equation}
\label{var-rad-cor}
 \delta R \; < \;  0 \; \; \; ,  \; \; \; - \; \frac{\delta \, R}{R} \; \ll \;
 \frac{\delta \, R_c}{R_c} \; \; .
\end{equation}
As is well known, because the external magnetic braking torque, pulsars slow down according to (e.g. see
Ref.~\cite{manchester:1977}):
\begin{equation}
\label{brak-eq}
 - \; \dot{\nu} \; \propto \;   \emph{m}^2 \; I^{-1} \; \nu^3 \; .
\end{equation}
So that:
\begin{equation}
\label{delta-brak-eq}
 \frac{\delta \dot{\nu} }{\dot{\nu}} \; = \; 2 \; \frac{\delta \emph{m}}{ \emph{m}}  \; - \; \frac{\delta I}{I} \;
 + \; 3 \;  \frac{\delta \nu}{\nu} \; \;  .
\end{equation}
From conservation of angular momentum we have:
\begin{equation}
\label{i-nu}
| \frac{\delta I}{I} | \; \; \simeq \; \;
 | \frac{\delta \nu}{\nu} | \; \;  .
\end{equation}
Moreover, from observational data it turns out that:
\begin{equation}
\label{nudot-nu}
| \frac{\delta  \dot{\nu}}{ \dot{\nu}} | \; \; \gg \; \;
 | \frac{\delta \nu}{\nu} | \; \;  ,
\end{equation}
so that Eq.~(\ref{delta-brak-eq}) becomes:
\begin{equation}
\label{delta-brak-eq-2}
 \frac{\delta \dot{\nu} }{\dot{\nu}} \; \simeq \; 2 \; \frac{\delta \emph{m}}{ \emph{m}}  \; \simeq \;
 2 \; \frac{\delta \, R_c}{R_c} \; \simeq \; - \; \frac{\delta \, B_c}{B_c} \; \; ,
\end{equation}
where we used Eqs.~(\ref{mom-var}) and~(\ref{magn-var-2}). Equation~(\ref{delta-brak-eq-2}) does  show that the variation
of the radius of the inner core leads to a glitch. \\
In rotation powered pulsar, starting from Eq.~(\ref{rad-var-1}) one can show that $ | \frac{\delta \nu}{\nu} | \simeq |
\frac{\delta \, R}{R} |$. So that, taking into account Eq.~(\ref{var-rad-cor}) we recover the phenomenological relation
Eq.~(\ref{nudot-nu}). A full account of glitches in radio pulsar will be presented elsewhere. Glitches in magnetars are
considered in the next Section, where we
show that, indeed,  Eqs.~(\ref{var-rad-cor}) and~(\ref{nudot-nu}) hold even in magnetars. \\
The most dramatic effect induced by glitches in magnetars is the release of a huge amount of magnetic energy in the
interior of the star and into the magnetosphere. To see this, let us consider the energy stored into the magnetic field in
the interior of the magnetar. We have:
\begin{equation}
\label{mag-ener-int}
E_{B}^{\emph{int}} \; = \; \frac{1}{6} \; R_c^3 \; B_c^2 \; + \; \frac{1}{8 \, \pi} \;  \int_{R_c}^{R} \;  \; d^3r \;
\left [ B_c \; \left (\frac{R_c}{r} \right )^3 \right ]^2 \;  \; \; ,
\end{equation}
where the first term is the energy stored into the core where the magnetic field is uniform. The variation of the magnetic
energy Eq.~(\ref{mag-ener-int}) caused by a glitch is easily evaluated. Taking into account Eq.~(\ref{magn-var-2}) and
$\left  ( \frac{R_c }{ R} \right )^3 \ll 1$, we get:
\begin{equation}
\label{var-mag-ener-int}
\delta \, E_{B}^{\emph{int}} \; \simeq \; - \; \frac{1}{3} \; R^3 \; B_S^2 \; \left  ( \frac{R}{ R_c} \right )^3 \;
\frac{\delta  R_c}{R_c} \; - \; \frac{1}{2} \;   R^3 \; B_S^2 \;  \frac{\delta  R_c}{R_c} \; \simeq \; - \; \frac{1}{3} \;
R^3 \; B_S^2 \; \left  ( \frac{R}{ R_c} \right )^3 \; \frac{\delta  R_c}{R_c}  \; .
\end{equation}
Equation~(\ref{var-mag-ener-int}) shows that there is a decrease of the magnetic energy. So that after a glitch in
magnetars a huge magnetic energy is released  in the interior of the star. We shall see that this energy is enough to
sustain the quiescent emission. On the other hand, the glitch induces also a sudden variation of the magnetic energy
stored into the magnetosphere. Indeed, from Eq.~(\ref{magn-core-surf}) we find:
\begin{equation}
\label{magn-surf-var}
\frac{\delta B_S}{B_S} \; = \; \frac{\delta B_c}{B_c} \; + \; 3 \; \frac{\delta R_c}{R_c} \;  -\; 3 \; \frac{\delta R}{R}
\; \simeq \; \frac{\delta R_c}{R_c} \;  -\; 3 \; \frac{\delta R}{R} \; \simeq \; \frac{\delta R_c}{R_c} \; > \; 0 \; \; .
\end{equation}
Thus, the magnetic energy stored into the magnetosphere:
\begin{equation}
\label{mag-ener-ext}
 E_{B}^{\emph{ext}} \; = \; \frac{1}{8 \, \pi} \;  \int_{R}^{\infty} \;  \; d^3r \;
\left [ B_S \; \left (\frac{R}{r} \right )^3 \right ]^2 \; = \; \frac{1}{6 } \; B_S^2 \; \; R^3 \; \; ,
\end{equation}
increases by:
\begin{equation}
\label{var-mag-ener-ext}
\delta  E_{B}^{\emph{ext}} \; = \;  \frac{1}{3} \; R^3 \; B_S^2 \; \left  ( \frac{\delta B_S}{ B_S} \; + \;
 \frac{3}{2}  \frac{\delta R}{R} \right ) \; \simeq \; \frac{1}{3} \; R^3 \; B_S^2 \; \frac{\delta B_S}{ B_S} \;
  > \;  0  \; \; .
\end{equation}
This magnetic energy is directly injected  into the magnetosphere, where it is dissipated by well defined physical
mechanism discussed in Section~\ref{bursts}, and it is responsible for bursts in soft gamma-ray repeaters and anomalous
$X$-ray pulsars. \\
To summarize,  in this Section we have found that dissipative phenomena in the inner core of a p-star tend to decrease the
strength of the core magnetic field. This, in turn, results in an increase of the radius of the core $\delta R_c > 0$, and
in a contraction of the surface of the star,  $\delta R < 0$. We have also show that the glitch releases an amount of
magnetic energy in the interior of the star and injects magnetic energy  into the magnetosphere, where it is completely
dissipated. Below we will show that these magnetic glitches are responsible for the quiescent emission and bursts in
gamma-ray repeaters and anomalous $X$-ray pulsars. Interestingly enough, in Ref.~\cite{Cheng:1995}  it was shown that SGR
events and earthquakes share several distinctive statistical properties, namely: power-law energy distributions,
log-symmetric waiting time distributions, strong correlations between waiting times of successive events, and weak
correlations between waiting times and intensities. These statistical properties of bursts can be easily understood if
bursts originate by  the release of a small amount of energy from a reservoir of stored energy. As a matter of fact, in
our theory the burst activity is accounted for by the release of a tiny fraction of magnetic energy stored in magnetars.
Even for giant bursts we find that the released energy is a few per cent of the magnetic energy. Moreover, the authors of
Ref.~\cite{Hurley:1994} noticed that there is a significantly  statistical similarity between the bursts from  {\it {SGR
1806-20}} and the microglitches  observed from the Vela pulsar with $ |\frac{\delta \nu}{\nu}| \thicksim 10^{-9}$. So that
we see that these early statistical studies of bursts are in complete agreement with our theory for bursts in magnetars.
Even more, we shall show that after a giant glitch there is an intense burst activity quite similar to the settling
earthquakes following a strong earthquake.
\subsection{\normalsize{BRAKING GLITCHES}}
\label{braking}
In Section~\ref{glitches} we found that magnetic glitches in p-stars lead to:
\begin{equation}
\label{delta-dot-nu}
 \frac{\delta \dot{\nu} }{\dot{\nu}} \; \; \simeq \; \; - \; \frac{\delta  B_c}{B_c} \; > \; 0 \; \; .
\end{equation}
Since there is variation of both the  inner core and the stellar radius, the moment of inertia of the star undergoes a
variation of $\delta I$. It is easy to see that the increase of the inner core leads to an increase of the moment of
inertia $I$; on the other hand, the reduction of the stellar radius implies $\delta I < 0$. In radio pulsar, where, by
neglecting the variation of the magnetic energy, the conservation of the mass leads to Eq.~(\ref{rad-var-2}), one can show
that:
\begin{equation}
\label{var-I-radio}
 \frac{\delta I}{I}  \; \; \simeq \; \; \frac{\delta  R}{R} \; < \; 0 \; \; .
\end{equation}
Moreover, from conservation of angular momentum:
\begin{equation}
\label{cons-ang}
 \frac{\delta I}{I}  \; \; = \; - \; \frac{\delta \nu}{\nu} \; \; ,
\end{equation}
it follows:
\begin{equation}
\label{radio-glitch}
0 \; < \;  \; \frac{\delta \nu}{\nu} \; \;  \simeq \; \; - \; \frac{\delta  R}{R} \; \ll \; - \; \frac{\delta  B_c}{B_c}
\; \;  \simeq \; \; \frac{\delta \dot{\nu} }{\dot{\nu}} \; \; .
\end{equation}
For magnetars, namely p-stars with super strong magnetic field, the variation of magnetic energy cannot be longer
neglected. In this case, since the magnetic energy decreases, we have that the surface contraction in magnetars is smaller
than in radio pulsars. That means that Eq.~(\ref{var-rad-cor}) holds even for magnetars. Moreover, since in radio pulsars
we known that:
\begin{equation}
\label{radio-glitch-max}
 \frac{\delta \nu}{\nu} \;  \;  = \; - \;  \frac{\delta I}{I}  \;\simeq \; \; - \; \frac{\delta  R}{R} \; \lesssim \;
 10^{-6} \; \; ,
\end{equation}
we see that in magnetars the following bound must hold:
\begin{equation}
\label{R-bound}
 - \; \frac{\delta  R}{R} \; \lesssim \;
 10^{-6} \; \; .
\end{equation}
As a consequence we may write:
\begin{equation}
\label{var-I-magnetar}
 \frac{\delta I}{I}  \; \; = \;  \; \left( \frac{\delta I}{I} \right )_{core} \; \; +  \; \;
 \left( \frac{\delta I}{I} \right )_{surf} \; \; , \; \;
 \left( \frac{\delta I}{I} \right )_{surf} \simeq \; \; \frac{\delta  R}{R} \; < \; 0 \; .
\end{equation}
As we show in a moment, the variation of the moment of inertia  induced by the core is positive. So that if the core
contribution overcomes the surface contribution to $\delta I$ we have a \emph{braking glitch} where $- \frac{\delta
\nu}{\nu} = \frac{\delta P}{P} > 0$. \\
We believe that the most compelling evidence in support to our proposal comes from the anomalous $X$-ray pulsar {\it{AXP
1E 2259+586}}. As reported in Ref.~\cite{Woods:2004a}, the timing data showed that a large glitch occurred  in  {\it{AXP
1E 2259+586}} coincident with the 2002 June giant burst. Remarkably, at the time of the giant flare on 1998 August 27, the
soft gamma ray repeater  {\it{ SGR 1900+14}} displayed a discontinuous spin-down consistent with a braking
glitch~\cite{Woods:1999}. Our theory is able to explain why {\it{AXP 1E 2259+586}} displayed a normal glitch, while {\it{
SGR 1900+14}} suffered a braking glitch. To see this, we recall the spin-down parameters of these pulsars:
\begin{equation}
\label{spin-par-magnetar}
\begin{split}
 \emph{SGR  1900+14}   &   \;  \; \; P  \;  \simeq \; 5.16 \; sec \;  ,  \;  \dot{P}  \;  \simeq \; 1.1 \; 10^{-10}
                       \; ,  \; B_S \;  \simeq \;  7.4 \; 10^{14} \; Gauss \; \; ,  \\
 \emph{AXP  1E 2259+586}  &  \;  \; \;  P \;  \simeq \; 6.98 \; sec \;  ,  \;  \dot{P}  \;  \simeq \; 2.0 \; 10^{-14}
                         \; , \; B_S \;  \simeq \;  1.2 \; 10^{13} \; Gauss \; \; .
\end{split}
\end{equation}
For canonical magnetars with  $M \, \simeq 1.4 \, M_{\bigodot}$ and radius  $R \, \simeq 10  \, Km$, we have $\sqrt{gH}
\simeq  0.55 \; GeV$. So that, using $ 1 \, GeV^2 \, \simeq \, 5.12 \, 10^{19} \, Gauss$, we rewrite
Eq.~(\ref{magn-core-surf}) as:
\begin{equation}
\label{magn-surf-num}
B_S\; \; \simeq \; \; 1.54 \;  10^{16}  \; \left (\frac{R_c}{R } \right )^3 \;  \; Gauss \; \; .
\end{equation}
Combining Eqs.~(\ref{spin-par-magnetar}) and~(\ref{magn-surf-num}) we get:
\begin{equation}
\label{R_c-R}
\begin{split}
\emph{SGR  1900+14}   &   \; \; \;\; \; \;  \;  \; \left (\frac{R_c}{R } \right )^3 \;  \; \simeq  \; 4.81 \; 10^{-2}
                          \; \; , \\
\emph{AXP  1E 2259+586} &  \; \; \;\;\; \; \;  \;   \left (\frac{R_c}{R } \right )^3 \; \; \simeq   \; 0.78 \; 10^{-3}
                          \; \; .
\end{split}
\end{equation}
According to Eqs.~(\ref{delta-dot-nu}), ~(\ref{cons-ang}) and~(\ref{var-I-magnetar}), to evaluate the sudden variation of
the frequency and frequency derivative,  we need $\delta  R$ and $\delta  R_c$. These parameters can be estimate from the
energy released during the giant bursts. In the case of {\it{AXP 1E 2259+586}},  the giant  2002 June burst followed an
intense burst activity which lasted for about one year. The authors of Ref.~\cite{Woods:2004a}, assuming a distance of $3
\; kpc$ to {\it{1E 2259+586}}, measured an energy release of $2.7 \, 10^{39} \, ergs$ and  $2.1 \,  10^{41} \, ergs$ for
the fast and slow decay intervals, respectively. Due to this uncertainty, we conservatively estimate the energy released
during the giant burst to be:
\begin{equation}
\label{GB-2259}
  \emph{AXP  1E 2259+586}   \; \; \; \;  \;  E_{\emph{burst}}  \;  \;    \; \; \simeq   \; 1.0 \; 10^{40}
                                             \; \; ergs \; \; .
\end{equation}
On 1998 August 27, a giant burst from the soft gamma ray repeater {\it{ SGR 1900+14}} was recorded. The estimate energy
released was:
\begin{equation}
\label{GB-1900}
  \emph{SGR 1900+14}   \; \; \; \;  \;  E_{\emph{burst}}  \;  \;    \; \; \simeq   \; 1.0 \; 10^{44}
                                             \; \; ergs \; \; .
\end{equation}
As we have already discussed in Sect.~\ref{glitches}, the energy released during a burst in magnetars is given by the
magnetic energy directly injected and dissipated into the magnetosphere, Eq.~(\ref{var-mag-ener-ext}). We rewrite
Eq.~(\ref{var-mag-ener-ext}) as
\begin{equation}
\label{var-mag-ener-ext-num}
\delta  E_{B}^{\emph{ext}}  \; \simeq \; \frac{1}{3} \; R^3 \; B_S^2 \; \frac{\delta B_S}{ B_S} \;
                             \simeq \;  2.6 \; 10^{44} \; ergs  \; \left ( \frac{B_S}{10^{14} \, Gauss} \right )^2
                             \; \frac{\delta B_S}{ B_S}  \; \; .
\end{equation}
So that, combining Eqs.~(\ref{var-mag-ener-ext-num}), (\ref{GB-1900}), (\ref{GB-2259})  and (\ref{spin-par-magnetar}) we
get:
\begin{equation}
\label{burst-param-1900-2259}
\begin{split}
\emph{SGR  1900+14}   &   \; \; \;\; \; \;  \;  \;  \frac{\delta B_S}{ B_S}  \; \simeq \; \frac{\delta R_c}{R_c} \;
                           \simeq \; 0.70 \; 10^{-2}    \; \;  \; , \\
\emph{AXP  1E 2259+586} &    \; \; \;\; \; \;  \;  \;  \frac{\delta B_S}{ B_S}  \; \simeq \; \frac{\delta R_c}{R_c} \;
                            \simeq \; 0.27 \; 10^{-2}    \; \; \; .
\end{split}
\end{equation}
Thus, according to Eq.~(\ref{delta-dot-nu}) we may estimate the sudden variation of $\dot{\nu}$:
\begin{equation}
\label{GB-delta-dot-nu}
 \frac{\delta \dot{\nu} }{\dot{\nu}} \; \; \simeq \; \;  2 \; \frac{\delta \, R_c}{R_c} \; \; \thicksim \; \; 10^{-2}
  \; \;,
\end{equation}
for both glitches. On the other hand we have:
\begin{equation}
\label{B-delta-I_c}
 \left( \frac{\delta I}{I} \right )_{core} \; \;  \simeq \; \; \frac{15}{2} \;
  \frac{\varepsilon_c - \varepsilon}{\overline{\varepsilon}}
  \; \left  ( \frac{R_c}{ R} \right )^5 \;   \frac{\delta \, R_c}{R_c} \; \; \simeq \; \;
  \frac{15}{2} \; \; \left  ( \frac{R_c}{ R} \right )^5 \;   \frac{\delta \, R_c}{R_c} \; \; ,
\end{equation}
leading to:
\begin{equation}
\label{B-delta-I_c-1900-2259}
\begin{split}
\emph{SGR  1900+14}   &   \; \; \;\; \; \;  \;  \;   \left( \frac{\delta I}{I} \right )_{core}
                           \simeq \;  3.34 \; \; 10^{-4}    \; \;  \; , \\
\emph{AXP  1E 2259+586} &    \; \; \;\; \; \;  \;  \;  \left( \frac{\delta I}{I} \right )_{core}
                            \simeq \;  1.34 \; \; 10^{-7}      \; \; \; .
\end{split}
\end{equation}
On the other hand, we expect that during the giant glitch $\left( \frac{\delta I}{I} \right )_{surf} \thicksim 10^{-6}$.
As a consequence, for {\it{AXP  1E 2259+586}} the core contribution is negligible with respect to  the surface
contribution to $\delta I$. In other words, {\it{AXP  1E 2259+586}} displays a normal glitch with $\frac{\delta \nu}{\nu}
\thicksim 10^{-6}$. On the contrary, Eq.~(\ref{B-delta-I_c-1900-2259}) indicates that {\it{SGR  1900+14}} suffered a
braking glitch with $\frac{\delta \nu}{\nu} \thicksim -  3.34 \; 10^{-4}$ giving:
\begin{equation}
\label{B-braking-1900}
 \Delta \; P  \; \;  \simeq \; \; 1.72  \; \; 10^{-3} \; \; sec \; \; .
\end{equation}
We would like to stress that our theory is in remarkable agreement with observations, for  a glitch of size $\frac{\delta
\nu}{\nu} =   4.24(11)  \; 10^{-6}$ was observed in {\it{AXP  1E 2259+586}} which preceded the burst
activity~\cite{Woods:2004a}. Moreover, our theory predicts a sudden increase of the spin-down torque  according to
Eq.~(\ref{GB-delta-dot-nu}). In Ref.~\cite{Woods:2004a} it is pointed out that it was not possible to give a reliable
estimate of the variation of the frequency derivative since the pulse profile was undergoing large changes, thus
compromising the phase alignment with the pulse profile template. Indeed, as discussed in Sect.~\ref{2259}, soon after the
giant burst {\it{AXP 1E 2259+586}} suffered an intense burst activity. Now, according to our theory, during the burst
activity there is both a continuous injection of magnetic energy into the magnetosphere and variation of the magnetic
torque explaining the anomalous timing noise observed in {\it{ 1E 2259+586}}. In addition, the authors of
Ref.~\cite{Woods:1999} reported a gradual increase of the nominal spin-down rate and a discontinuous spin down event
associated with the 1998 August 27 flare from {\it{SGR  1900+14}}. Extrapolating the long-term trends found before and
after August 27, they found evidence of a braking glitch with $\Delta P \,  \simeq \,  0.57(2)  \; 10^{-3} \;  sec$. In
view of our theoretical uncertainties, the agreement with our Eq.~(\ref{B-braking-1900}) is rather good. \\
We feel that it is worthwhile to point out that the standard magnetar theory is completely unable to predict the
remarkable evidence of braking glitches. As a matter of fact, to our knowledge, the only attempt to explain the braking
glitch observed in {\it{SGR  1900+14}} is done in Ref.~\cite{Thompson:2000} where it is suggested that violent August 27
event involved a glitch. However,  the magnitude of the glitch was  estimated by scaling to the largest glitches in young,
active pulsars with similar spin-down ages and internal temperature. In this way they deduced the estimate $|\frac{\Delta
P}{P}| \thicksim 10^{-5}$ to $10^{-4}$. In our opinion, this can hardly be considered a valid explanation for the braking
glitch. First, the authors of Ref.~\cite{Thompson:2000} overlooked the well known fact that radio pulsars display normal
glitches and no braking glitches. Second, these authors cannot explain why  {\it{AXP 1E 2259+586}}  displayed a normal
glitch instead of a braking glitch. \\
Let us conclude this Section by briefly discussing the 2004 December 27 giant flare from {\it{SGR  1806-20}}. During this
tremendous outburst {\it{SGR  1806-20}} released a huge amount of energy, $ E_{\emph{burst}} \thicksim 10^{46} \; ergs$.
Using the spin-down parameters reported in Ref.~ \cite{Mereghetti:2005a}:
\begin{equation}
\label{spin-par-1806}
 \emph{SGR  1806-20}    \;  \; \; P  \;  \simeq \; 7.55 \; sec \;  ,  \;  \dot{P}  \;  \simeq \; 5.5 \; 10^{-10}
                       \; ,  \; B_S \;  \simeq \;  2.0 \; 10^{15} \; Gauss \; \; ,
\end{equation}
we find:
\begin{equation}
\label{burst-param-1806}
 \emph{SGR  1806-20}  \; \;\; \; \;   \frac{\delta B_S}{ B_S}  \; \simeq \; \frac{\delta R_c}{R_c} \;
                           \simeq \; 9.6 \; 10^{-2}    \; \; .
\end{equation}
Thus, we predict that {\it{SGR  1806-20}} should display a gigantic braking glitch with $\frac{\Delta P}{P} \simeq
 2.4 \; 10^{-2}$, or :
\begin{equation}
\label{B-braking-1806}
 \Delta \; P  \; \;  \simeq \; \; 1.8  \; \; 10^{-1} \; \; sec \; \; .
\end{equation}
\subsection{\normalsize{QUIESCENT LUMINOSITY}}
\label{quiescent}
The basic mechanism to explain the quiescent $X$-ray emission in our magnetars is the internal dissipation of magnetic
energy. Our mechanism is basically the same as in the standard magnetar model based on neutron star~\cite{Duncan:1996}.
Below we shall critically compare our proposal with the standard theory. \\
In Section~\ref{glitches} we showed that during a glitch there is a huge amount of magnetic energy released into the
magnetar:
\begin{equation}
\label{delta-ener-int}
 - \; \delta \, E_{B}^{\emph{int}}  \; \; \simeq \;  \; \frac{1}{3} \;
R^3 \; B_S^2 \; \left  ( \frac{R}{ R_c} \right )^3 \; \frac{\delta  R_c}{R_c}  \; .
\end{equation}
As in previous Section, we use {\it {SGR 1900+14}} and  {\it {AXP 1E 2259+586}} as prototypes for soft gamma ray repeaters
and anomalous $X$-ray pulsars, respectively. Using the results of Sect.~\ref{braking}, we find:
\begin{equation}
\label{delta-ener-1900-2259}
\begin{split}
\emph{SGR  1900+14}   &   \; \; \;\; \; \;  \;  \; - \; \delta \, E_{B}^{\emph{int}}  \; \; \simeq \; \;  3.0 \; \; 10^{47}
                          \;  \;  erg  \; \;  \frac{\delta  R_c}{R_c} \;  \; , \\
\emph{AXP  1E 2259+586} &   \; \; \;\; \; \;  \;  \; - \; \delta \, E_{B}^{\emph{int}}  \; \; \simeq \; \;  4.8 \; \;
                           10^{45}  \;  \;  erg  \; \;  \frac{\delta  R_c}{R_c} \;  \; .
\end{split}
\end{equation}
This release of magnetic energy is dissipated leading to observable surface luminosity. To see this, we need a thermal
evolution model which calculates the interior temperature distribution. In the case of neutron stars such a calculation
has been performed in Ref.~\cite{VanRiper:1991}, where it is showed that the isothermal approximation is a rather good
approximation in the range of inner temperatures of interest. The equation which determines the thermal history of a
p-star has been discussed in Ref.~\cite{cea:2003} in the isothermal approximation:
\begin{equation}
\label{cooling}
 C_V \; \frac{d T}{d t}
\; = \; - \; (L_{\nu} \; + \; L_{\gamma}) \; ,
\end{equation}
where $L_{\nu}$ is the neutrino luminosity, $L_{\gamma}$ is the photon luminosity and $C_V$ is the specific heat. Assuming
blackbody photon emission from the surface at an effective surface temperature $T_S$ we have:
\begin{equation}
\label{blackbody}
 L_{\gamma} \; = \; 4 \, \pi \, R^2 \, \sigma_{SB} \, T_S^4 \; ,
\end{equation}
where $\sigma_{SB}$ is the $Stefan-Boltzmann$ constant. In Ref.~\cite{cea:2003} we assumed that the surface and interior
temperature were related by:
\begin{equation}
\label{surface}
 \frac{T_S}{T} \; = \; 10^{-2} \; a  \; \; .
\end{equation}
Equation~(\ref{surface}) is relevant for a p-star which is not bare, namely for p-stars which are endowed with a thin
crust. It turns out~\cite{cea:2004} that p-stars have a sharp edge of thickness of the order of about $ 1 \; fermi$. On
the other hand, electrons which are bound by the coulomb attraction, extend several hundred \emph{fermis} beyond the edge.
As a consequence, on the surface of the star there is a positively charged layer which is able to support a thin crust of
ordinary matter. The vacuum gap between the core and the crust, which is of order of several hundred \emph{fermis}, leads
to a strong suppression of the surface temperature with respect to the core temperature. The precise relation between
$T_S$ and $T$ could be obtained by a careful study of the crust and core thermal interaction. In any case, our
phenomenological relation Eq.~(\ref{surface}) allows a wide variation of $T_S$, which encompasses the neutron star
relation (see, for instance, Ref.~\cite{gudmundsson:1983}). Moreover, our cooling curves display a rather weak dependence
on the parameter $a$ in Eq.~(\ref{surface}). Since we are interested in the quiescent luminosity, we do not need to known
the precise value of this parameter.  So that, in the following we shall assume $a \thicksim 1$. In other words, we
assume:
\begin{equation}
\label{surface-new}
  T_S \;  \;  \simeq \; \; 10^{-2} \; \; T \; \; \; .
\end{equation}
Obviously, the parameter $a$ is relevant to evaluate the surface temperature once the core temperature is given. Note
that, in the relevant range of core temperature $ T \thicksim 10^{8}$ $\, {}^\circ K$, our Eq.~(\ref{surface-new}) is
practically identical to the parametrization adopted in Ref.~\cite{Duncan:1996} within the standard magnetar model:
\begin{equation}
\label{Temp-D-T}
  T_S \;  \;  \simeq \; \; 1.3 \; \; 10^6 \; {}^\circ K \; \; \left ( \frac{T}{10^8 \, {}^\circ K} \right )^{\frac{5}{9}}
                \; \; \; .
\end{equation}
The neutrino luminosity $L_{\nu}$ in Eq.~(\ref{cooling}) is given by the direct $\beta$-decay quark reactions, the
dominant cooling processes by neutrino emission. From the results of Ref.~\cite{cea:2003}, we find:
\begin{equation}
\label{luminosity}
 L_{\nu} \; \; \simeq \; \; 1.22 \; \; 10^{36} \; \;  \frac{erg}{s} \; \;  T_9^8 \; \; \;  ,
\end{equation}
where $T_9$ is the temperature in units of $10^9$ $\, {}^\circ K$.  Note that the neutrino luminosity $L_{\nu}$ has the
same temperature dependence as the neutrino luminosity by  modified URCA reactions in neutron stars (see, for instance
Ref.~\cite{shapiro:1983}), but it is more than two order of magnitude smaller.
From the cooling curves reported in Ref.~\cite{cea:2003} we infer that the surface and interior temperature are almost
constant up to time $\tau \thicksim 10^{5} \, years$. Observing that magnetars candidates are rather young pulsar with
$\tau_{age} \lesssim 10^{5} \, years$, we may estimate the average surface luminosities as:
\begin{equation}
\label{Luminosity}
 L_{\gamma} \; \; \simeq \; \frac{ - \; \delta \, E_{B}^{\emph{int}} }{\tau_{age}}  \; \; \; \; \; .
\end{equation}
We assume $ \tau_{age} \thickapprox \tau_c$ for {\it {SGR 1900+14}}. On the other hand, as discussed in
Section~\ref{Introduction}, we known that for {\it {AXP 1E 2259+586}}  $ \; \; \tau_{age} \thicksim 10^3 \, years \ll
\tau_c$. We get:

\begin{equation}
\label{lum-1900-2259}
\begin{split}
\emph{SGR  1900+14}   &   \; \; \;\; \; \;  \;  \;  L_{\gamma}  \; \; \simeq \; \; 1.3 \; \; 10^{37}
                          \;  \; \frac{erg}{s}  \; \;  \frac{\delta  R_c}{R_c} \;  \; , \\
\emph{AXP  1E 2259+586} &   \; \; \;\; \; \;  \; \;  L_{\gamma} \; \; \simeq \; \;  1.5 \; \;
                           10^{35}  \;  \; \frac{erg}{s} \; \;  \frac{\delta  R_c}{R_c} \;  \; .
\end{split}
\end{equation}
So that it is enough to assume that {\it {SGR 1900+14}} suffered in the past a glitch with $\frac{\delta  R_c}{R_c}
\thicksim 10^{-2}$ to sustain the observed luminosity $ L_{\gamma} \thicksim 10^{35} \; \frac{erg}{s}$ (assuming a
distance of about $10 \; kpc$). In the case of {\it {AXP 1E 2259+586}}, assuming a distance of  about $3 \; kpc$, the
observed luminosity is $ L_{\gamma} \thicksim 10^{34} \; \frac{erg}{s}$, so that we infer that this pulsar had suffered in
the past a giant glitch with  $\frac{\delta  R_c}{R_c} \thicksim 10^{-1}$, quite similar to the recent {\it {SGR 1806-20}}
glitch.
\\
Let us discuss the range of validity of our approximation. Equation~(\ref{Luminosity}) is valid as long as  $L_{\gamma}$
dominates over  $L_{\nu}$,  otherwise the star is efficiently cooled by neutrino emission and the surface luminosity
saturates to $L_{\gamma}^{max}$. We may quite easily evaluate this limiting luminosity from $L_{\gamma}^{max} \simeq
L_{\nu}$. Using Eq.~(\ref{surface-new}) and $R \simeq 10 \; Km$, we get:
\begin{equation}
\label{Lum-max}
 L_{\gamma}^{max} \; \; \simeq \;  \;   4.2 \; \; 10^{37}  \; \frac{erg}{s}   \; \; \; \; .
\end{equation}
Note that, since our neutrino luminosity is reduced by more than two order of magnitude with respect to neutron stars,
$L_{\gamma}^{max}$ is about two order of magnitude greater than the maximum allowed surface luminosity in neutron
stars~\cite{VanRiper:1991}. Thus, while our theory allows to account for observed luminosities  up to $10^{36} \;
\frac{erg}{s}$, the standard model based on neutron stars is in embarrassing contradiction with observations. \\
Let us, finally, comment on the quiescent thermal spectrum in our theory. As already discussed, the origin of the
quiescent emission is the huge release of magnetic energy in the interior of the magnetar. Our previous estimate of the
quiescent luminosities assumed that the interior temperature distribution was uniform. However, due to the huge magnetic
field, the thermal conductivity is enhanced along the magnetic field. This comes out to be the case for both electron and
quarks, since we argued that the magnetic and  chromomagnetic fields are aligned . As a consequence, we expect that the
quiescent spectrum should be parameterized  as two blackbodies with parameter $R_1 \, , \, T_1$ and $R_2 \, , \, T_2$,
respectively. Since the blackbody luminosities $L_{\gamma1}$ and $L_{\gamma2}$ are naturally of the same order, our
previous estimates for the quiescent luminosities are unaffected. Moreover, since the thermal conductivity is enhanced
along the magnetic field, the high temperature blackbody, with temperature $T_2$, originates  from the heated polar
magnetic cups. Thus we have:
\begin{equation}
\label{black-bodies}
\begin{split}
 &  R_1  \; \; \lesssim \;  \; R  \; \; \; , \; \; \; R_2 \lesssim \; 1 \; Km \; \; , \; \; \\
 &  T_2  \; \; >  \; \; T_1 \; \;  \; , \; \; R_1  \; \; >  \; \; R_2 \;  \; , \; \; L_{\gamma1} \; \; \simeq \; \;
    L_{\gamma2} \; \; .
\end{split}
\end{equation}
Note that there is a natural  anticorrelation between  blackbody radii and temperatures.
Customary, the quiescent spectrum of anomalous $X$-ray pulsars and soft gamma ray repeaters is fitted in terms of
blackbody plus power law. In particular, it is assumed that the power law component extends to energy greater than an
arbitrary cutoff energy $E_{cutoff} \simeq 2 \, KeV$. It is worthwhile to stress that these parameterizations of the
quiescent spectra are in essence phenomenological fits, for there are not sound physical motivations. Indeed, within the
standard magnetar model~\cite{Duncan:1996} the power law should be related to hydromagnetic wind accelerated by Alfven
waves. However,  any physical justification for the arbitrary cutoff energy $E_{cutoff}$ is lacking. Moreover, the
luminosity of the wind emission should increase with magnetic field strength as $L_{wind} \simeq L_{PL} \thicksim B_S^2$.
On the other hand, the blackbody luminosities should scale as $B_S^{4.4}$~\cite{Duncan:1996}. So that the ratio
$L_{PL}/L_{BB}$ decreases with increasing  magnetic field strengths, contrary to observations~\cite{Marsden:2001}.
Finally, observations of a small energy dependence of pulsed fraction in some anomalous $X$-ray pulsars requires ad hoc
tuning of the blackbody and power law components. Thus, we see that the standard magnetar model is in
striking contradictions with observations. \\
On the contrary, in our theory well defined physical arguments lead to the two blackbody representation of the quiescent
spectra, whose parameters are constrained by our Eq.~(\ref{black-bodies}). As a matter of fact, we have checked in
literature that the quiescent spectra of both anomalous $X$-ray pulsars and soft gamma ray repeaters could be accounted
for by two balckbodies. For instance, in Ref.~\cite{Morii:2003} the quiescent spectrum of {\it {AXP 1E 1841-045}} is well
fitted with the standard power law plus blackbody (reduced $\chi^2/dof \simeq 1.11$), nevertheless the two blackbody model
gives also a rather good fit (reduced $\chi^2/dof \simeq 1.12$). Interestingly enough the blackbody parameters:
\begin{equation}
\label{black-bodies-1841}
\begin{split}
 &  R_1  \; =  \; 5.7 \; \; ^{+0.6}_{-0.5} \; \; \; \; Km \; \; \; , \; \; \;   T_1  \; =  \; 0.47 \; \pm \; 0.02  \; \;
             KeV \; \; \; , \\
 &  R_2  \; =  \; 0.36 \; ^{+0.08}_{-0.07} \; \; Km \; \; \; , \; \; \;   T_2  \; =  \; 1.5 \; ^{+0.2}_{-0.1}  \; \;  KeV
 \; \; \; ,
\end{split}
\end{equation}
are in agreement with  Eq.~(\ref{black-bodies}). Moreover, assuming that the power law component in the standard
parametrization of quiescent spectra account for the hot blackbody component in our parametrization, we find that the
suggestion  $L_{\gamma1} \; \simeq \; L_{\gamma2}$ in Eq.~(\ref{black-bodies}) is in agreement  with
observations~\cite{Marsden:2001}. It should be stressed, however, that the two blackbodies are not the whole story. In
Appendix we show that thermal photons originating from the hot polar cups are reprocessed by electrons trapped above the
polar cups. These electrons, which are responsible for the faint low energy spectrum, could result in broad spectral
features in the quiescent spectrum. These
spectral features, in turn, could result in observable deviations from the two blackbody fit. \\
Another interesting consequence of the anisotropic distribution of the surface temperature due to strong magnetic fields
is that the thermal surface blackbody radiation will be modulated by the stellar rotation. As a matter of fact, in
Ref.~\cite{Ozel:2002} it is argued that the observed properties of anomalous $X$-ray pulsars can be accounted for by
magnetars with a single hot region. It is remarkable that our interpretation explains naturally the observed change in
pulse profile of {\it {SGR 1900+14}} following the 1998 August 27 giant flare. In addition, the thermal radiation
reprocessed by electrons near the polar cups could result in an effective description with two hot spots. It seem that our
picture is in fair qualitative agreement with several observations. However, any further discussion of this matter goes
beyond the aim of the present paper.
\subsection{\normalsize{BURSTS}}
\label{bursts}
In the present Section we discuss how glitches in our magnetars give rise to the  burst activity from soft gamma-ray
repeaters and anomalous $X$-ray pulsars. We said in Sect.~\ref{glitches} that the energy released during a burst in a
magnetar is given by the magnetic energy directly injected into the magnetosphere, Eq.~(\ref{var-mag-ener-ext}). Before
addressing the problem of the dissipation of this magnetic energy in the magnetosphere, let us discuss what are the
observational signatures at the onset of the burst. Observations indicate that at the onset of giant bursts the flux
displays  a spike with a very short rise time $t_1$ followed by a rapid but more gradual decay time $t_2$. According to
our previous discussion, the onset of bursts is due to the positive variation of the surface magnetic field $\delta B_S$,
which in turn implies an sudden increase of the magnetic energy stored in the magnetosphere. Now, according to
Eq.~(\ref{mag-ener-ext}) we see that almost all the magnetic energy is stored in the region:
\begin{equation}
\label{4.54}
 R \;  \lesssim \; r  \;  \lesssim \; 10 \; R  \; \;  \; .
\end{equation}
So that the rise time is essentially the time needed to propagate in the magnetosphere the information that the surface
magnetic field is varied. So that we are lead to:

\begin{equation}
\label{4.55}
  t_1 \; \simeq \; 9 \; R  \; \simeq \; 3 \; 10^{-4}  \; sec  \; \; ,
\end{equation}
which indeed is in agreement with observations. On the other hand, in our proposal the decay time $t_2$ depends on the
physical properties of the magnetosphere. Indeed, it is natural to identify $t_2$ with the time needed to the system to
react to the sudden variation of the magnetic field. In other words, we may consider the magnetosphere as a huge electric
circuit which is subject to a sudden increase of power from some external device. The electric circuit reacts to the
external injection of energy within  a transient time. So that, in our case the time $t_2$ is a function of the geometry
and the conducting properties of the magnetosphere. In general, it is natural to expect that $t_1 \, \ll \, t_2$ so that
the time extension of the initial spike is:
\begin{equation}
\label{4.56}
 \delta \; t_{spike} \;  \simeq \; t_2 \; - \; t_1 \; \simeq \;  t_2  \;  \; \; .
\end{equation}
Remarkably, observations shows that the observed giant bursts are characterized by almost the same $\delta \; t_{spike}$:
\begin{equation}
\label{4.57}
 \delta \; t_{spike} \;  \simeq  \;  t_2  \; \simeq \; 0.1  \; sec  \; \; ,
\end{equation}
signalling that the structure of the magnetosphere of soft gamma-ray repeaters and anomalous $X$-ray pulsars are very
similar. Since the magnetic field is varied by  $\delta B_S$ in a time $\delta  t_{spike}$, then from Maxwell equations it
follows that it must be an induced electric field. To see this, let us consider the dipolar magnetic field in polar
coordinate:
\begin{equation}
\label{mag-polar}
\begin{split}
 B_r & = \; - \frac{2 \; B_S \; R^3 \; \cos \theta}{r^3}  \; \;  ,  \\
 B_\theta & = \; - \frac{ B_S \; R^3 \; \sin \theta}{r^3}  \; \;  ,  \\
B _\varphi & =  \; 0 \; ,
\end{split}
\end{equation}
Thus, observing that $\frac{\delta B_S}{\delta t_{spike}}$ is the time derivative of the magnetic field it is easy to find
the induced azimuthal electric field:
\begin{equation}
\label{electric}
E_\varphi  =  + \frac{\delta B_S}{\delta t_{spike}} \frac{ R^3  \; \sin \theta}{r^2}  \; \; \; , \; \; \; r \; \geq \; R
\; \; \; .
\end{equation}
To discuss the physical effects of the induced azimuthal electric field Eq.~(\ref{electric}), it is convenient to work in
the co-rotating frame of the star. We assume that the magnetosphere contains a neutral plasma. Thus, we see that charges
are suddenly accelerated by the huge induced azimuthal electric field $E_\varphi$ and thereby acquire an azimuthal
velocity $v_\varphi \simeq 1$ which is directed along the electric field for positive charges and in the opposite
direction for negative charges. Now, it is well known that relativistic charged particles moving in the magnetic field
$\vec{B}(\vec{r})$, Eq.~(\ref{mag-polar}), will emit synchrotron radiation~\cite{wallace:1977}. As we discuss below, these
processes are able to completely dissipate the whole magnetic energy injected into the magnetosphere following a glitch.
However, before discuss this last point in details, we would like to point out some general consequences which lead to
well defined observational features. As we said before, charges are accelerated by the electric field $E_\varphi$ thereby
acquiring a relativistic azimuthal velocity. As a consequence, they are subject to the drift Lorentz force $\vec{F} \, =
\, q \, \vec{v}_\varphi \, \times \, \vec{B}(\vec{r})$, whose radial component is:
\begin{equation}
\label{F_r}
 F_r  \; = \;  -  q \, v_\varphi  B_\theta \;
   \simeq \;  + q \, v_\varphi B_S
  \sin \theta  \left ( \frac{ R}{r} \right )^3  ,
\end{equation}
while the $\theta$ component is:
\begin{equation}
\label{F_theta}
F_\theta \;  =  \; + q \, v_\varphi  B_r \;  \simeq \; - \, 2 \, q \,  v_\varphi
 B_S  \cos \theta  \left ( \frac{R}{r} \right )^3  .
\end{equation}
The radial component $F_r$ pushes both positive and negative charges radially outward. Then, we see that the plasma in the
outermost part of the magnetosphere is subject to a intense radial centrifugal force, so that the plasma must flow
radially outward giving rise to a blast wave. On the other hand,  $F_\theta$ is centripetal in the upper hemisphere and
centrifugal in the lower hemisphere. As a consequence, in the lower hemisphere charges are pushed towards the magnetic
equatorial plane $\cos \theta = 0$, while in the upper hemisphere (the north magnetic pole) the centripetal force gives
rise to a rather narrow jet along the magnetic axis. As a consequence, at the onset of the giant burst there is an almost
spherically symmetric outflow from the pulsar together with a collimated jet from the north magnetic pole. Interestingly
enough, a fading radio source has been seen from {\it {SGR 1900+14}} following the  August 27 1998 giant
flare~\cite{Frail:1999}. Indeed, the radio afterglow is consistent with an outflow expanding subrelativistically into the
surrounding medium. This is in agreement with our model once one takes into account that the azimuthal electric field is
rapidly decreasing with the distance from the star, so that $v_\varphi \lesssim 1$ for the plasma in the outer region of
the magnetosphere. However, we believe that the most compelling evidence in favour of our proposal comes from the detected
radio afterglow following the 27 December 2004 gigantic flare from
{\it {SGR 1806-20}}~\cite{Gaensler:2005,Cameron:2005,Wang:2005,Granot:2005,Gelfand:2005}. Indeed, the fading radio source
from {\it {SGR 1806-20}} has similar properties as that observed from {\it {SGR 1900+14}}, but much higher energy. The
interesting aspect is that in this case the spectra of the radio afterglow showed clearly the presence of the expected
spherical non relativistic expansion together with a sideways expansion of a jetted explosion (see Fig.~1 of
Ref.~\cite{Gaensler:2005} and Fig.~1 of Ref.~\cite{Cameron:2005}), in striking agreement with our theory. Note that the
standard magnetar model is unable to account for these observed features of the radio afterglow. Finally, the lower limit
of the outflow $E \gtrsim 10^{44.5} \; ergs$~\cite{Gelfand:2005} implies that the blast wave and the jet dissipate only a
small fraction of the burst energy which is about $10^{46} \; ergs$ (see Section~\ref{braking}). Thus, we infer that
almost all the burst energy must be dissipated into the magnetosphere. In the co-rotating frame of the star the plasma, at
rest before the onset of the burst, is suddenly accelerated by the induced electric field thereby acquiring an azimuthal
velocity $v_\varphi \simeq 1$. Now, relativistic charges are moving in the dipolar magnetic field of the pulsar. So that,
they will lose energy by emitting synchrotron radiation until they come at rest. Of course, this process, which involves
charges that are distributed in the whole magnetosphere, will last for a time $t_{dis}$ much longer that $\delta
t_{spike}$. Actually, $t_{dis}$ will depend on the injected energy, the plasma distribution and the magnetic field
strength. Moreover, one should also take care of repeated charge and photon scattering. So that it is not easy to estimate
$t_{dis}$ without a precise knowledge of the pulsar magnetosphere. At the same time, the fading of the luminosity with
time, the so-called light curve $L(t)$, cannot be determined without a precise knowledge of the microscopic dissipation
mechanisms. However, since the dissipation of the magnetic energy involves the whole magnetosphere, it turns out that we
may  accurately reproduce the time variation of $L(t)$ without a precise knowledge of the microscopic dissipative
mechanisms. Indeed, in Sect.~\ref{light} we develop an effective description where our ignorance on the microscopic
dissipative processes is encoded in few macroscopic parameters, which allows us to determine the light curves. In the
remaining of the present Section we investigate the spectral properties of the luminosity. To this end, we need to
consider the synchrotron radiation spectral distribution. Since radiation from electrons is far more important than from
protons, in the following we shall focus on electrons. It is well known that the synchrotron radiation will be mainly at
the frequency~\cite{schwinger:1949}(see also Ref.~\cite{wallace:1977}):
\begin{equation}
\label{omega_m}
\omega_m \; \simeq \; \gamma^2 \; \frac{e  B}{ m_e} \; ,
\end{equation}
where $\gamma$ is the electron Lorentz factor. Using Eq.~(\ref{mag-polar}) we get:
\begin{equation}
\label{cyclo}
 \omega(r)  \; \simeq \; \gamma^2 \;   \frac{e  B_S}{ m_e} \;  \left ( \frac{ R}{r} \right )^3 \; \; , \; \;
                R \;  \lesssim \; r  \;  \lesssim \; 10 \; R  \; \;  \; .
\end{equation}
It is useful to numerically estimate the involved frequencies. To this end, we consider the giant flare of 1998, August 27
from {\it {SGR 1900+14}}:
\begin{equation}
\label{1900-par}
 B_S \; \simeq \; 7.4 \; 10^{14} \; Gauss \; \; \; , \; \; \; \frac{\delta B_S}{B_S} \; \simeq \; 10^{-2}  \; \;  \; .
\end{equation}
So that, from Eq.~(\ref{cyclo}) it follows:
\begin{equation}
\label{4.62}
  \omega(r)  \; \simeq \; \gamma^2 \; 8.67 \; MeV \;  \left ( \frac{ R}{r} \right )^3 \; \; , \; \;
                R \;  \lesssim \; r  \;  \lesssim \; 10 \; R  \; \;  \; ,
\end{equation}
or
\begin{equation}
\label{4.63}
  \omega_1  \, \simeq \, \gamma^2 \, 8.67 \; KeV  \; \lesssim \; \; \omega \; \; \lesssim \;
        \omega_2  \, \simeq \, \gamma^2 \, 8.67 \; MeV   \; \;  \; .
\end{equation}
The power injected into the magnetosphere is supplied by the azimuthal electric field during the initial hard spike. So
that to estimate the total power we need to evaluate the power supplied by the azimuthal electric field. Let us consider
the infinitesimal volume $dV=r^2 \sin \theta dr d\theta d\varphi$; the power supplied by the induced electric field
$E_\varphi$ in $dV$:
\begin{equation}
\label{inf-pow}
 d \dot{W}_{E_\varphi} \; \simeq \;  n_e \; e \; \frac{\delta B_S}{\delta t_{spike}}  \; v_\varphi \;  R^3  \;
 \sin^2 \theta dr d\theta d\varphi  \; ,
\end{equation}
where $n_e$ is  the electron number density. Since the magnetosphere is axially symmetric it follows that $n_e$ cannot
depend on $\varphi$. Moreover, within our theoretical uncertainties we may neglect the dependence on $\theta$. So that,
integrating over $\theta$ and $\varphi$ we get:
\begin{equation}
\label{4.68}
 d \dot{W}_{E_\varphi} \; \simeq \; 2 \; \pi^2 \;  n_e \; e \; \frac{\delta B_S}{\delta t_{spike}}  \; v_\varphi \;  R^3  \;
 dr   \; .
\end{equation}
In order to determine the spectral distribution of the supplied power, we note that to a good approximation all the
synchrotron radiation is emitted at $\omega_m$, Eq.~(\ref{omega_m}). So that, we may use Eq.~(\ref{4.62}) to get:
\begin{equation}
\label{dr-dnu}
 - \; dr \;  \simeq \; \frac{R}{3} \;  \gamma^{\frac{2}{3}} \; \left ( \frac{e  B_S}{m_e}  \right )^{1/3}
 \frac{1}{\omega^{\frac{4}{3}}} \; d\omega \; \; .
\end{equation}
Inserting Eq.~(\ref{dr-dnu}) into Eq.~(\ref{4.68}) we obtain the  spectral power:
\begin{equation}
\label{spect-pow}
 F(\omega) \; d\omega \; \simeq \;  \frac{2 \pi^2}{3} \; n_e \; e \; \frac{\delta B_S}{\delta t_{spike}}  \; v_\varphi \;  R^4  \;
  \; \gamma^{\frac{2}{3}} \; \left ( \frac{e  B_S}{m_e}  \right )^{1/3}
 \frac{1}{\omega^{\frac{4}{3}}} \; d\omega \; ,
\end{equation}
while the total luminosity is given by:
\begin{equation}
\label{lum-tot}
L  \; = \; \int_{\omega_1}^{\omega_2} \; F(\omega) \; d\omega \; \; .
\end{equation}
Note that $L$ is the total luminosity injected into the magnetosphere during the initial hard spike. So that, since the
spike lasts $\delta \; t_{spike}$, we have:
\begin{equation}
\label{4.72}
E_{\emph{burst}}   \; \simeq  \; \delta \; t_{spike} \; L  \; \; ,
\end{equation}
where $E_{\emph{burst}}$ is the total burst energy. In the case of the 1998 August 27 giant burst from {\it {SGR 1900+14}}
the burst energy is given by Eq.~(\ref{GB-1900}). Thus, using Eqs.~(\ref{4.72}) and ~(\ref{4.57}) we have:
\begin{equation}
\label{4.73}
L  \; \simeq \; 10^{45}  \; \frac{ergs}{sec} \; \; \; ,
\end{equation}
which, indeed, is in agreement with observations. It is worthwhile to estimate the electron number density needed to
dissipate the magnetic energy injected in the magnetosphere. To this end, we assume an uniform number density. Thus, using
Eqs.~(\ref{lum-tot}), (\ref{spect-pow}) and (\ref{4.63}) we get:
\begin{equation}
\label{4.74}
L  \; \simeq \; 18 \; \pi^2 \; n_e \; e \; \frac{\delta B_S}{\delta t_{spike}} \;  R^4   \; \; ,
\end{equation}
where we used $v_\varphi \simeq 1$. Specializing to the  August 27 giant burst we find:
\begin{equation}
\label{4.75}
 n_e \; \simeq \; 2.0  \; 10^{14}  \; cm^{-3}  \;  \; \; ,
\end{equation}
indeed quite a reasonable value. Soon after the initial spike, the induced azimuthal electric field vanishes and the
luminosity decreases due to dissipative processes in the magnetosphere. As thoroughly discussed in Sect.~\ref{light}, it
is remarkable that the fading of the luminosity can be accurately reproduced without a precise knowledge of the
microscopic dissipative mechanisms. So that combining the time evolution of the luminosity $L(t)$, discussed in
Sect.~\ref{light}, with the spectral decomposition we may obtain the time evolution of the spectral components. In
particular, firstly we show that starting from Eq.~(\ref{spect-pow}) the spectral luminosities can be accounted for by two
blackbodies and a power law. After that, we discuss the time evolution of the three different spectral components. \\
The spectral decomposition Eq.~(\ref{spect-pow}) seems to indicate that the synchrotron radiation follows a power law
distribution. However,  one should take care of reprocessing effects which redistribute the spectral distribution. To see
this, we note that photons with energy $\omega \geq 2 \, m_e$ quickly will  produce copiously almost relativistic
$e^{\pm}$ pairs. Now, following Ref.~\cite{Duncan:1995}, even if the particles are injected steadily in a time $\delta
t_{spike}$, it is easy to argue that the energy of relativistic particles is rapidly converted due to comptonization to
thermal photon-pair plasma. Since the pair production is quite close to the stellar surface, we may adopt the rather crude
approximation of an uniform magnetic filed $B \simeq B_S$ throughout the volume $V_{plasma} \simeq 12 \, \pi \, R^3$.
Since typical magnetic fields in magnetars are well above $B_{QED}$, electrons and positrons sit in the lowest Landau
levels. In this approximation we deal with an almost one dimensional pair plasma whose energy density
is~\cite{Duncan:1995}:
\begin{equation}
\label{4.76}
 u_e \; \simeq \; m_e \; (n_{e^+} \; + n_{e^-}) \; \simeq \frac{2}{(2 \pi)^{\frac{3}{2}}}  \; e B_S \;  m_e^2 \;
         \left ( \frac{T_{plasma}}{m_e} \right )^{\frac{1}{2}} \exp (-  \frac{m_e}{T_{plasma}})  \;  \; \; ,
\end{equation}
for $T_{plasma} < m_e$, $T_{plasma}$ being the plasma temperature. Thus, the total energy density of the thermal
photon-pair plasma is:
\begin{equation}
\label{4.77}
 u \; = \; u_e \; + \; u_{\gamma} \; \simeq \;  u_e \; + \; \frac{\pi^2}{15} \; T_{plasma}^4 \; \; \; .
\end{equation}
The plasma temperature is determined by equaling the thermal energy Eq.~(\ref{4.77}) with the fraction of burst energy
released in the spectral region $\omega \geq 2 \, m_e$. It is easy to find:
\begin{equation}
\label{4.78}
 E_{pairs} \; \simeq \; 0.147 \; E_{burst} \; \; \; ,
\end{equation}
where for the numerical estimate we approximated $\omega_1 \simeq 10 \, KeV$ and $\omega_2 \simeq 10 \, MeV$,
corresponding to mildly relativistic electrons in the magnetosphere. So that we have:
\begin{equation}
\label{4.79}
 \frac{2}{(2 \pi)^{\frac{3}{2}}}  \; e B_S \;  m_e^2 \;
         \left ( \frac{T_{plasma}}{m_e} \right )^{\frac{1}{2}} \exp (-  \frac{m_e}{T_{plasma}})  \;  + \;
 \; \frac{\pi^2}{15} \; T_{plasma}^4 \; \simeq \; \frac{ E_{pairs}}{V_{plasma}} \; \; \; .
\end{equation}
In the case of August 27 giant burst from {\it {SGR 1900+14}} we find:
\begin{equation}
\label{4.80}
 \sqrt{x} \;  \exp (-  \frac{1}{x}) \; +  \; 0.311 \; x^4  \simeq \; 1.32 \; 10^{-2} \; \; , \; \;
 x \; = \; \frac{T_{plasma}}{m_e} \; \; \; ,
\end{equation}
whose solution gives $T_{plasma} \simeq 135 \, KeV$. However, this is not the end of the whole story. Indeed, our thermal
photon-pair plasma at temperature  $T_{plasma}$ will be reprocessed by thermal electrons on the surface which are at
temperature of the thermal quiescent emission $T_Q  \lesssim 1 \, KeV$. So that, photons at temperature $T_{plasma} \gg
T_Q$ are rapidly cooled by Thompson scattering off electrons in the stellar atmosphere, which extends over several
hundreds \emph{fermis} beyond the edge of the star. The rate of change of the radiation energy density is given
by~\cite{Peebles:1993}:
\begin{equation}
\label{4.81}
 \frac{1}{u_{\gamma}} \; \frac{d u_{\gamma}}{d t} \;  \simeq \; \frac{4 \, \sigma_T \, n_Q}{m_e} \;
 ( T_Q - T_{plasma} ) \; \; \; ,
\end{equation}
where $n_Q$ is the number density of electrons in the stellar atmosphere. The electron number density in the atmosphere of
a p-star is of the same order as in strange stars, where $n_Q \simeq 10^{33} \, cm^{-3}$ (see for instance
Ref~\cite{alcock:1986}). So that, due to the very high electron density of electrons near the surface of the star, the
thermal photon-pair plasma is efficiently cooled to a final temperature much smaller than $T_{plasma}$. At the same time,
the energy transferred to the stellar surface leads to an increase of the effective quiescent temperature. Therefore we
are lead to conclude that during the burst activity the quiescent luminosity must increase. Let $T_1$ be the final plasma
temperature, then we see that the thermal photon-pair plasma contribution to the luminosity can be accounted for with an
effective blackbody with temperature $T_1$ and radius $R_1$ of the order of the stellar radius. As a consequence the
resulting blackbody luminosity is:
\begin{equation}
\label{4.82}
 L_{1} \; = \; 4 \, \pi \, R_1^2 \, \sigma_{SB} \, T_1^4 \; \; \; , \; \; \; R_1 \; \lesssim \; R \; \; .
\end{equation}
In general, the estimate of the effective blackbody temperature $T_1$ is quite difficult. However, according to
Eq.~(\ref{4.78}) we known that $L_1$ must account for about  $0.147$ of the total luminosity. So that we have:
\begin{equation}
\label{4.83}
 L_{1}(t) \; \simeq \; 0.147 \; L(t) \; \; .
\end{equation}
This last equation allow us to determine the blackbody temperature. For instance, soon after the hard spike we have $L(0)
\simeq \frac{E_{\emph{burst}}}{\delta  t_{spike}} \simeq 10^{45}  \; \frac{ergs}{sec}$ for the giant burst from {\it {SGR
1900+14}}. Thus, using $R_1  \simeq  R$, from Eq.~(\ref{4.83}) we get:
\begin{equation}
\label{4.84}
 T_1(0) \;  \simeq \; 61 \; KeV \; \; \; ,
\end{equation}
with surface luminosity $L_1(0) \simeq  10^{44}  \; \frac{ergs}{sec}$. \\
Let us consider the remaining spectral power with $\omega \lesssim 2 m_e$. We recall that the spectral power
Eq.~(\ref{spect-pow}) originates from the power supplied by the induced electric field Eq.~(\ref{4.68}). It is evident
from Eq.~(\ref{4.68}) that, as long as $v_\varphi \simeq 1$, the power supplied by the electric field $E_\varphi$ does not
depend on the mass of accelerated charges. Since the plasma in the magnetosphere is neutral, it follows that protons
acquire the same energy as electrons. On the other hand, since the protons synchrotron frequencies are reduced by a factor
$\frac{m_e}{m_p}$, the protons will emit synchrotron radiation near $\omega_1$. As a consequence, photons with frequencies
near $\omega_1$ suffer resonant synchrotron scattering, which considerably redistribute the available energy over active
modes. On the other hand, for $\omega \gg \omega_1$ the spectral power will follows the power law Eq.~(\ref{spect-pow}).
Thus, we may write:
\begin{equation}
\label{4.85}
 F(\omega) \;  \sim \;  \frac{1}{\omega^{\frac{4}{3}}} \; \; \; \; \; , \; \; \; \; \; 5 \,
            \omega_1 \; \lesssim \; \omega \; \lesssim \;  2 \, m_e \; \; \; ,
\end{equation}
where we have somewhat arbitrarily assumed the low energy cutoff $\thicksim 5 \, \omega_1$. On the other hand, for $\omega
\; \lesssim \; 5 \, \omega_1$ the redistribution of the energy by resonant synchrotron scattering over electron and proton
modes lead to an effective description of the relevant luminosity as thermal blackbody with effective temperature and
radius $T_2$ and $R_2$, respectively. Obviously, the blackbody radius $R_2$ is fixed by the geometrical constrain that the
radiation is emitted in the magnetosphere at distances $r \lesssim 10 \, R$. So that we have:
\begin{equation}
\label{4.86}
 R_2  \;  \lesssim  \; 10 \; R \; \;  \; .
\end{equation}
The effective blackbody temperature $T_2$ can be estimate by observing that the  integral of the spectral power up to $5
\, \omega_1$ account for about the $60 \; \% $ of the total luminosity. Thus, we have:
\begin{equation}
\label{4.87}
 L_{2}(t) \; \simeq \; 0.60 \; L(t) \; \; ,
\end{equation}
where
\begin{equation}
\label{4.88}
 L_{2} \; = \; 4 \, \pi \, R_2^2 \, \sigma_{SB} \, T_2^4 \; \; \; , \; \; \; R_2 \; \lesssim \; 10 \; R \; \; .
\end{equation}
Equations~(\ref{4.87}) and (\ref{4.88}) can be used to  to determine the effective blackbody temperature. If we consider
again  the giant burst from {\it {SGR 1900+14}}, soon after the hard spike, assuming  $R_2 \simeq 10 \, R$, we readily
obtain:
\begin{equation}
\label{4.89}
 T_2(0) \;  \simeq \; 27 \; KeV \; \; \; .
\end{equation}
To summarize, we have found that  the spectral luminosities can be accounted for by two blackbodies and a power law. In
particular for the blackbody components we have:
\begin{equation}
\label{4.90}
\begin{split}
 &  R_1  \; \lesssim \; R  \; \; , \; \; R_2 \lesssim \; 10 \; R \; \; ; \; \;
    T_2  \;  <  \; T_1 \; \; , \; \;  R_1  \; <  \; R_2 \;  \; \\
 &  \; \; \; \;  \; \; \; \; L_{1} \; \simeq \; 0.15 \; L \; \; \;, \; \; \;
    L_{2} \; \simeq \; 0.60 \; L \; \; \; .
\end{split}
\end{equation}
Interestingly enough, Eq.~(\ref{4.90}) displays an anticorrelation between blackbody radii and temperatures, in fair
agreement with observations. Moreover, the remaining $25 \%$ of the total luminosity is accounted for by a power law
leading to the high energy tail of the spectral flux:
\begin{equation}
\label{4.91}
\frac{d N}{d E}  \;  \sim \;   E^{- \alpha}  \; \; \; , \; \;  \; \alpha \; \simeq \; 2.33 \; \; \; \; ,
\end{equation}
extending up to $E \simeq 2 m_e \simeq 1 \, MeV$. Indeed, the high energy power law tail is clearly displayed in the giant
flare from {\it {SGR 1900+14}} (see Fig.~3 in Ref.~\cite{Feroci:1999}), and in the recent gigantic flare from {\it {SGR
1806-20}} (see Fig.~4 in Ref.~\cite{Hurley:2005}). \\
It is customary to fit the spectra with the sum of a power law and an optically thin thermal bremsstrahlung. It should be
stressed that the optically thin thermal bremsstrahlung model is purely phenomenological and without a physical basis. In
view of this, a direct comparison of our proposal with data is problematic. Fortunately, the authors of
Ref.~\cite{Feroci:2001} tested several spectral functions to the observed spectrum in the afterglow of the giant outburst
from {\it {SGR 1900+14}}. In particular they found that, in the time interval $68 \, sec \, \lesssim \, t \, \lesssim \,
195 \, sec$, the minimum $\chi^2$ spectral model were composed by two blackbody laws plus a power law. By fitting the time
averaged spectra they reported~\cite{Feroci:2001}:
\begin{equation}
\label{4.92}
 T_2 \;  \simeq \; 9.3 \; KeV \; \;  , \; \; T_1 \;  \simeq \; 20.3 \; KeV \; \; \; , \; \; \; \alpha \; \simeq \; 2.8 \; \; .
\end{equation}
Moreover, it turns out that the power law accounts for approximately $10 \%$ of the total energy above $25 \; KeV$, while
the low temperature blackbody component accounts for about  $85 \%$ of the total energy above $25 \; KeV$. In view of our
neglecting  the contribution to energy from protons, we see that our proposal is in accordance with the observed energy
balance. Unfortunately, in Ref.~\cite{Feroci:2001} the blackbody radii are not reported. To compare our estimate of the
blackbody temperatures with the fitted values in Eq.~(\ref{4.92}), we note that our values reported in Eqs.~(\ref{4.84})
and (\ref{4.89}) correspond to the blackbody temperatures soon after the initial hard spike. Thus, we need to determine
how the blackbody temperatures evolve with time. To this end, we already argued that  soon after the initial spike the
luminosity decreases due to dissipative processes in the magnetosphere. In Sect.~\ref{light} we show that the fading of
the luminosity can be accurately reproduced without a precise knowledge of the microscopic dissipative mechanisms. In
particular, the relevant light curve is given by Eqs.~(\ref{5.6}) and (\ref{5.10}). At $t=0$ we have seen that the total
luminosity is well described by three different spectral components. During the fading of the luminosity, it could happens
that microscopic dissipative processes modify the different spectral components. However, it is easy to argue that this
does not happens. The crucial point is that the three spectral components originate from emission by a macroscopic part of
the magnetosphere; moreover the time needed to modify a large volume of magnetosphere by microscopic processes is much
larger than the dissipation time $t_{dis} \, \sim \, 10^2 \, sec$. Then we conclude that, even during the fading of the
luminosity, the decomposition of the luminosity into three different spectral components retain its validity. Now, using
the results in Sect.~\ref{light}, we find:
\begin{equation}
\label{4.93}
 \frac{L(t \, \simeq \, 68 \, sec)}{L(0)}  \; \simeq \; 3.67 \; 10^{-2} \; \; \; , \; \; \;
\frac{L(t \, \simeq \, 195 \, sec)}{L(0)}  \; \simeq \; 1.67 \; 10^{-2} \; \; \; .
\end{equation}
Combining Eq.~(\ref{4.93}) with Eqs.~(\ref{4.82}), (\ref{4.83}), (\ref{4.87}) and (\ref{4.88}) we obtain:
\begin{equation}
\label{4.94}
\begin{split}
 &  T_2(t \, \simeq \, 68 \, sec) \;  \simeq \; 11.8 \; KeV \; \; , \; \; T_2(t \, \simeq \, 195 \, sec)
        \;  \simeq \; 9.7 \; KeV  \; \; \; ,  \\
 &  T_1(t \, \simeq \, 68 \, sec) \;  \simeq \; 26.7 \; KeV \; \; , \; \; T_1(t \, \simeq \, 195 \, sec)\;  \simeq \; 21.9 \; KeV
     \; ,
\end{split}
\end{equation}
in reasonable agreement with Eq.~(\ref{4.92}). Finally, let us comment on the time evolution of the spectral exponent
$\alpha$ in the power law Eq.~(\ref{4.91}). From Eq.~(\ref{spect-pow}) it follows that high energy modes have less energy
to dissipate. Accordingly, once a finite amount of energy is stored into the magnetosphere, the modes with higher energy
become inactive before the lower energy modes. As a consequence, the effective spectral exponent will increases with time
and the high energy tail of the emission spectrum becomes softer, in perfect agreement with observations. This explains
also why the fitted spectral exponent $\alpha$ in Eq.(\ref{4.92}) is slightly higher than our estimate in Eq.(\ref{4.91}).
\subsection{\normalsize{HARDNESS RATIO}}
\label{hardness}
\begin{figure}[ht]
\includegraphics[width=0.90\textwidth,clip]{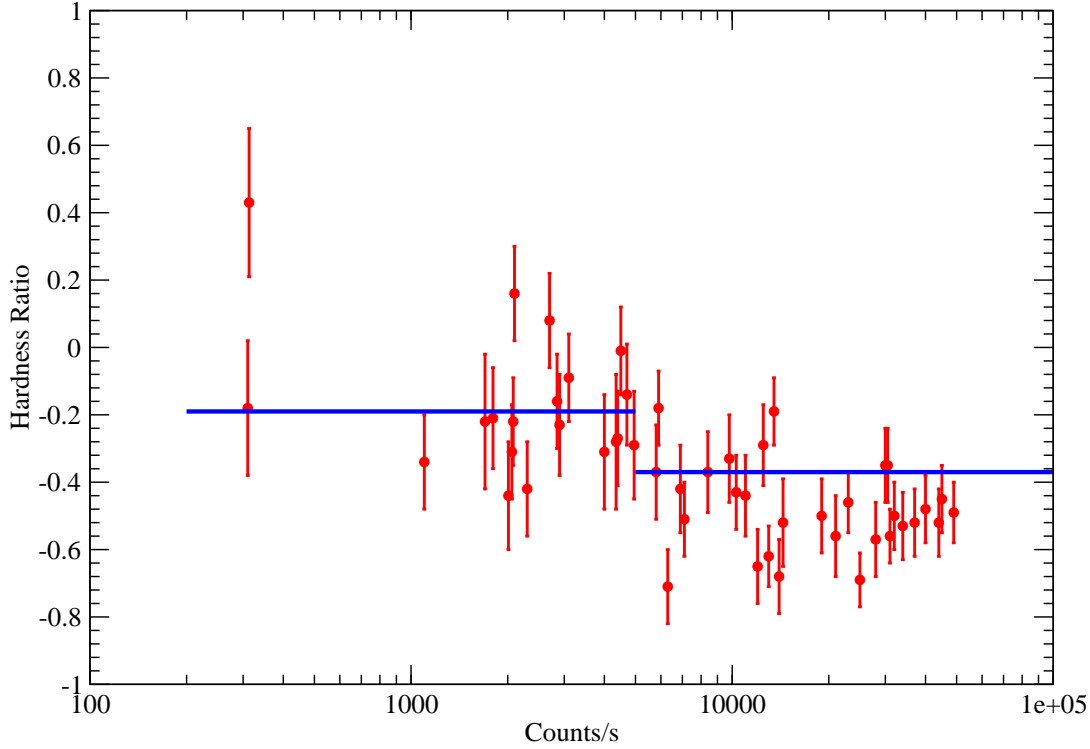}
\caption{\label{fig_4}
Hardness-intensity plot of the time resolved hardness ratio, Eq.~(\ref{4.95}). Data have been extract from Fig.~3 of
Ref.~\cite{Gotz:2004}. Blue lines are our Eqs.~(\ref{4.100}) and (\ref{4.102}).}
\end{figure}
Recently, it has been reported evidence for a hardness-intensity anti correlation within bursts from {\it {SGR
1806-20}}~\cite{Gotz:2004}. Indeed, the authors of Ref.~\cite{Gotz:2004} reported observations of the soft gamma ray
repeaters {\it {SGR 1806-20}} obtained in October 2003, during a period of bursting activity. They found that some bursts
showed a significant spectral evolution. However, in the present Section we focus on the remarkable correlation between
hardness ratio and count rate. Following Ref.~\cite{Gotz:2004} we define the hardness ratio as:
\begin{equation}
\label{4.95}
 HR \;  = \frac{H \; - \; S}{H \; + \; S} \; \; \; ,
\end{equation}
where $H$ and $S$ are the background subtracted counts in the ranges $40-100 \; KeV$ and $15-40 \; KeV$ respectively. In
Figure~\ref{fig_4} we report the hardness ratio data extracted from Fig.~3 of Ref.~\cite{Gotz:2004}. A few comments are in
order. First, the hardness ratio becomes negative for large enough burst intensities. Moreover, there is  a clear decrease
of the hardness ratio with increasing burst intensities. Note that, no detailed predictions are available within the
standard magnetar model. On the other hand, within our approach we are able to explain why the hardness ratio is negative
and decreases with increasing burst intensities. To see this, we note that the hardness ratio Eq.~(\ref{4.95}) is defined
in terms of total luminosities in the relevant spectral intervals. Thus, to determine the total luminosity in the spectral
interval $\omega_1-\omega_2$ we may use:
\begin{equation}
\label{4.96}
L (\omega_1-\omega_2) \; = \; \int_{\omega_1}^{\omega_2} \; F(\omega) \; d\omega \; \; ,
\end{equation}
where $ F(\omega)$ is given by Eq.~(\ref{spect-pow}). A straightforward integration gives:
\begin{equation}
\label{4.97}
L(\omega_1-\omega_2) \; \simeq \; 2 \; \pi^2 \; n_e \; e \; \frac{\delta B_S}{\delta t_{spike}} \;  R^4 \;
                      \left ( \frac{r_1}{R} \; - \; \frac{r_2}{R} \right ) \; \; ,
\end{equation}
where $r_1$ and $r_2$ are given by:
\begin{equation}
\label{4.98}
  \omega_{1,2}  \; \simeq \; \gamma^2 \; \frac{e  B_S}{ m_e} \; \left ( \frac{ R}{r_{1,2}} \right )^3 \; \; \; .
\end{equation}
Assuming $\gamma \thicksim 1$, we may rewrite Eq.~(\ref{4.98}) as:
\begin{equation}
\label{4.99}
  \omega_{1,2}  \; \simeq \; 10 \; MeV \;  \left ( \frac{ R}{r_{1,2}} \right )^3 \; \; \; .
\end{equation}
Using Eqs.~(\ref{4.97}) and ~(\ref{4.99}) it is easy to determine the hardness ratio:
\begin{equation}
\label{4.100}
 HR \;  = \frac{L(40-100 \; KeV) \; - \; L(15-40 \; KeV)}{L(40-100 \; KeV) \; + \; L(15-40 \; KeV)} \; \simeq \;
          \frac{1.66 \; - \; 2.44}{ 1.66 \; + \; 2.44} \; \simeq \; - \; 0.19 \; \; .
\end{equation}
In Figure~\ref{fig_4} we display our estimate of the hardness ratio Eq.~(\ref{4.100}). We see that data are in quite good
agreement with Eq.~(\ref{4.100}) at least up to count rate $\thicksim \, 5 \, 10^3 \; \frac{counts}{sec}$. For larger
count rates data seem to lie below our value. We believe that, within our approach, there is a natural explanation for
this effect. Indeed, for increasing count rates we expect that the hard tail $\omega \gtrsim 2 \, m_e \simeq 1 \, MeV$ of
the spectrum will begin to contribute to the luminosity. According to the discussion in Sect.~\ref{bursts} these hard
photons are reprocessed leading to an effective blackbody with temperature $T_1$. Now, for small and intermediate bursts
the blackbody temperature $T_1$ is considerably smaller than Eq.~(\ref{4.84}), so that the effective blackbody contributes
mainly to the soft tail of the spectrum. Obviously, the total luminosity of the effective blackbody is:
\begin{equation}
\label{4.101}
L(1-10 \; MeV) \; \simeq \; 2 \; \pi^2 \; n_e \; e \; \frac{\delta B_S}{\delta t_{spike}} \;  R^4 \;
                      \left ( 2.15 \; - \; 1 \right ) \; \; .
\end{equation}
Since this luminosity contributes to the soft part of the emission spectrum, Eq.~(\ref{4.100}) gets modified as:
\begin{equation}
\label{4.102}
 HR \;  \simeq \;  \frac{1.66 \; - \; 3.59}{ 1.66 \; + \; 3.59} \; \simeq \; - \; 0.37 \; \; .
\end{equation}
Equation~(\ref{4.102}) is displayed in Fig.~\ref{fig_4} for rates $\gtrsim  5 \, 10^3 \; \frac{counts}{sec}$. Note that we
did not take into account the proton contribution to the luminosity. Observing that protons contribute mainly to
luminosities at low energy $\omega \lesssim 10 \,  KeV$, we see that adding the proton contributions leads to smaller
hardness ratios bringing our estimates to a better agreement with data. In any case, we see that our theory allows to
explain in a natural way the puzzling anti correlation between hardness ratio and intensity.
\section{\normalsize{LIGHT CURVES}}
\label{light}
In our magnetar theory the observed burst activities are triggered by glitches which inject magnetic energy into the
magnetosphere where, as discussed in previous Sections, it is dissipated. As a consequence the observed luminosity is time
depended. In this Section we set up an effective description which allows us to determine the light curves, i.e. the time
dependence of the luminosity. In general, the  energy injected into the magnetosphere after the glitch decreases due to
dissipative effects described in Sect.~\ref{bursts}, leading to the luminosity  $L(t) = - \frac{d E(t)}{dt}$. Actually,
the precise behavior of $L(t)$ is determined once the dissipation mechanisms are known. However, since the dissipation of
the magnetic energy involves the whole magnetosphere, we may accurately reproduce the time variation of $L(t)$ without a
precise knowledge of the microscopic dissipative mechanisms. Indeed, on general grounds we have that the dissipated energy
is given by:
\begin{equation}
\label{5.1}
L(t) \; \; = \; \; - \; \frac{d E(t)}{dt} \; \; = \; \kappa(t) \; E^\eta \; \; , \; \; \eta \; \leq \; 1 \; \; \; ,
\end{equation}
where $\eta$ is the efficiency coefficient. Obviously the parameter $\kappa(t)$ does depend on the physical parameters of
the magnetosphere. For an ideal system, where the initial injected energy is huge, the linear regime, where $\eta = 1$, is
appropiate. Moreover, we expect that the dissipation time $ \thicksim \frac{1}{\kappa}$ is much smaller than the
characteristic time needed to macroscopic modifications of the magnetosphere. Thus, we may safely assume $\kappa(t) \simeq
\kappa_0$ constant. So that we get:
\begin{equation}
\label{5.2}
L(t) \; \; = \; \; - \; \frac{d E(t)}{dt} \; \; \simeq \; \kappa_0 \; E \; \; .
\end{equation}
It is straightforward to solve Eq.~(\ref{5.2}):
\begin{equation}
\label{5.3}
 E(t)  \; = \; E_0 \; \exp(- \frac{t}{\tau_0}) \; \; \; \; , \; \; \;
 L(t) \;  = \;  L_0 \; \exp(- \frac{t}{\tau_0}) \; \; , \; \; L_0 \; = \; \frac{E_0}{\tau_0} \; \; , \; \;
 \tau_0 \; = \; \frac{1}{\kappa_0} \; \; \; .
\end{equation}
Note that the dissipation time $\tau_0  = \frac{1}{\kappa_0}$ encodes all the physical information on the microscopic
dissipative phenomena. Since the injected energy is finite, the dissipation of energy degrades with the decreasing of the
available energy. Thus, the relevant equation is Eq.~(\ref{5.1}) with $\eta < 1$. In this case, solving Eq.~(\ref{5.1}) we
find:
\begin{equation}
\label{5.4}
 L(t) \;  = \;  L_0 \; \left ( 1 - \frac{t}{t_{dis}} \right )^{\frac{\eta}{1-\eta}}  \; \; ,
\end{equation}
where we have introduced the dissipation time:
\begin{equation}
\label{5.5}
t_{dis}  \; \; = \; \; \frac{1}{\kappa_0} \; \frac{E_0^{1-\eta}}{1-\eta}  \; \; .
\end{equation}
Then, we see that the time evolution of the luminosity is linear up to some time $t_{break}$, after that we have a break
from the linear regime $\eta = 1$ to a non linear regime with $\eta < 1$. If we indicate with $t_{dis}$ the total
dissipation time,  we get:
\begin{equation}
\label{5.6}
\begin{split}
L(t) \; &    = \;  L_0 \; \exp(- \frac{t}{\tau_0}) \; \; \; \;  \; \; ,  \; \; \; \; \; \; \; \; \; \; \; \; \; \; \; \;
                   \; \; \; \; \; \; \; \; \; \; \; \; \; \; \; \;\; \; 0 \; < \; t \; < \; t_{break}  \; \;  ,  \\
L(t) \; &    = \;  L(t_{break}) \;  \left ( 1 - \frac{t-t_{break}}{t_{dis}-t_{break}} \right )^{\frac{\eta}{1-\eta}}  \;
\; ,
               \; \; \;  \; \; \; \;  \; \; t_{break}  \; < \; t \; < \; t_{dis}   \; \;  .
\end{split}
\end{equation}
Equation~(\ref{5.6}) is relevant to describe the light curve after a giant burst, where there is a huge amount of magnetic
energy dissipated into the magnetosphere. It is interesting to compare our light curves, Eq.~(\ref{5.6}), with the
standard magnetar model. The decay of the luminosity in the standard magnetar model is due to the evaporation by a
fireball formed after a giant burst and trapped onto the stellar surface~\cite{Duncan:1996,Duncan:2001}. Indeed, the
authors of Ref.~\cite{Feroci:2001} considered the light curves after the giant flare of 1998, August 27 from {\it {SGR
1900+14}}, and the giant flare of 1979 March 5 from {\it {SGR 0526-66}}. Assuming that the luminosity varies as a power of
the remaining fireball energy $L \thicksim E^a$, they found:
\begin{equation}
\label{5.7}
 L(t) \;  = \;  L_0 \; \left ( 1 - \frac{t}{t_{evap}} \right )^{\frac{a}{1-a}}  \; \; ,
\end{equation}
where $t_{evap}$ is the time at which the fireball evaporates, and the index $a$ accounts for the geometry and the
temperature distribution of the trapped fireball. For a spherical fireball of uniform temperature $a = \frac{2}{3}$, so
that the index $a$ must satisfies the constrain:
\begin{equation}
\label{5.8}
a \;  \leqslant \frac{2}{3}  \;   \; \; .
\end{equation}
Note that our Eq.~(\ref{5.6}) reduces to Eq.~(\ref{5.7}) if $t_{break} = 0$ and $\eta = a$. However, we stress that our
efficiency exponent  must satisfy the milder constraint $\eta \leqslant 1$.

\begin{figure}[ht]
\includegraphics[width=0.9\textwidth,clip]{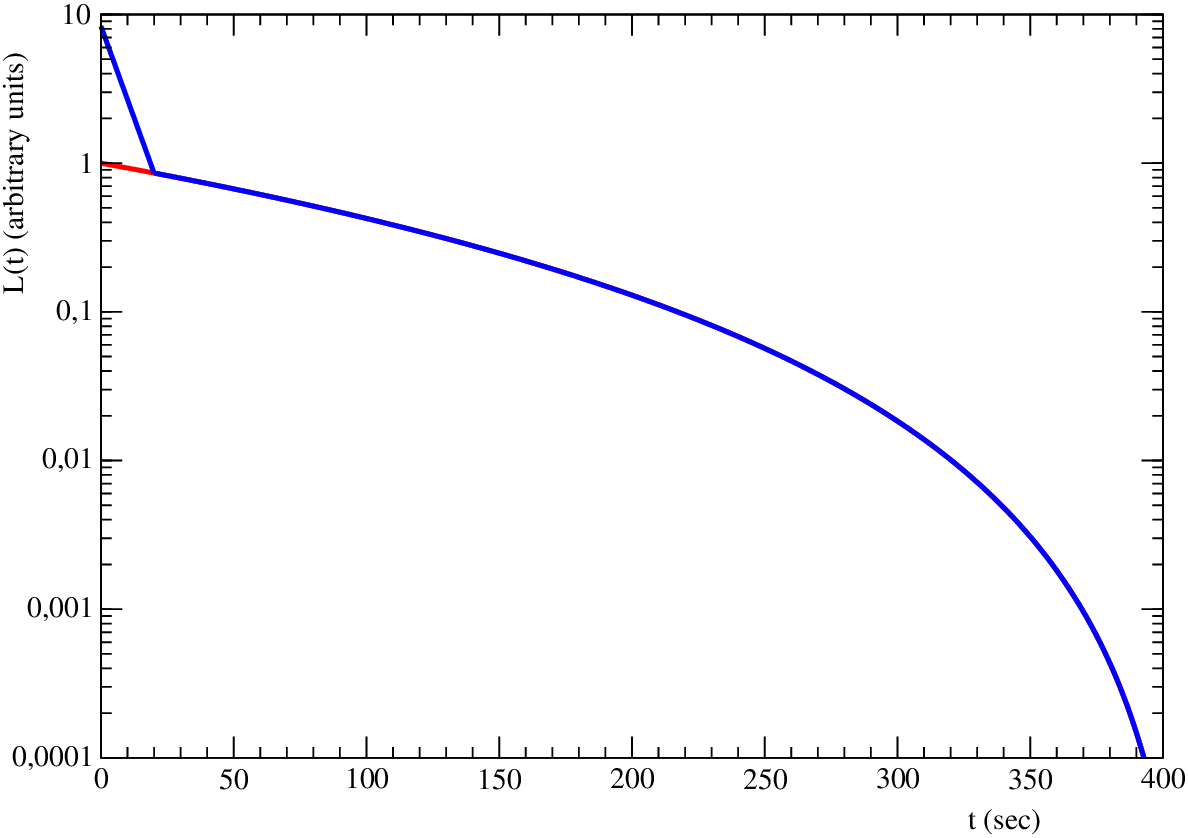}
\caption{\label{fig_5}
Light curve after the giant flare of 1998, August 27 from {\it {SGR 1900+14}}.  Red continuous line is the light curve in
the standard magnetar model, Eq.~(\ref{5.7}) with parameters given in Eq.~(\ref{5.9}). Blue line is our light curve
Eq.~(\ref{5.6}) with parameters in  Eq.~(\ref{5.10})}
\end{figure}
\begin{figure}[ht]
\includegraphics[width=0.9\textwidth,clip]{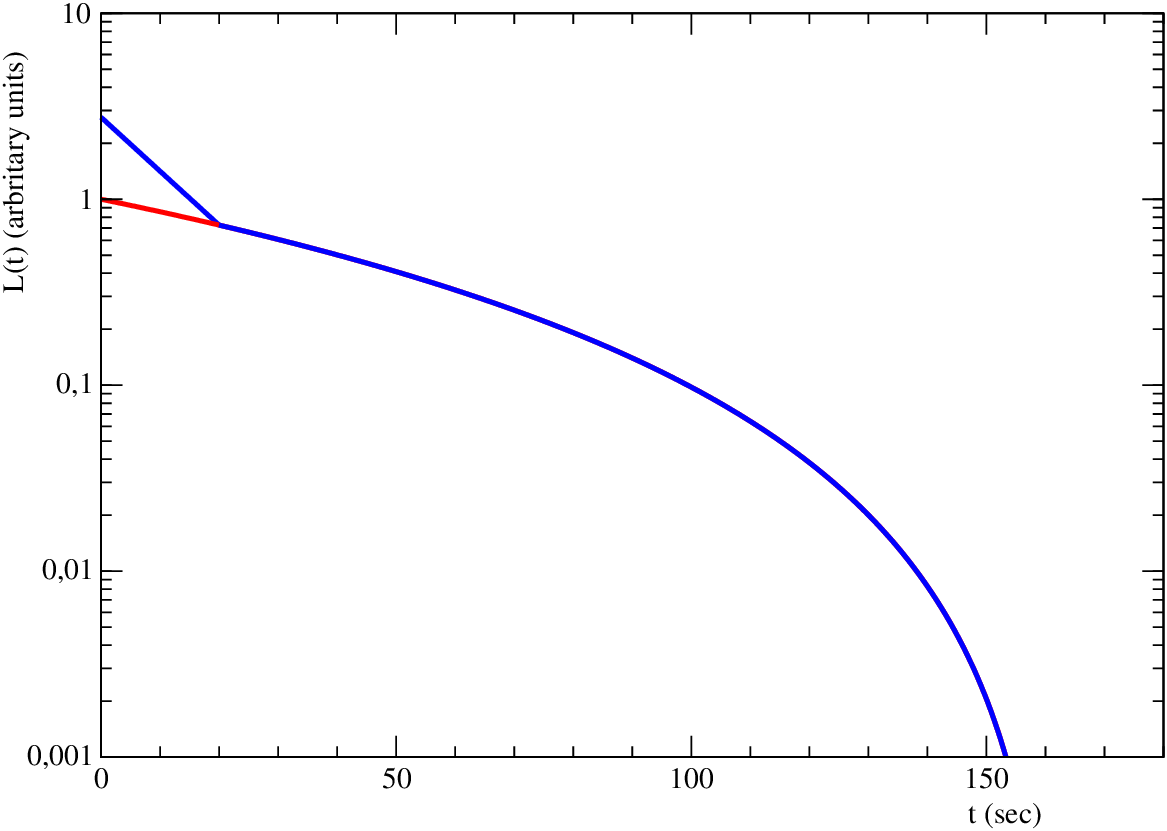}
\caption{\label{fig_6}
Light curve after the giant flare of 1979 March 5 from {\it {SGR 0526-66}}.  Red continuous line is the light curve in the
standard magnetar model, Eq.~(\ref{5.7}) with parameters given in Eq.~(\ref{5.11}). Blue line is our light curve
Eq.~(\ref{5.6}) with parameters in  Eq.~(\ref{5.12})}
\end{figure}
The authors of Ref.~\cite{Feroci:2001} performed a best fit of the light curve of the August 27 flare, background
subtracted and binned to $ 5 \, sec$, to Eq.~(\ref{5.7}) and found:
\begin{equation}
\label{5.9}
a \;  = \; 0.756 \; \pm \; 0.003 \; \; \; , \; \; \; t_{evap} \; = \;  414 \; sec  \; \; \;  .
\end{equation}
Indeed, from Fig.~2 in Ref.~\cite{Feroci:2001} one sees that the trapped fireball light curve account for the decay trend
of the experimental light curve and  matches the sudden final drop of the flux. However, it should stressed that the
fitting parameter $a$ in Eq.~(\ref{5.9}) does not satisfy the physical constraint Eq.~(\ref{5.8}).  Even more, any
deviations from spherical geometry or uniform temperature distribution lead to  parameters $a$ smaller than the upper
bound $\frac{2}{3}$. Moreover, the trapped fireball light curve underestimate by about an order of magnitude the measured
flux during the first stage of the outburst. We interpreted the different behavior of the flux during the initial phase of
the outburst as a clear indication of the linear regime described by our Eq.~(\ref{5.3}). As a matter of fact, we find
that the measured light curve could be better described by Eq.~(\ref{5.6}) with parameters (see Fig.~\ref{fig_5}):
\begin{equation}
\label{5.10}
\tau_0 \; = \; 8.80  \; sec \; \; , \; \; t_{break}  \; = \; 20 \; sec \; \; , \; \; \; \eta  \;  = \; 0.756 \; \; \; ,
 \; \; \;  t_{dis} \; = \;  414 \; sec  \; \; \; .
\end{equation}
The same criticisms apply to the fit within the standard magnetar model of the light curve after the giant flare of 1979
March 5 from {\it {SGR 0526-66}}. The trapped fireball light curve fit in Ref.~\cite{Feroci:2001} gives:
\begin{equation}
\label{5.11}
a \;  = \; 0.71 \; \pm \; 0.01 \; \; \; , \; \; \; t_{evap} \; = \; 163 \; \pm \; 5 \; sec  \; \; \;  .
\end{equation}
Again the parameter $a$ exceeds the bound Eq.~(\ref{5.8}), and the fit underestimates the flux during the first stage of
the outburst (see Fig.~14, Ref.~\cite{Feroci:2001}). Fitting our Eq.~(\ref{5.6}) to the measured flux reported in Fig.~14
of Ref.~\cite{Feroci:2001}, we estimate:
\begin{equation}
\label{5.12}
\tau_0 \; = \; 15  \; sec \; \; , \; \; t_{break}  \; = \; 20 \; sec \; \; , \; \; \; \eta  \;  = \; 0.71 \; \; \; ,
 \; \; \;  t_{dis} \; = \;  163 \; sec  \; \; \; .
\end{equation}
In Figs.~\ref{fig_5} and \ref{fig_6} we compare our light curves Eqs.~(\ref{5.6}), (\ref{5.10}) and (\ref{5.12}) with the
best fits performed in Ref.~\cite{Feroci:2001}. Obviously, both light curves agree for $t > t_{break}$, while in the
linear regime $t < t_{break}$, where the trapped fireball light curves underestimate the flux, our light curve follow the
exponential decay and seem to be in closer agreement with observational data.

Several observations indicate that after a giant burst there are smaller and recurrent bursts. According to our theory
these small and recurrent bursts are the effect of several small glitches following the giant glitch. We may think about
these small bursts like the seismic activity following a giant earthquake. These seismic glitches are characterized by
light curves very different from the giant burst light curves. In the standard magnetar model these light curves are
accounted for with an approximate $t^{-0.7}$ decay~\cite{Lyubarsky:2002}. Below we compare this prediction with
observations and argue that the standard theory is unable to adequately describe observational data. On the other hand,
within our theory there is a natural way to describe the seismic burst activity.  Indeed, during these seismic bursts,
that we shall call settling bursts, there is an almost continuous injection of energy into the magnetosphere which tends
to sustain an almost constant luminosity. This corresponds to an effective $\kappa$ in Eq.~(\ref{5.1}) which decreases
smoothly with time. The simplest choice is:
\begin{equation}
\label{5.13}
\kappa(t) \; = \; \frac{\kappa_0}{1 + \kappa_1 t}  \; \; \; .
\end{equation}
Inserting into  Eq.~(\ref{5.1}) and integrating, we get:
\begin{equation}
\label{5.14}
 E(t)  \; = \; \left [ E_0^{1-\eta} \; - (1 - \eta) \;  \frac{\kappa_0}{\kappa_1} \; \ln (1 + \kappa_1 t)
  \right ]^{\frac{\eta}{1-\eta}} \; \; \; .
\end{equation}
So that the luminosity is:
\begin{equation}
\label{5.15}
 L(t)  \; = \; \frac{L_0}{(1 + \kappa_1 t)^{\eta}} \; \left [1  \; -  (1 - \eta) \;  \frac{\kappa_0}{\kappa_1 E_0^{1-\eta}}
          \; \ln (1 + \kappa_1 t) \right ]^{\frac{\eta}{1-\eta}} \; \; \; .
\end{equation}
After defining the dissipation time:
\begin{equation}
\label{5.16}
 \ln (1 + \kappa_1 t_{dis}) \; =  \;  \frac{\kappa_1}{\kappa_0 }\; \frac{E_0^{1-\eta}}{1 - \eta} \; \; \; ,
\end{equation}
we rewrite Eq.~(\ref{5.15}) as
\begin{equation}
\label{5.17}
 L(t)  \; = \; \frac{L_0}{(1 + \kappa_1 t)^{\eta}} \; \left [ 1  \; -   \;  \frac{\ln (1 + \kappa_1 t)}{\ln (1 + \kappa_1
           t_{dis})}  \right ]^{\frac{\eta}{1-\eta}} \; \; \; .
\end{equation}
Note that the light curve Eq.~(\ref{5.17}) depends on two characteristic time constants $\frac{1}{\kappa_1}$ and
$t_{dis}$. We see that $\kappa_1 t_{dis}$, which is roughly the number of small bursts occurred in the given event, gives
an estimation of the seismic burst intensity. Moreover, since during the seismic bursts the injected energy is much
smaller than in giant bursts, we expect that fitting Eq.~(\ref{5.17}) to the observed light curves will result in values
of $\eta$ smaller than the typical values in giant bursts. In the following Sections we show that, indeed, our light
curves Eq.~(\ref{5.17}) are in good agreement with several observations.
\subsection{\normalsize{AXP 1E 2259+586}}
\label{2259}
On 2002, June 18 SGR-like bursts was recorded from {\it{AXP 1E 2259+586}}. Coincident with the burst activity were gross
changes in the pulsed flux, persistent flux, energy spectrum, pulse profile and spin down of the
source~\cite{Woods:2004a}. As discussed in previous Sections, these features are naturally accounted for within our
theory. However, we believe that the most remarkable and compelling evidence for our proposal comes from the observed
coincidence of the burst activity with a large glitch. Moreover, the time evolution of the unabsorbed flux from {\it{AXP
1E 2259+586}} following the 2002 June outburst reported in Ref.~\cite{Woods:2004a} can be explained naturally within our
theory, but it is completely unreachable within the standard magnetar theory. So that we consider the June 18 SGR-like
bursts from {\it{AXP 1E 2259+586}} the Rosetta Stone for our magnetar theory. \\
The temporal decay of the flux during the burst activity displays a rapid initial decay which lasted about $1 \; days$,
followed by a more mild decay during the year following the onset of the burst activity. Indeed, the authors of
Ref.~\cite{Woods:2004a} splitted the data into two segments, and fit each independently to a power law:
\begin{equation}
\label{5.19}
\begin{split}
F(t) \; &  \; \;  \thicksim \; \;  t^{\alpha_1}  \;   \; , \; \alpha_1 \; = \; - \; 4.8 \; \pm \; 0.5 \; \; \; \; \;
                  \; , \; \;  t  \;  \lesssim 1 \; \; days \;  \; \;  ,  \\
F(t) \; &   \; \;  \thicksim \; \;  t^{\alpha_2}  \;   \; , \; \alpha_2 \; = \; - \; 0.22 \; \pm \; 0.01 \; \; \, ,
            \; \; t  \;   \gtrsim 1 \; \; days \;  \; \;    .
\end{split}
\end{equation}
\begin{figure}[ht]
\includegraphics[width=0.9\textwidth,clip]{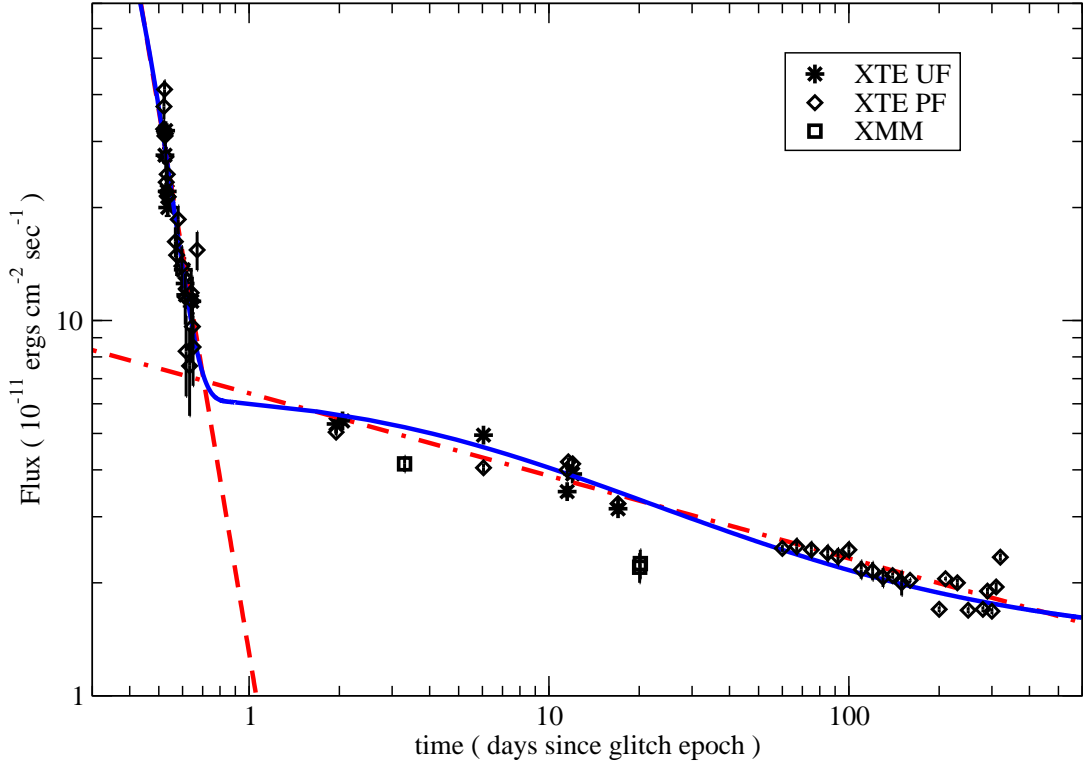}
\caption{\label{fig_7}
The time evolution of the unabsorbed flux from {\it{AXP 1E 2259+586}} following the 2002 June outburst. Data have been
extracted from Fig.~13 in Ref.~\cite{Woods:2004a}. Red dashed and dot-dashed lines are the phenomelogical power law fits,
Eq.~(\ref{5.19}). Blue continuous line is our light curve Eq.~(\ref{5.21}), with parameters in Eq.~(\ref{5.24}).}
\end{figure}
It is evident from Eq.~(\ref{5.19}) that the standard magnetar model is completely unable to reproduce the phenomelogical
power law fit. On the other hand, even the phenomenological parametrization cannot account for the time evolution of the
flux. Indeed, if we assume the power law Eq.~(\ref{5.19}) for the decay of the flux, then we cannot explain why and how
the source returns in its quiescent state with quiescent flux~\cite{Woods:2004a}:
\begin{equation}
\label{5.20}
F_Q  \;  \;  \simeq  \; \;  1.53 \;  10^{-11} \; \frac{ergs}{ cm^{2} \, sec}  \; \; .
\end{equation}
Note that adding the quiescent flux to the power law decay does not resolve the problem, for in that case the fits worst
considerably. Our interpretation of the puzzling light curve displayed in Fig.\ref{fig_7} is that {\it{AXP 1E 2259+586}}
has undergone a giant burst at the glitch epoch, and soon after the pulsar has entered into a intense seismic burst
activity. Accordingly, the flux can be written as:
\begin{equation}
\label{5.21}
 F(t) \; = \;  F_{GB}(t) \;  + \; F_{SB}(t)  \; +  \; F_Q \; \; \; \; ,
\end{equation}
where $F_Q$ is the quiescent flux, Eq.~(\ref{5.20}), $F_{GB}(t)$ is the giant burst contribution to the  flux given by
Eq.~(\ref{5.6}), and $F_{SB}(t)$ is the seismic  burst contribution  given by Eq.~(\ref{5.17}). Since during the first
stage of the outburst there are no available data, we may parameterize the giant burst contribution as:
\begin{equation}
\label{5.22}
F_{GB}(t) \; = \;  F_{GB}(0) \;  \left ( 1 - \frac{t}{t_{GB}} \right )^{\frac{\eta_{GB}}{1-\eta_{GB}}} \; \; ,
               \; \; \;  0 \; < \; t \; < \; t_{GB}  \; \;  ,
\end{equation}
while $F_{SB}(t)$ is given by:
\begin{equation}
\label{5.23}
 F_{SB}(t)  \; = \; \frac{F_{SB}(0)}{(1 + \kappa_1 t)^{\eta_{SB}}} \;
          \left [ 1  \; -   \;  \frac{\ln (1 + \kappa_1 t)}{\ln (1 + \kappa_1
           t_{SB})}  \right ]^{\frac{\eta_{SB}}{1-\eta_{SB}}}  \; \; ,
               \; \; \;  0 \; < \; t \; < \; t_{SB} \; \; \; ,
\end{equation}
where $t_{GB}$ and $t_{SB}$ are the dissipation time for giant and seismic bursts respectively. In Fig.~\ref{fig_7} we
display our light curve Eq.~(\ref{5.21}) with the following parameters:
\begin{equation}
\label{5.24}
\begin{split}
F_{GB}(0) \; &  \simeq \; 1.5 \;  10^{-8} \; \frac{ergs}{ cm^{2} \, sec} \; \; , \; \;   \eta_{GB} \; \simeq \;
             0.828 \; \; , \; \; t_{GB} \;  \simeq \;  0.91 \; days \;   \\
F_{SB}(0) \; &  \simeq \; 5.0 \;  10^{-11} \; \frac{ergs}{ cm^{2} \, sec} \;  , \; \eta_{SB} \; \simeq \; 0.45
                 \; ,  \; t_{SB} \;  \simeq \;  10^3 \; days \;  ,  \; \kappa_1  \; \simeq \; 0.20 \; days^{-1}  \; .
\end{split}
\end{equation}
A few comments are in order. First, the agreement with data is rather good. Second, our efficiency exponent $\eta_{GB}$ is
consistent with the values found in the giant bursts from  {\it {SGR 1900+14}} and {\it {SGR 0526-66}}, and cannot be
justify in the standard magnetar model. On the other hand, quite consistently, we have $ \eta_{SB} < \eta_{GB}$. Finally,
we stress that from our interpretation of the light curve it follows that the onset of the intense seismic burst activity
($\kappa_1  t_{SB} \sim 200$) did not allow a reliable estimation of  $\frac{\delta \dot{\nu} }{\dot{\nu}}$, which we
predicted to be of order $10^{-2}$. \\
\begin{figure}[ht]
\includegraphics[width=0.9\textwidth,clip]{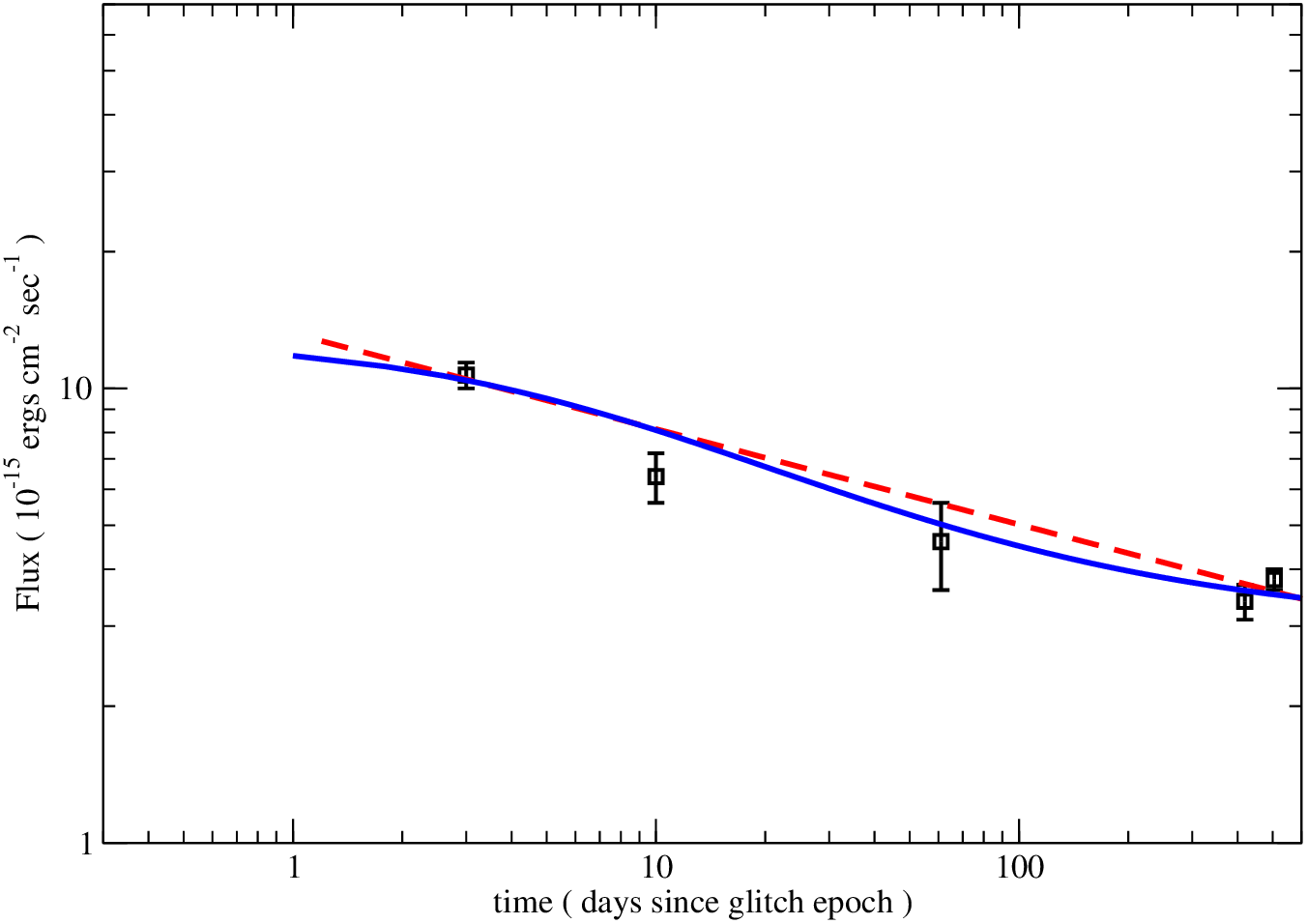}
\caption{\label{fig_8}
The time evolution of the unabsorbed IR flux from {\it{AXP 1E 2259+586}} following the 2002 June outburst. Data have been
extracted from Fig.~1 in Ref.~\cite{Tam:2004}. Red dashed line is the phenomelogical power law fit $t^{-0.21 \pm 0.01}$.
Blue continuous line is our light curve Eq.~(\ref{5.25}).}
\end{figure}
Interestingly enough, following the 2002 June outburst it was detected a infrared flux changes correlated with the $X$-ray
flux variability~\cite{Tam:2004}. Since the observations began three days after the 2002 June outburst, according to our
theory the infrared flux is parameterized as:
\begin{equation}
\label{5.25}
 F_{SB}^{IR}(t)  \; = \; \frac{F_{SB}^{IR}(0)}{(1 + \kappa_1 t)^{\eta_{SB}}} \;
          \left [ 1  \; -   \;  \frac{\ln (1 + \kappa_1 t)}{\ln (1 + \kappa_1
           t_{SB})}  \right ]^{\frac{\eta_{SB}}{1-\eta_{SB}}}  \; \; + \; \; F_{Q}^{IR} \; \; ,
               \; \; \;  0 \; < \; t \; < \; t_{SB} \; \; ,
\end{equation}
with the same parameters as in Eq.~(\ref{5.23}). Indeed, assuming $F_{SB}^{IR}(0) \simeq  9.5 \;  10^{-15}
\frac{ergs}{cm^{-2} sec^{-1}}$ and $F_{Q}^{IR} \simeq  3.3 \; 10^{-15} \frac{ergs}{cm^{-2} sec^{-1}}$, we found that our
light curve Eq.~(\ref{5.25}) is in remarkable good agreement with data (see Fig.~\ref{fig_8}). The strong correlation
between infrared and $X$-ray flux decay observed after the 2002 June outburst from {\it{AXP 1E 2259+586}} strongly suggest
a physical link between the origin of both type of radiation. In particular, this unambiguously implies that the infrared
flux originates from the magnetosphere. Indeed, in Appendix we argue that the soft emission from  gamma-ray repeaters,
anomalous $X$-ray pulsars, and isolated $X$-ray pulsars originates from thermal photons reprocessed by electrons trapped
above the polar cups. On the other hand, we have already shown that the burst activity produces an increases of the
thermal flux from the pulsar surface. So that, we see that in our theory the observed correlation between infrared and
$X$-ray flux decays finds a natural explanation.
\subsection{\normalsize{SGR 1900+14}}
\label{1900}
Soon after the 1998, August 27 giant burst, the soft gamma repeater {\it {SGR 1900+14}}  entered a remarkable phase of
activity. On August 29 an unusual burst from {\it {SGR 1900+14}} was detected~\cite{Ibrahim:2001} which lasted for a long
time $\thicksim 10^{3} sec$. As discussed in Ref.~\cite{Ibrahim:2001}, on observational grounds it can be ruled out
extended afterglow tails following ordinary bursts. Moreover, the standard magnetar model predicts faint transient
afterglows on a time scale comparable to the duration of the bright $X$-ray emission of the burst peak. So that such
mechanism cannot explain the long extended afterglow tail of August 29.
\begin{figure}[ht]
\includegraphics[width=0.9\textwidth,clip]{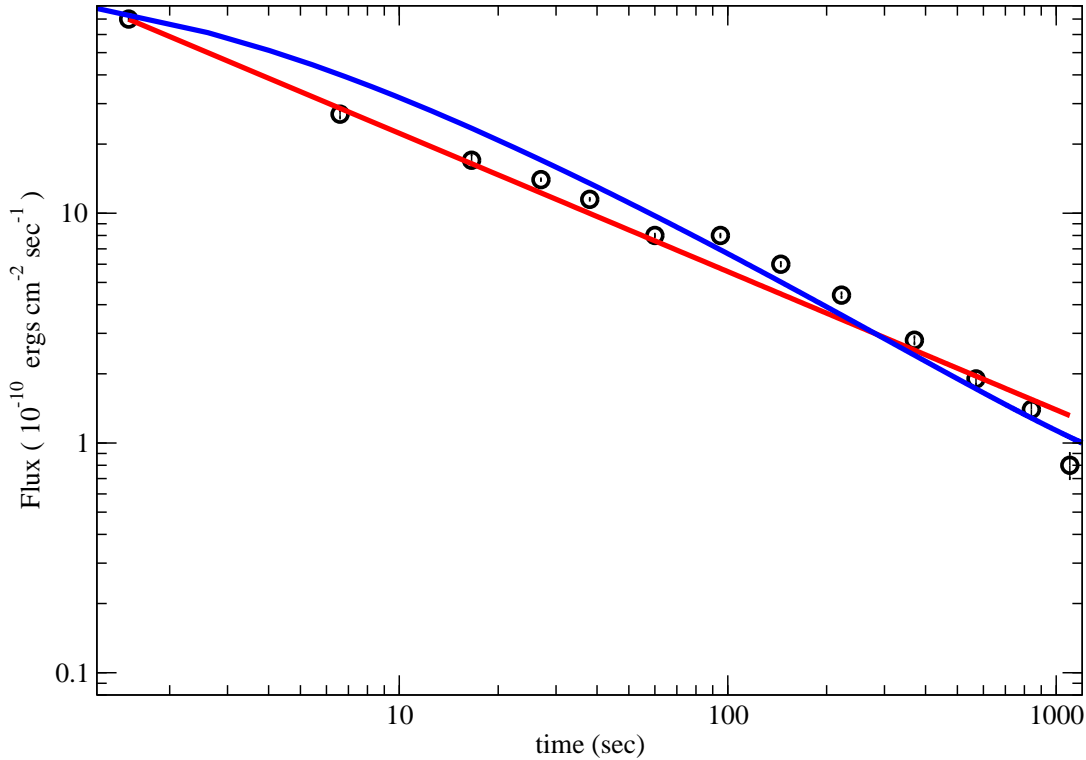}
\caption{\label{fig_9}
Flux evolution of the August 29 burst from {\it {SGR 1900+14}}. Data has been extracted from Fig.~4, panel (d), of
Ref.~\cite{Ibrahim:2001}. Red line is the phenomenological power law fit Eq.~(\ref{5.26}); blue line is our light curve
Eqs.~(\ref{5.28}) and (\ref{5.29}).}
\end{figure}
\begin{figure}[ht]
\includegraphics[width=0.9\textwidth,clip]{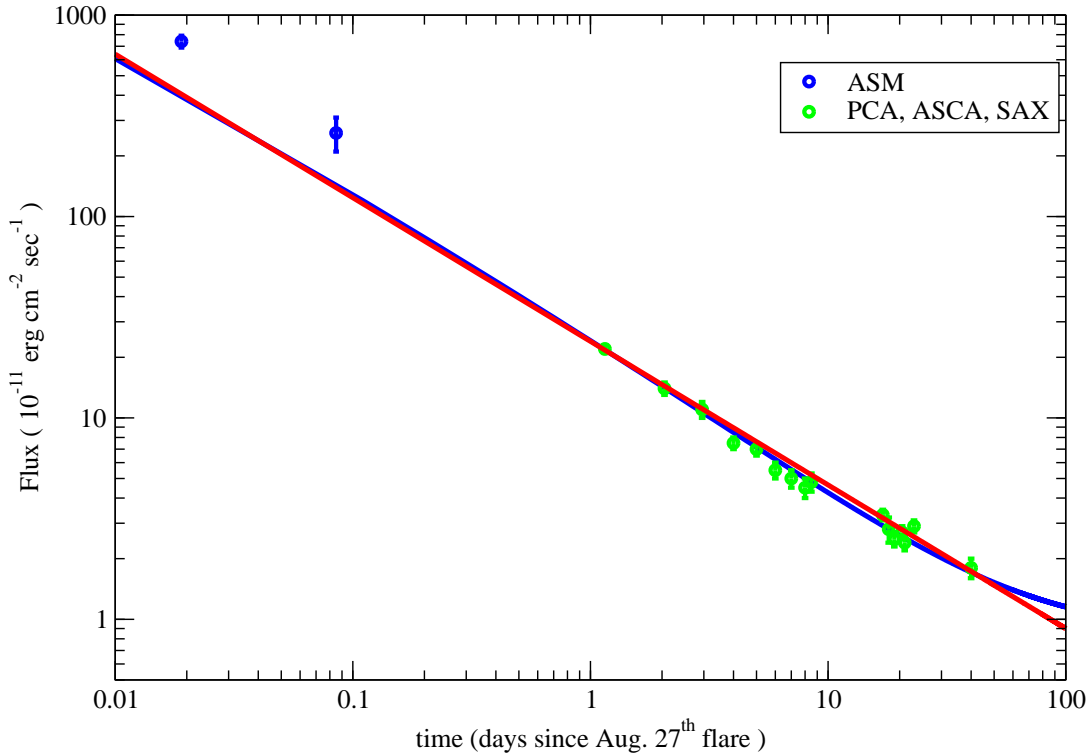}
\caption{\label{fig_10}
The time evolution of the unabsorbed flux from {\it{SGR 1900+14 }} following the 1998 August outburst. Data has been
extracted from Fig.~2 of Ref.~\cite{Woods:2001}. Red line is the power law best fit $F(t) \thicksim t^{-(0.713 \pm
0.025)}$. Blu line is our light curve Eqs.~(\ref{5.28}) and (\ref{5.30}).}
\end{figure}
\begin{figure}[ht]
\includegraphics[width=0.9\textwidth,clip]{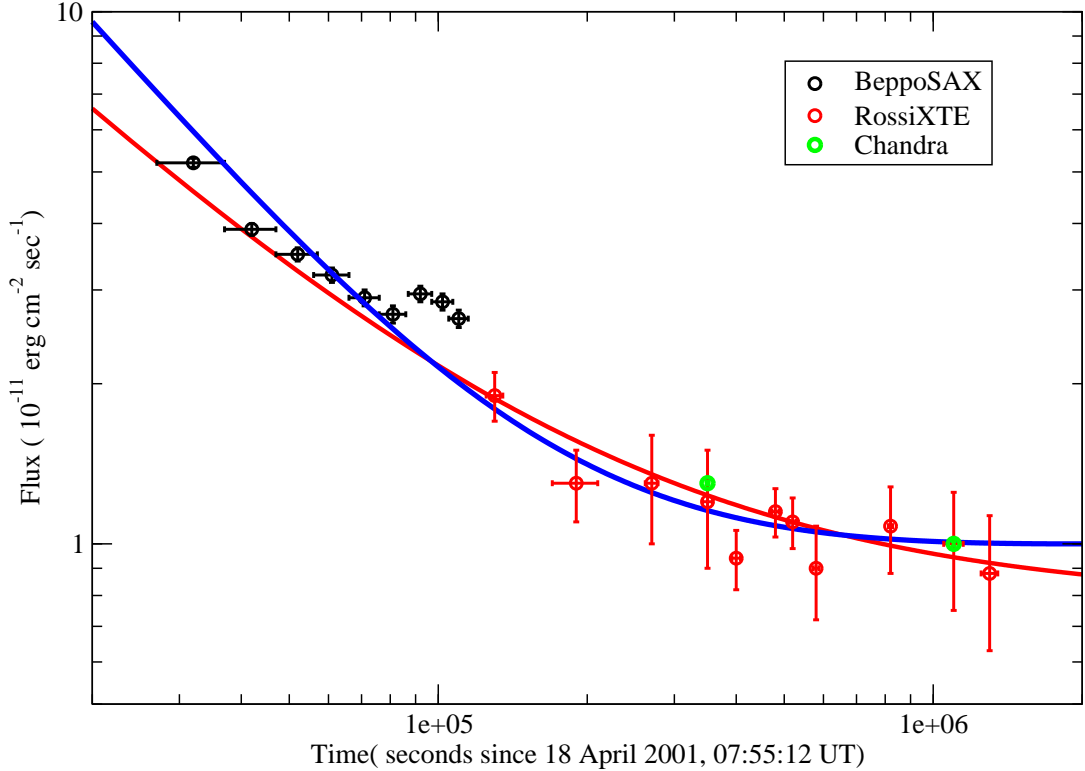}
\caption{\label{fig_11}
Temporal behavior of the $X$-ray flux from {\it {SGR 1900+14}} in the aftermath of the 2001 April 18 flare. Data have been
extracted from Fig.~2 of Ref.~\cite{Feroci:2003}. Red line is the power law best fit Eqs.~(\ref{5.31}) and (\ref{5.32}).
Blu line is our light curve Eqs.~(\ref{5.28}) and (\ref{5.33}).}
\end{figure}
\begin{figure}[ht]
\includegraphics[width=0.9\textwidth,clip]{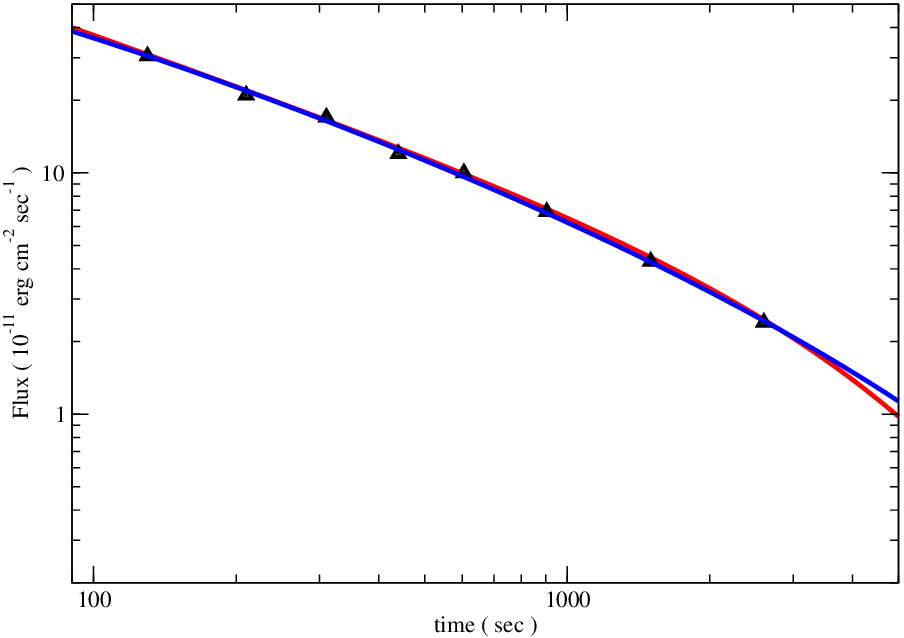}
\caption{\label{fig_12}
The temporal decay of the flux from the 2001 April 28 burst from {\it {SGR 1900+14}} in the energy band $2 - 20 \; KeV$.
The data has been extracted from Fig.~5 of Ref.~\cite{Lenters:2003}. Red continuous line is the phenomenological best-fit
power law times exponential function adopted in Ref.~\cite{Lenters:2003} to describe data, Eqs.~(\ref{5.34}) and
(\ref{5.35}). Blue continuous line is our light curve Eqs.~(\ref{5.28}) and (\ref{5.36}).
 }
\end{figure}
\begin{figure}[ht]
\includegraphics[width=0.9\textwidth,clip]{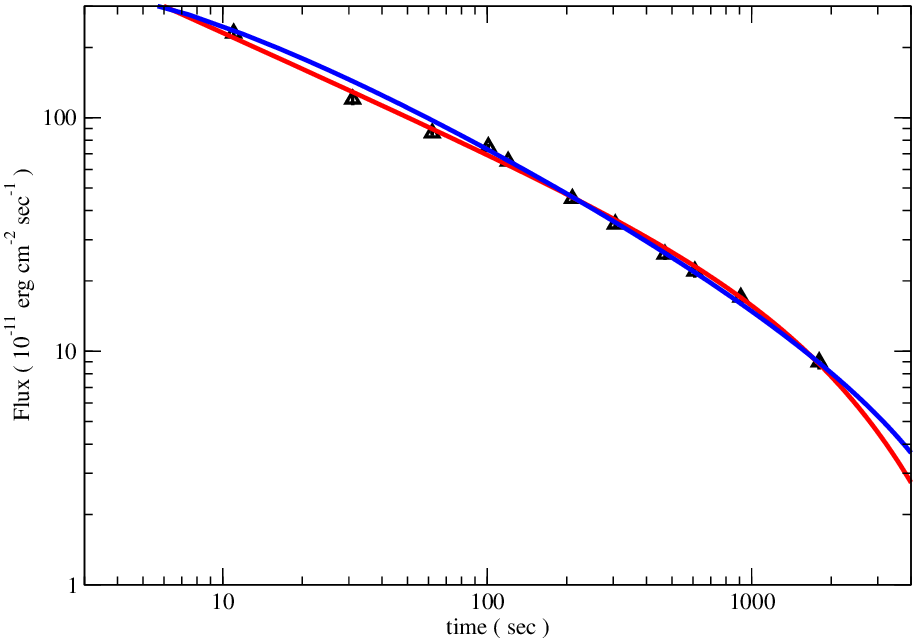}
\caption{\label{fig_13}
The temporal decay of the flux from the 1998 August 29 burst from {\it {SGR 1900+14}} in the energy band $2 - 20 \; KeV$.
The data has been extracted from Fig.~10 of Ref.~\cite{Lenters:2003}. Red continuous line is the phenomenological best-fit
power law times exponential function adopted in Ref.~\cite{Lenters:2003} to describe data, Eqs.~(\ref{5.34}) and
(\ref{5.37}). Blue continuous line is our light curve Eqs.~(\ref{5.28}) and (\ref{5.38}).
 }
\end{figure}
\begin{figure}[ht]
\includegraphics[width=0.9\textwidth,clip]{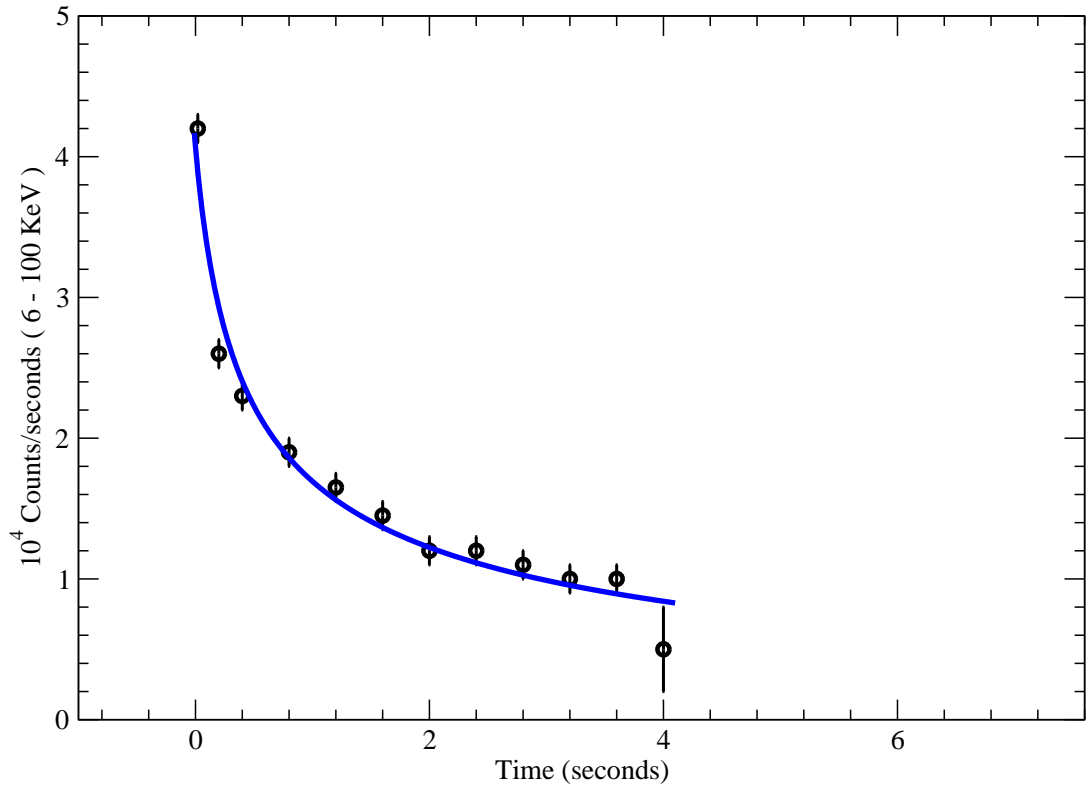}
\caption{\label{fig_14}
Time history of the 2001 July 2 burst from {\it {SGR 1900+14}} in the energy band $7 - 100 \; KeV$ as observed by FREGATE.
The data has been extracted from Fig.~1 of Ref.~\cite{Olive:2004} by binning the light curve histogram. Blue continuous
curve is our light curve Eqs.~(\ref{5.28}) and (\ref{5.39}).
}
\end{figure}

In Figure~\ref{fig_9} we display the flux decay after the August 29 burst. Data has been extracted from
Ref.~\cite{Ibrahim:2001}. In Ref.~\cite{Ibrahim:2001} the temporal behavior of the flux decay has been parameterized as a
power law (red line in Fig.~\ref{fig_9}):
\begin{equation}
\label{5.26}
F(t)   \;  \;  = \; \; (89.16 \; \pm \; 1.34)  \; 10^{-10}  \; \frac{ergs}{ cm^{2} \, sec}  \;  t^{-(0.602 \pm 0.025)}\; .
\end{equation}
As already stressed, the phenomenological power law decay cannot explain the return of the source in its quiescent state
where the flux is~\cite{Hurley:1999a}:
\begin{equation}
\label{5.27}
F_Q  \;  \;  =  \; \;  0.96 \; \pm 0.07 \;  10^{-11} \; \frac{ergs}{ cm^{2} \, sec}  \; \; .
\end{equation}
On the other hand, we may easily account for the observed flux decay by our light curve:
\begin{equation}
\label{5.28}
 F(t)  \; = \; \frac{F(0)}{(1 + \kappa_1 t)^{\eta}} \;
          \left [ 1  \; -   \;  \frac{\ln (1 + \kappa_1 t)}{\ln (1 + \kappa_1
           t_{dis})}  \right ]^{\frac{\eta}{1-\eta}}  \; \; + \; \; F_Q \; \; ,
\end{equation}
where $F_Q$ is fixed by Eq.~(\ref{5.27}). Indeed, in Fig.~\ref{fig_9} we compare our light curve Eq.~(\ref{5.28}) with
observational data. The agreement is quite satisfying if we take:
\begin{equation}
\label{5.29}
F(0) \;  \simeq \; 1.05 \;  10^{-9} \; \frac{ergs}{ cm^{2} \, sec} \;  , \; \eta \; \simeq \; 0.5
                 \; ,  \; t_{dis} \;  \simeq \; 1.2 \; 10^3 \; sec \;  ,  \; \kappa_1  \; \simeq \; 0.50 \; sec^{-1}  \; .
\end{equation}

The authors of Ref.~\cite{Woods:2001} have analyzed a large set of $X$-ray observations of {\it{SGR 1900+14 }} in order to
construct a more complete flux history. They found that the flux level was more than an order of magnitude brighter than
the level during quiescence. This transient flux enhancement lasts about $40 \; days$ after the giant flare. Unlike the
authors of Ref.~\cite{Woods:2001}, which argued that this enhancement was an artifact of the August 27 flare, we believe
that the flux history can be adequately described as seismic burst activity of the source. In Fig.~\ref{fig_10} we report
the flux light curve extracted from Fig.2 of Ref.~\cite{Woods:2001} together with their power law best fit. Again we find
the the flux history is accounted for quite well by our light curve Eq.~(\ref{5.28}) with the following parameters:
\begin{equation}
\label{5.30}
F(0) \;  \simeq \; 4.8 \;  10^{-8} \; \frac{ergs}{ cm^{2} \, sec} \;  , \; \eta \; \simeq \; 0.55
                 \; ,  \; t_{dis} \;  \simeq \; 200 \; days \;  ,  \; \kappa_1  \; \simeq \;2 \; 10^3  \; days^{-1}  \; .
\end{equation}
The agreement between our light curve Eqs.~(\ref{5.28}) and (\ref{5.30}) with the power law best fit is striking. Moreover
we see that our curve deviates from the power law fit for $t > 60 \; days$ tending  to $F_Q$ at $t = t_{dis}$.
The authors of Ref.~\cite{Woods:2001} noted that extrapolating the fit to the August 27 $X$-ray light curve back toward
the flare itself, one finds that the expected flux level lies below the ASM flux measurements (blue open points in
Fig.~\ref{fig_10}). Moreover, these authors observed that the discrepancy reduces somewhat when one pushes forward the
reference epoch to about 14 minutes after the onset of the flare. However, unlike the authors of Ref.~\cite{Woods:2001} we
believe that the discrepancy is due to a true physical effect, namely the observed discrepancy from extrapolated light
curve and ASM measurements is a clear indication that the surface luminosity increases after the burst activity. In
particular, soon after the  August 27 giant flare we have seen in  Sect.~\ref{bursts} that the surface temperature
increases up to $\thicksim 61 \; KeV$ and the surface luminosity reaches $\thicksim 10^{44} \; \frac{erg}{sec}$. Almost
all the deposited energy is dissipated within the dissipation time of the giant flare $\thicksim 400 \; sec$.
Nevertheless, it is natural to expect a more gradual afterglow where a small fraction of the energy deposited onto the
star surface is gradually dissipated. As a matter of fact, we find that the observed level of luminosity  $L_X \,
\thicksim \, 10^{38} \; \frac{erg}{sec}$ (assuming a distance $d = 10 \; kpc$) at about $0.01 \; days$ since the August 27
giant flare, is consistent with the gradual afterglow scenario.

On 2001 April 18 the soft gamma ray repeater {\it{SGR 1900+14 }} emitted an intermediate burst. The light curve of this
event did not show any initial hard spike and was clearly spin-modulated. Moreover, the energetics appeared to be
intermediate in the $40-700 \, KeV$ range, with a total emitted energy of about $1.9 \; 10^{42} \;
ergs$~\cite{Feroci:2003}. In Fig.~\ref{fig_11} we report the temporal behavior of the $X$-ray ($2 - 10 \; KeV$) flux from
{\it{SGR 1900+14 }} in the aftermath of the 2001 April 18 flare. Data has been extracted from Fig.~2 of
Ref.~\cite{Feroci:2003}. The authors of Ref.~\cite{Feroci:2003} attempted a simple power law function to the flux data:
\begin{equation}
\label{5.31}
 F(t)  \; \thicksim \; t^{-\alpha} \; \; + \; \; K \; \; ,
\end{equation}
where the constant $K$ should take care of the quiescent luminosity. Indeed, the authors of Ref.~\cite{Feroci:2003}
fitting Eq.~(\ref{5.31}) to the data found:
\begin{equation}
\label{5.32}
\alpha \; = 0.89(6) \; \; \; , \; \;  K \; = \; 0.78(5) \;  10^{-11} \; \; \frac{ergs}{ cm^{2} \, sec} \;  \; .
\end{equation}
As it is evident from Fig.~\ref{fig_11}, the power law  globally fits the data quite nicely. However, the reduced $\chi^2$
turns out to be in excess to $3$, mainly due to the bump occurring in the light curve at $t \thicksim 10^5 \,
sec$~\cite{Feroci:2003}. Indeed, after excluding the bump they get  a good fit with $\chi^2/dof \simeq 1$ without an
appreciable variation of the fit parameters~\cite{Feroci:2003}. However, there is still a problem with the phenomelogical
power law decay of the flux. As a matter of fact, Eq.~(\ref{5.32}) shows that the power law fit underestimates the
quiescent luminosity. In our opinion this confirms that the phenomenological power law decay of the flux is not adeguate
to describe the time variation of the flux. On the other hand, we find that our light curve Eq.~(\ref{5.28}), with
quiescent luminosity fixed to the observed value Eq.~(\ref{5.27}), furnishes a rather good description of the flux decay
once the parameters are given by:
\begin{equation}
\label{5.33}
F(0) \;  \simeq \; 2.6 \;  10^{-7} \; \frac{ergs}{ cm^{2} \, sec} \;  , \; \eta \; \simeq \; 0.68
                 \; ,  \; t_{dis} \;  \simeq \; 3 \; 10^6 \; sec \;  ,  \; \kappa_1  \; \simeq \; 0.25 \; sec^{-1}  \; .
\end{equation}
Within our interpretation, Eq.~(\ref{5.33}) shows that the flux decay in the aftermath of the April 18 flare is
characterized by a very large seismic burst activity ($\kappa_1 t_{dis} \thicksim 10^6$), which lasts for about $10^6 \;
sec$. So that the bump in the flux at  $t \thicksim 10^5 \, sec$ is naturally explained as  fluctuations in the intensity
of the seismic bursts.

In Ref.~\cite{Lenters:2003} it is reported the spectral evolution and temporal decay of the $X$-ray tail of a burst from
 {\it{SGR 1900+14 }} recorded on 2001 April 28, 10 days after the intense April 18 event. In Fig.~\ref{fig_12} we display
the temporal decay of the flux from the 2001 April 28 burst in the energy band $2 - 20 \; KeV$. The data has been
extracted from Fig.~5 of Ref.~\cite{Lenters:2003}. These authors attempted several functional forms to fit the decay of
the flux. They reported that the decay was equally well fitted by a either power law times exponential or broken power
law. We stress that both fits are phenomenological parametrization of the observational data, and that both fits are
unable to recover the quiescent flux. For definitiveness, we shall compare our light curve with the power law times
exponential fit:
\begin{equation}
\label{5.34}
 F(t)  \; \thicksim \; t^{-\alpha} \; \exp(-\frac{t}{\tau}) \; \; \; \; ,
\end{equation}
The best fit to the temporal decay of the flux from the 2001 April 28 burst in the energy band $2 - 20 \; KeV$
gives~\cite{Lenters:2003}:
\begin{equation}
\label{5.35}
\alpha \; = 0.68 \; \pm \; 0.04 \; \; \; \; , \; \; \; \; \tau \; = \; 5 \; \pm \; 1 \; 10^3 \; sec \; \; \; \; .
\end{equation}
In Figure~\ref{fig_12} we compare the phenomenological best fit Eqs.~(\ref{5.34}) and (\ref{5.35}) with our light curve
Eq.~(\ref{5.28}), where the quiescent luminosity is fixed to the observed value Eq.~(\ref{5.27}), and the parameters are
given by:
\begin{equation}
\label{5.36}
F(0) \;  \simeq \; 1.4 \;  10^{-9} \; \frac{ergs}{ cm^{2} \, sec} \;  , \; \eta \; \simeq \; 0.5
                 \; ,  \; t_{dis} \;  \simeq \; 5.5 \; 10^3 \; sec \;  ,  \; \kappa_1  \; \simeq \; 0.06 \; sec^{-1}  \; .
\end{equation}
Again, we see that our light curve gives a quite satisfying description of the flux decay.

Interestingly, the authors of Ref.\cite{Lenters:2003} analyzed with the same techniques the 1998 August 29 burst from {\it
{SGR 1900+14}}. The fit to the decay of the flux in the energy band $2 - 20 \; KeV$ with Eq.~(\ref{5.34}) resulted in:
\begin{equation}
\label{5.37}
\alpha \; =  \;  0.510 \; \pm \; 0.008 \; \; \;  ,  \; \; \; \tau \; = \; 2.9 \; \pm \; 0.2 \; 10^3 \; sec \; \; \; \; .
\end{equation}
Even in this case our light curve is able to follow the time decay of the flux in a satisfying way. In Fig.~\ref{fig_13}
we compare our light curve with parameters:
\begin{equation}
\label{5.38}
F(0) \;  \simeq \; 5.0 \;  10^{-9} \; \frac{ergs}{ cm^{2} \, sec} \;  , \; \eta \; \simeq \; 0.45
                 \; ,  \; t_{dis} \;  \simeq \; 7.0 \; 10^3 \; sec \;  ,  \; \kappa_1  \; \simeq \; 0.25 \; sec^{-1}  \; ,
\end{equation}
and the phenomelogical power law times exponential fit Eqs.~(\ref{5.34}) and (\ref{5.37}).

Let us consider, finally, the light curve for the intermediate burst from {\it {SGR 1900+14}} occurred on 2001 July
2~\cite{Olive:2004}. In Figure~\ref{fig_14} we display the time decay of the flux after the July 2 burst. The data have
been extracted from Fig.~1 of Ref.~\cite{Olive:2004} by binning the light curve histogram. The displayed errors are our
estimate, so that the data should be considered as purely indicative of the decay of the flux. We find that Fig.~1 of
Ref.\cite{Olive:2004} is very suggestive, for it seem to indicate that the burst results from several small bursts, i.e.
according to our theory the burst is a seismic burst. As a consequence we try the fit with our light curve
Eq.~(\ref{5.28}). In this case the quiescent flux has been fixed to $F_Q \simeq 0$, for the observational data has been
taken in the energy range $7 - 100 \; KeV$ where the quiescent flux is very small. Attempting the fit to the data we find:
\begin{equation}
\label{5.39}
F(0) \;  \simeq \; 4.07 \;  10^{4} \; \frac{counts}{sec} \;  , \; \eta \; \simeq \; 0.36
                 \; ,  \; t_{dis} \;  \simeq \; 40 \;  sec \;  ,  \; \kappa_1  \; \simeq \; 5.0 \; sec^{-1}  \; .
\end{equation}
The resulting light curve is displayed in Fig.~\ref{fig_14}. The peculiarity of this burst resides in the fact that the
burst activity terminates suddenly at $t \simeq 4 \; sec$ well before the natural end at $t_{dis}  \simeq  40 \; sec $.
\subsection{\normalsize{SGR 1627-41}}
\label{1627}
\begin{figure}[ht]
\includegraphics[width=0.9\textwidth,clip]{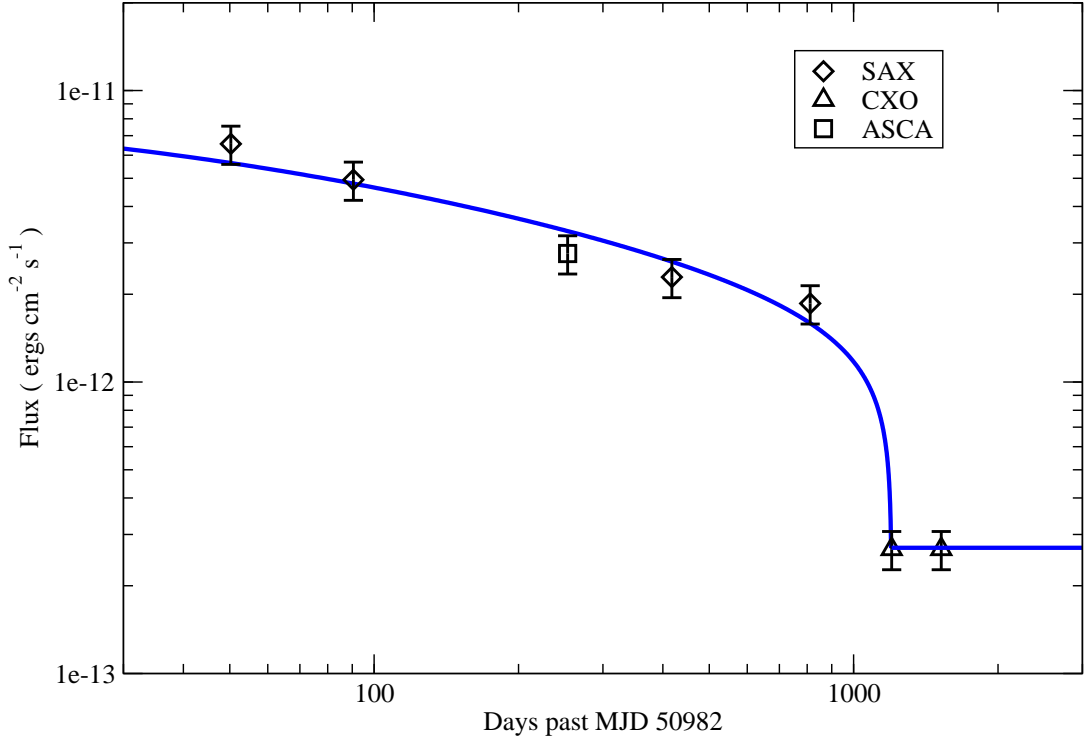}
\caption{\label{fig_15}
Time decay of the flux from {\it {SGR 1627-41}}. The data has been taken from Table~1 of Ref.~\cite{Kouveliotou:2003}.
Blue continuous line is our best fitted light curve Eqs.~(\ref{5.28}), (\ref{5.40}) and (\ref{5.41}).
 }
\end{figure}
{\it {SGR 1627-41}} was discovered with the Burst And Transient Source Experiment (BATSE) on the Compton Gamma-Ray
Observatory (CGRO) in June 1998~\cite{Kouveliotou:1998,Woods:1999a} when it emitted over 100 bursts within an interval of
6 weeks. The authors of Ref~\cite{Kouveliotou:2003} presented the results of the monitoring of the flux decay of the
$X$-ray counterpart of {\it {SGR 1627-41}} spanning an interval of roughly five years. Moreover, these authors attempted
to understand the three year monotonic decline of {\it {SGR 1627-41}} as cooling after a single deep crustal heating event
coinciding with the burst activity in 1998 within the standard magnetar model. They assumed an initial energy injection to
the crust of the order of $10^{44} \; ergs$. However, it must be pointed out that this assumption is highly unrealistic,
for the estimate of the total energy released in bursts during the activation of {\it {SGR 1627-41}} range between $ 4 \;
10^{42} - 2 \; 10^{43} \; ergs$. In addition, since gamma rays was not detected, they assume that the conversion
efficiency of the total energy released during the activation into soft gamma rays were considerable less than $ 100 \;
\%$. They also assumed that the core temperature is low, i.e. the core cools via the direct URCA
process~\cite{shapiro:1983}. Notwithstanding these rather ad hoc assumptions, the authors of Ref.~\cite{Kouveliotou:2003}
was unable to explain the March 2003 data point, which clearly showed that the flux did not decay further (see
Fig.~\ref{fig_15}). In other words, the levelling of the flux during the third year followed by its sharp decline is a
feature that is challenging that standard magnetar model based on neutron stars, and that beg for an explanation within
that model. On the other hand, we now show that the peculiar {\it {SGR 1627-41}} light curve find a natural interpretation
within our theory. In Fig.~\ref{fig_15} we display the time decay of the flux. The data has been taken from Table~1 of
Ref.~\cite{Kouveliotou:2003}. In this case we are able to  best fit our light curve Eq.~(\ref{5.28}) to available data.
Since the number of observations is rather low, to get a sensible fit we have fixed the dissipation time to $1200 \; days$
and the quiescent luminosity to the levelling value at $t \gtrsim 1200 \; days$:
\begin{equation}
\label{5.40}
F_Q \;  \simeq \; 2.7 \;  10^{-13} \; \frac{ergs}{ cm^{2} \, sec} \; \; \; , \; \;  \; t_{dis} \;  \simeq \; 1200 \; days
          \;  \; .
\end{equation}
The best fit of our light curve to data gives:
\begin{equation}
\label{5.41}
F(0) \;  = \; 0.83(11) \;  10^{-11} \; \frac{ergs}{ cm^{2} \, sec} \;  , \; \eta \; = \; 0.25(8)
                 \; ,   \; \kappa_1  \; = \; 0.04(1) \; days^{-1}  \; .
\end{equation}
with a reduced $\chi^2 \, \thicksim \, 1$. From Fig.~\ref{fig_15}, where we compare our best fitted light curve with data,
we see that our theory allow a quite satisfying description of the three year monotonic decline of the flux.
\subsection{\normalsize{SGR 1806-20}}
\label{1806}
\begin{figure}[th]
\includegraphics[width=0.9\textwidth,clip]{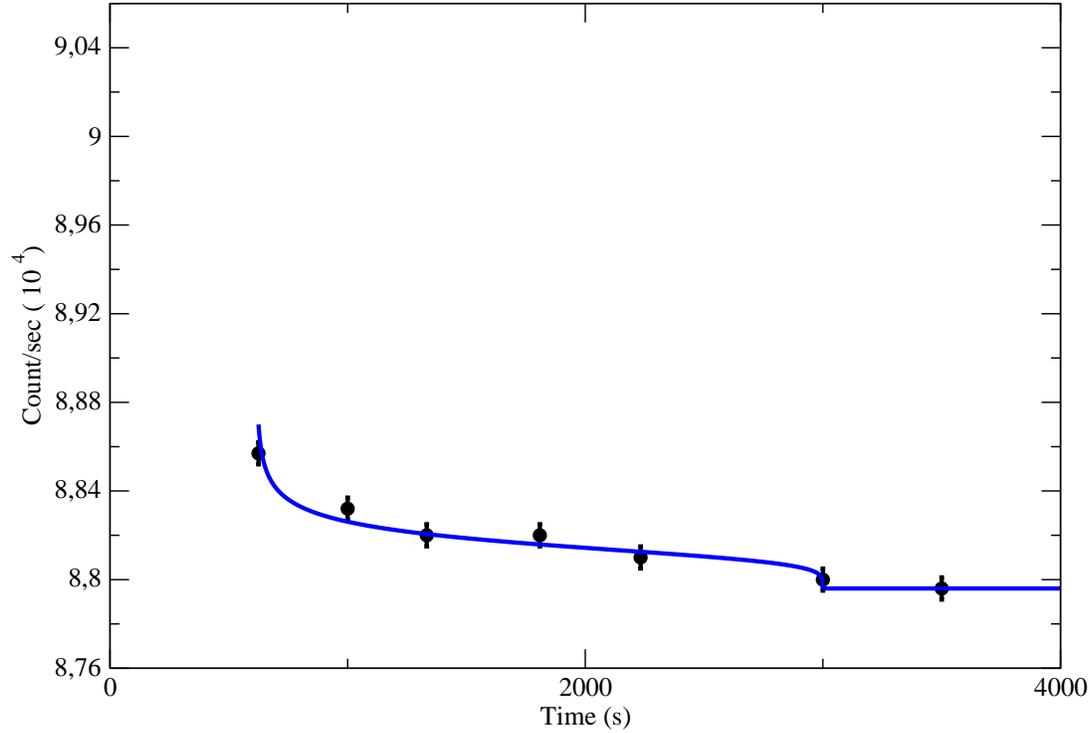}
\caption{\label{fig_16}
Time history of the second component after the 2004 December 27 giant burst from {\it {SGR 1806-20}}. The data has been
extracted from Fig.~5 of Ref.~\cite{Mereghetti:2005b} by binning the light curve histogram. Blue continuous curve is our
light curve Eqs.~(\ref{5.28}), (\ref{5.43}) and (\ref{5.44}). }
\end{figure}
{\it {SGR 1806-20 }} entered an active phase in 2003, culminating in a gigantic flare on 2004 December 27, with energy
greatly exceeding that of all previous events. In Figure~3 of Ref.~\cite{Hurley:2005} it is reported the time history of
the flux averaged over the rotation period of the pulsar soon after the giant flare. These authors fitted the light curve
within the standard magnetar model based on the evaporation of a fireball formed after the giant burst and trapped onto
the stellar surface, Eq.~(\ref{5.7}). The fit of the rotation smoothed curve to the fireball function
gives~\cite{Hurley:2005}:
\begin{equation}
\label{5.42}
 a  \;  = \; 0.606 \;  \pm \; 0.003 \; \; \; ,
 \; \; \;  t_{evap} \; = \;  382 \; \pm \; 3 \; \; sec  \; \; \; .
\end{equation}
However, from Figure~3 of Ref.~\cite{Hurley:2005} it is evident that fireball light curve underestimate the luminosity for
$t \lesssim 30 \; sec$. Thus we see that the light curve can be better accounted for by our light curve Eq.~(\ref{5.6})
with $t_{break} \simeq 30 \; sec$, quite close to the values found for the giant bursts from {\it {SGR 1900+14 }} and {\it
{SGR 0526-66}}. A more precise determination of the parameters of our light curve, however, must await for more precise
data in the initial phase of the afterglow. Instead, in the present Section we discuss the light curve of a second,
separate component after the giant burst reported in Ref.~\cite{Mereghetti:2005b}. The authors of
Ref.~\cite{Mereghetti:2005b} found evidence for a separate component in the light curve starting at $t \thicksim 400 \;
sec$ from the onset of the giant burst,  forming a peak at $t \thicksim 600 \; sec$ and ending at $t \thicksim 3000 \;
sec$ (see Fig.~5 of Ref.~\cite{Mereghetti:2005b}). As already discussed, in our theory it is expected that there is a
intense seismic burst activity following a giant burst. In Figure~\ref{fig_16} we display the flux history starting from
the giant flare. We show a few point of the second component extracted from Fig.~5 of Ref.~\cite{Mereghetti:2005b} by
binning the light curve histogram. The displayed errors are our estimate, so that the data should be considered as purely
indicative of the decay of the flux. We fit the data with our light curve Eqs.~(\ref{5.28}) assuming:
\begin{equation}
\label{5.43}
F_Q \;  \simeq \; 8.796 \;  10^{4} \; \frac{count}{sec} \; \;  , \; \; t_{dis} \; =  \; t_{end} \; -\; t_{start} \; \; ,
\; \; t_{start} \; \simeq \; 625 \; sec  \;  \; , \; \; t_{end} \; \simeq \; 3000 \; sec  \;  \; .
\end{equation}
The best fit of our light curve to data gives:
\begin{equation}
\label{5.44}
F(0) \;  = \; 0.074 \;  10^{4} \; \frac{count}{sec} \;  , \; \eta \; = \; 0.18
                 \; ,   \; \kappa_1  \; = \; 0.10 \; sec^{-1}  \; .
\end{equation}
Indeed, in Fig.~\ref{fig_16} we compare our  light curve with data and  find that our theory allow a quite satisfying
description of the time history of the flux.
\section{\normalsize{CONCLUSIONS}}
\label{conclusion}
Let us summarize the main results of the present paper. We have discussed p-stars endowed with super strong dipolar
magnetic field. We found a well defined criterion to distinguish rotation powered pulsars from magnetic powered pulsars
(magnetars). We showed that glitches, that in our magnetars are triggered by magnetic dissipative effects in the inner
core, explain both the quiescent emission and bursts from soft gamma-ray repeaters and anomalous $X$-ray pulsars. In
particular, we were able to account for the braking glitch from {\it {SGR 1900+14}} and the normal glitch from {\it {AXP
1E 2259+586}} following a giant burst. We accounted for the observed puzzling anti correlation between hardness ratio and
intensity. Within our magnetar theory we were able to account quantitatively for light curves for both gamma-ray repeaters
and anomalous $X$-ray pulsars. In particular we explained the light curve after the June 18, 2002 giant burst from {\it
{AXP 1E 2259+586}}. Finally, in Appendix we discussed the origin of the soft emission from soft gamma-ray repeaters,
anomalous $X$-ray pulsars, isolated $X$-ray pulsars. \\
We believe that anomalous $X$-ray pulsars and soft gamma-ray repeaters are two class of intriguing objects that  are
challenging the standard paradigm based on neutron stars. In the present paper we convincingly argued that the standard
magnetar theory is completely unable to account for the observational properties of  anomalous $X$-ray pulsars and soft
gamma-ray repeaters. On the other hand, we feel that the ability of our p-star theory to reach a complete understanding of
several observational features of soft gamma-ray repeaters and anomalous $X$-ray pulsars strongly supports our proposal
for a drastic revision of the standard paradigm of
relativistic astrophysics. \\
Let us conclude by briefly addressing the theoretical foundation of our theory. As a matter of fact, our proposal for
p-stars stems from recent numerical lattice results in QCD~\cite{cosmai:2003a,cosmai:2003b}, which suggested that the
gauge system gets deconfined in strong enough chromomagnetic field. This leads us to consider the new class of compact
quark stars made of almost massless deconfined up and down quarks immersed in a chromomagnetic field in
$\beta$-equilibrium. Our previous studies showed that these compact stars are more stable than neutron stars whatever the
value of the chromomagnetic condensate. This remarkable result indicates that the true ground state of QCD in strong
enough gravitational field is not realized by hadronic matter, but by p-matter. In other words, the final collapse of an
evolved massive star leads inevitably to the formation of a p-star.

\appendix
\section{\normalsize{ORIGIN OF THE SOFT EMISSION IN $X$-RAY PULSARS}}
\label{appendix}
A number of anomalous $X$-ray pulsars have recently been detected in the optical-infrared wavelengths (for a recent
review, see Ref.~\cite{Israel:2003}). Since pulsar surface emission cannot account for the observed soft emission spectra,
the emission must be  magnetospheric in origin. This is clearly demonstrated by the  correlation of the infrared flux with
the bursting activity recently observed for the anomalous $X$-ray pulsar {\it{AXP 1E 2259+586}}~\cite{Tam:2004} (see
Sect.~\ref{2259}). Remarkably, there is compelling evidence for an excess of the flux in the infrared band in anomalous
$X$-ray pulsars  with respect to the thermal  component extrapolated from $X$-ray data. Up to now the standard magnetar
model, is unable to account for the observed infrared emission or variability. In any case, irrespective to the actual
mechanism responsible for the soft emission there is no doubt that the optical-infrared emission originates in the
magnetosphere. Looking at the broad band energy spectrum of anomalous $X$-ray pulsars one realizes quickly that the
spectra are amazingly similar to the ones of isolated $X$-ray pulsars, like {\it {RXJ 1856.5-3754}} or {\it {RXJ
0720.4-3125}}. This strongly suggest that it must exist some natural mechanism capable to generate the soft tail of the
spectrum of isolated pulsar, anomalous $X$-ray pulsars and soft gamma ray repeaters. In Refs.~\cite{cea:2003,cea:2004} we
already advanced the proposal that the faint emissions from {\it {RXJ 1856.5-3754}} and {\it {RXJ 0720.4-3125}} originate
in the magnetosphere from synchrotron radiation emitted by electrons. That proposal was based on the observational fact
that the faint emission can be parameterized quite well by a non thermal power law. However, we did not address the
problem of the physical mechanism which is at the heart of the electron energy spectrum needed to generate the power law
emission. In this Appendix we shall discuss a fair natural mechanism which is able to explain the faint emission for
isolated pulsars as well as anomalous $X$-ray pulsars and soft gamma ray repeaters. In our mechanism the power law
emission in the infrared-optical band is due to thermal radiation reprocessed in the
magnetosphere by electrons trapped near the magnetic polar cups. \\
To start with, let us consider the motion of a charged particle in the pulsar magnetic dipolar field. We are assuming the
presence of neutral plasma formed by electrons and protons with number densities $n_e = n_p$. Now, as is well known, for
strong enough magnetic fields the plasma near the stellar surface will be channelled toward the magnetic pole. So that we
are led to consider the motion of charged particles which are drifting toward the magnetic polar cups. Further, we may
consider a polar region with area small enough such that the dipolar magnetic field depends only on the distance from the
surface. If the magnetic polar axis is in direction $z$, then charged particles moving towards the stellar surface will
feel an almost uniform magnetic field having z-component $B(z)$. Thus we are led to consider the motion of charge particle
in the magnetic field:
\begin{equation}
\label{A.1}
 B(z) \;  = \; - \;  B_S \; \frac{R^3}{z^3}  \; \;  , \; z \; \geq \; R \; \; ,
\end{equation}
where, for the sake of definitiveness we are considering the north magnetic pole where the magnetic field is entering into
the stellar surface.
Let us consider, firstly, electron with charge $- e$ and mass $m_e$. The electron wave function $\psi(x,y,z)$ can be
obtained by solving the Schr\"odinger equation in presence of the magnetic field $B(z)$. We are interested in the physical
problem where an electron, starting from a distance $z_0$ from the star, is approaching the stellar surface. Usually, it
is assumed that the magnetic field is almost constant, so that our problem reduces to the well known motion in an uniform
magnetic field. In this case one gets:
\begin{equation}
\label{A.2}
\psi(x,y,z) \;  = \; \exp ( - i p_z z) \; \phi_n(x,y)  \; \; \; ,
\end{equation}
with energy eigenvalues:
\begin{equation}
\label{A.3}
\varepsilon_{n,p_z} \;  = \;  \frac{p_z^2}{2 m_e}  \; + \; \frac{e B_S}{m_e} \; (n \,+ \frac{1}{2} \, \pm \, \frac{1}{2})
\; \; \; \; \; n \, = \, 0, 1, .... \; .
\end{equation}
However, if $z_0 \gg R$ the uniform field approximation is no longer valid, but the weakly varying field is more
appropriate. In this case we may write:
\begin{equation}
\label{A.4}
\psi(x,y,z) \;  = \; \zeta(z) \; \phi_n(x,y)  \; \; \; ,
\end{equation}
where $\phi_n(x,y)$ is the solution of the Schr\"odinger equation in the weakly varying  $B(z)$:
\begin{equation}
\label{A.5}
\frac{1}{2 m_e} \left [ - \, \frac{\partial^2}{\partial x^2} \; + \; e^2 B^2(z) y^2 \; + \;
 2 i B(z) \frac{\partial}{\partial x} \;  - \, \frac{\partial^2}{\partial y^2} \;
\right ] \phi_n(x,y) \; = \; \varepsilon_{n}(z) \phi_n(x,y) \; ,
\end{equation}
\begin{equation}
\label{A.6}
\varepsilon_{n}(z) \;  = \;   \frac{e B(z)}{m_e} \; (n \,+ \frac{1}{2} \, \pm \, \frac{1}{2}) \; \; \; \; \; n \, = \, 0,
1, .... \; .
\end{equation}
So that, writing the total energy of electrons drifting from $z_0$ toward the star as:
\begin{equation}
\label{A.7}
 E \;  = \;  \varepsilon_{n}(z_0) \; + \; \varepsilon_{drift} \; \; \;  \; ,
\end{equation}
the wave function $\zeta(z)$ satisfies the effective Schr\"odinger equation:
\begin{equation}
\label{A.8}
\frac{1}{2 m_e} \left [ - \frac{d^2}{d z^2} \; + \; V_B(z) \right ] \zeta(z) \; = \; \varepsilon_{drift} \; \zeta(z) \; ,
\end{equation}
where:
\begin{equation}
\label{A.9}
 V_B(z) \; = \;  \frac{e B_S}{m_e} \; \frac{R^3}{z_0^3} \; (n \,+ \frac{1}{2} \, \pm \, \frac{1}{2})
\; \;  \left [ \left ( \frac{z_0}{z} \right )^3 \; - \; 1 \right ] \; \; \; \; R \; \leq \; z \; \leq \; z_0 \; \; .
\end{equation}
Inspection of Eq.~(\ref{A.9}) shows that, as long as the total momentum is not exactly parallel to the magnetic field,
electrons will feel a repulsive barrier which at the stellar surface is:
\begin{equation}
\label{A.10}
 V_0 \; = \;  \frac{e B_S}{m_e} \; \simeq 11.6 \;  KeV \; B_{12} \; \; \; , \; \; \; B_{12} \; = \;
\frac{B_S}{10^{12} \; Gauss}   \; \; ,
\end{equation}
at least. This effect is the quantum mechanical counterpart of the well known fact that classical charges moving towards
regions with increasing magnetic fields are subject to a repulsive force. On the other hand, when we consider protons the
minimal height of the magnetic barrier is:
\begin{equation}
\label{A.11}
 V_0 \; = \;  \frac{e B_S}{m_p} \; \simeq 6.3 \; B_{12} \; eV \; \;  \; \; .
\end{equation}
In other words,  protons drifting towards the surface almost do not feel any barrier. As a consequence, there is an
accumulation of positive charge on the surface. Note that on the surface of the star there is a positively charged layer
which is able to support a thin crust of ordinary matter. Thus, protons do not dissolve into the star, but they are
trapped in the atmosphere of electrons which extends over a distance $\delta \thicksim 10^3 \; \emph{fermis}$ beyond the
edge. Let $n$ be the number density of trapped protons, then on the surface of the star there is a surface charge density
$\sigma \simeq \delta \, n$ which, in turn, gives rise to an uniform electric field. It follows that electrons that are
moving toward the star are subject to both the repulsive magnetic potential and the attractive electric potential. So that
the effective potential is:
\begin{equation}
\label{A.12}
 V(z) \; = \;  \frac{e B_S}{m_e} \; \frac{R^3}{z_0^3} \; (n \,+ \frac{1}{2} \, \pm \, \frac{1}{2})
\; \;  \left [ \left ( \frac{z_0}{z} \right )^3 \; - \; 1 \right ] \; + 4 \, \pi \, e^2 \, n \, \delta \, z \; , \; \; R
\; \leq \; z \; \leq \; z_0 \; \; .
\end{equation}
\begin{figure}[t]
\includegraphics[width=0.9\textwidth,clip]{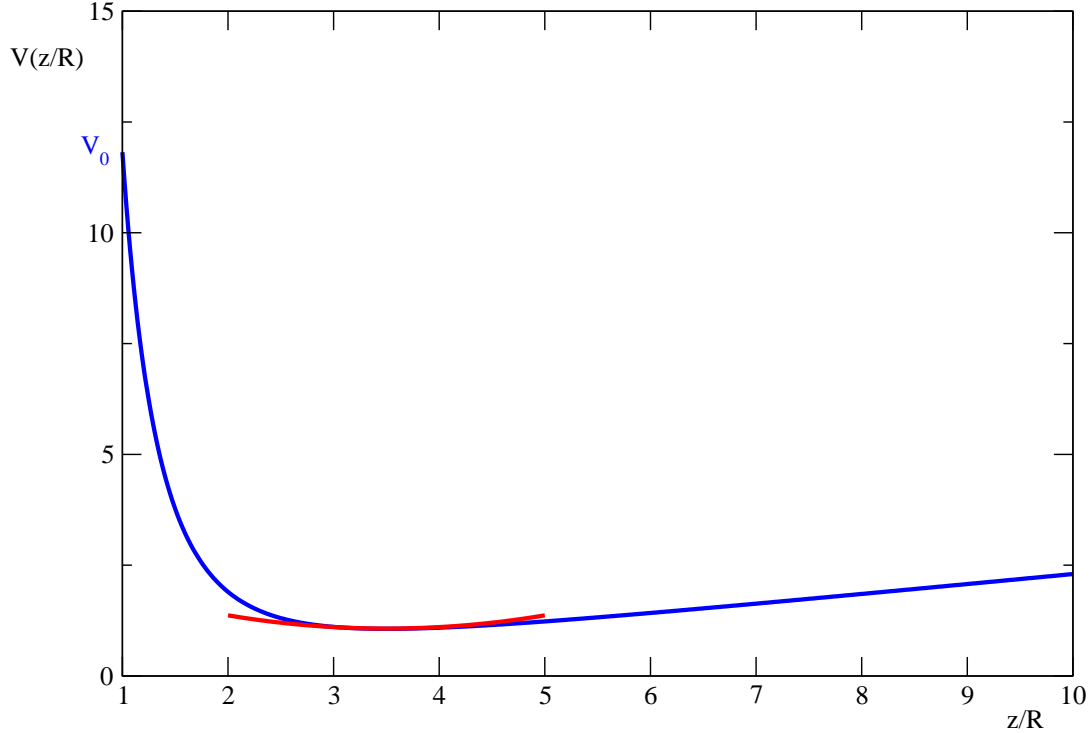}
\caption{\label{fig_17}
Blue line is the effective potential Eq.~(\ref{A.13}) for electrons drifting toward the stellar surface in the polar cup
regions. The red line is the harmonic approximation to the potential.}
\end{figure}
In the case of minimal magnetic barrier, using Eq.~(\ref{A.10}) we rewrite Eq.~(\ref{A.12}) as:
\begin{equation}
\label{A.13}
 V(z) \; \simeq \; 11.6 \;  KeV \; B_{12}  \; \frac{R^3}{z_0^3}  \;  \left [ \left ( \frac{z_0}{z} \right )^3 \; - \; 1 \right ] \;
 + \; 0.23  \;  KeV \; n_{13} \; \frac{z}{R} \; , \; \; 1
\; \leq \; \frac{z}{R} \; \leq \; \frac{z_0}{R} \; \; ,
\end{equation}
where $n_{13} = \frac{n}{10^{13} \, cm^{-3}} \, $.
In Fig.~\ref{fig_17} we display the effective potential Eq.~(\ref{A.13}) assuming $z_0 \simeq 10 \, R$. We see that the
effective potential $V(z)$ displays a minimum at $z = \bar{z}$ where the repulsive magnetic force is balanced by the
attractive electric force. Then, electron are trapped above the polar cup at a distance of order $\bar{z}$. To determine
the energy spectrum we need to solve the Schr\"odinger equation with the effective potential:
\begin{equation}
\label{A.14}
\frac{1}{2 m_e} \left [ - \frac{d^2}{d z^2} \; + \; V(z) \right ] \zeta(z) \; = \; \varepsilon_{drift} \; \zeta(z) \; .
\end{equation}
We may adopt the harmonic approximation to the potential by expanding around $\bar{z}$. A straightforward calculation
gives:
\begin{equation}
\label{A.15}
\bar{z} \; \simeq \; 3.5 \; R \;B_{12}^{\frac{1}{4}} \; n_{13}^{- \frac{1}{4}} \; \; , \; \; \bar{V} \; \equiv \;
V(\bar{z}) \; \simeq \; 1.08 \; KeV \;B_{12}^{\frac{1}{4}} \; n_{13}^{\frac{3}{4}} \; \; \; .
\end{equation}
Moreover:
\begin{equation}
\label{A.16}
 R^2 \; V^{''}(\bar{z}) \simeq \; 0.261 \; KeV \;B_{12}^{-\frac{1}{4}} \; n_{13}^{\frac{5}{4}} \; \;  \; .
\end{equation}
Note that, as long as $z_0 \gg R$, $\bar{z}$, $\bar{V}$ and $V^{''}(\bar{z})$ do not depend on $z_0$. In this
approximation $\zeta(z)$ satisfies the harmonic oscillator equation centered at $\bar{z}$ with frequency:
\begin{equation}
\label{A.17}
 \omega_m \simeq \; 0.46 \; 10^{-12} \; eV \;B_{12}^{-\frac{1}{8}} \; n_{13}^{\frac{5}{8}} \; \;  \; .
\end{equation}
Thus, we find for the drift energy the quasi continuum spectrum:
\begin{equation}
\label{A.18}
 \varepsilon_{drift,j} \; = \; \bar{V} \; + \; \omega_m \; (j \; + \; \frac{1}{2}) \; \; , \; \; j \; \geq \; 0 \; \; \; .
\end{equation}
\begin{figure}[t]
\includegraphics[width=0.9\textwidth,clip]{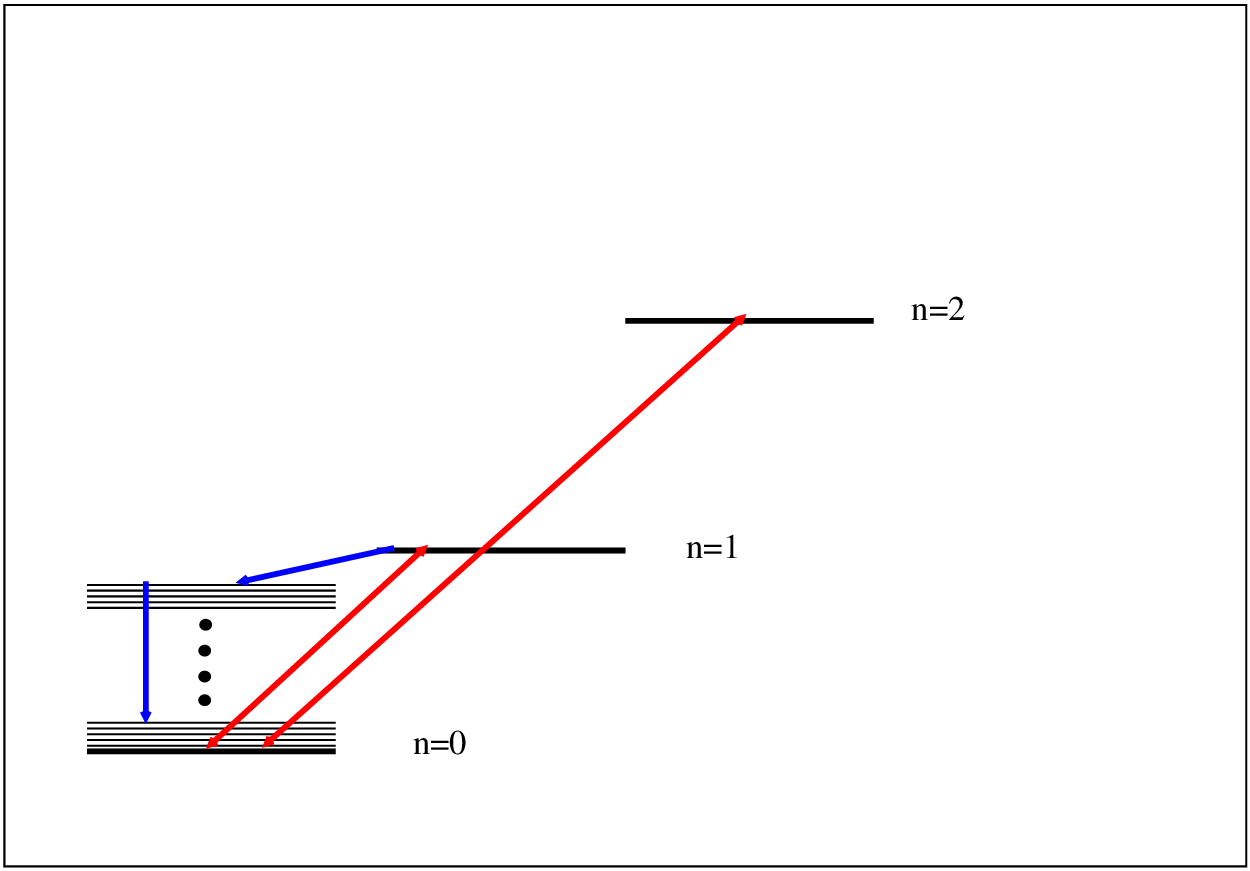}
\caption{\label{fig_18}
Energy spectrum for electrons trapped in the magnetosphere near the polar cups. Heavy lines are the Landau levels with
cyclotron frequency $\omega_B$. Light lines are the quasi continuum spectrum Eq.~(\ref{A.18}). Red and blue arrows
indicate  possible radiative transitions.}
\end{figure}
Let us pause to briefly summarize our results. Our quantum mechanical treatment of charges which are drifting toward the
star in the  weakly varying dipolar magnetic field has shown that electrons feel a huge magnetic barrier. On the other
hand, the magnetic barrier is reduced by a factor $\frac{m_e}{m_p}$ for protons. As a consequence, protons are free to
reach the stellar surface, where they are trapped in the electron atmosphere, while electrons are repelled into the
magnetosphere. The resulting charge separation produces an electric field which trapped electron at a distance $\sim
\bar{z}$. Obviously, the neutrality of the plasma implies that the number densities of trapped electrons and protons are
equal. The resulting picture is very interesting. We have electrons which oscillate around $\bar{z}$ along the magnetic
field. On the other hand, these electrons feel the magnetic field $B(\bar{z})$, which now is truly almost constant. So
that, if $\omega_B$ is the cyclotron frequency at $\bar{z}$, then using Eq.~(\ref{A.15}) we get:
\begin{equation}
\label{A.19}
 \omega_B \; = \;  \frac{e B(\bar{z})}{m_e} \simeq \; 0.268 \; KeV \; B_{12}^{\frac{1}{4}} \; n_{13}^{\frac{3}{4}} \; \; \; .
\end{equation}
Then, the energy spectrum of these electrons is:
\begin{equation}
\label{A.20}
 E_{n,j} \; = \; \bar{V} \; + \;  \omega_B \; n \; + \;  \omega_m \; (j \; + \; \frac{1}{2}) \; \;  , \; \; \; \; n \; ,
         \;  j \; \geq \; 0 \; \; \; .
\end{equation}
In Figure~\ref{fig_18} we  illustrate schematically the spectrum Eq.~(\ref{A.20}). According to Eq.~(\ref{A.20}) the
spectrum comprises discrete Landau levels with cyclotron frequency $\omega_B$ and an almost continuum associated to the
drifting motion. As discussed below, electrons  perform radiative transition between Landau levels. These transitions are
responsible for absorption and spectral features observed in isolated $X$-ray pulsars, anomalous $X$-ray pulsars and soft
gamma ray repeaters. Moreover, we see that there are also transitions where electrons absorb thermal photons with
frequency $\omega \sim \omega_B $ and emit photons with frequencies $\omega^{'} \ll \omega$. These radiative transitions
give rise to the soft spectrum. Before addressing the problem of the soft spectrum, let us discuss the puzzling absorption
features detected in the isolated $X$-ray pulsar {\it {1E 1207.4-5209}}~\cite{Bignami:2003}. The spectrum of  {\it {1E
1207.4-5209}} shows three distinct features, regularly spaced at $0.7 \; KeV$, $1.4 \; KeV$ and $2.1 \; KeV$, plus
possibly a fourth at $2.1 \; KeV$. These features vary in phase with the star rotation. Indeed, it turns out that the
$X$-ray source pulsation is largely due to the phase variation of the lines with the pulsar rotation. The most natural and
logical explanation for the observed features is cyclotron resonant absorption. In this case the fundamental cyclotron
frequency is $0.7 \; KeV$. Thus, the inferred magnetic field is $\sim 8 \, 10^{10} \; Gauss$ for electrons and $\sim 2 \,
10^{14} \; Gauss$ for protons. However, the magnetic field inferred from the spin parameters turns out to $B_S \simeq 2.4
\, 10^{12} \; Gauss$. Obviously, as noted in Ref.~\cite{Bignami:2003} electron cyclotron scattering at a distance $3 - 4
\; R$ above the pulsar surface would fit all the observations. Remarkably, using the inferred magnetic field $B_S$ and
Eq.~(\ref{A.19}) we find:
\begin{equation}
\label{A.21}
 \omega_B \; \simeq \; 0.7 \; KeV  \; \; , \; \;  \;B_{S,12} \; \simeq \; 2.4 \; \; \; \;
            n_{13} \;  \simeq \; 2.69 \;  \; \; .
\end{equation}
Moreover from Eq.~(\ref{A.15}) we get $\bar{z} \; \simeq \; 3.4 \; R$, in perfect accord with observations. Finally, we
stress that the presence of the almost continuum spectrum associated to the drifting motion along the magnetic field leads
naturally to rather broader spectral features in accord with several observations. Even though it is beyond the aim of the
present Section an accurate comparison with available data, we would like to stress that our proposal is in gratifying
qualitative agreement with observations. Indeed, absorption features have been found in the thermal emission of several
isolated $X$-ray pulsars which range in $0.2 - 0.7 \; KeV$. The thermal spectrum of isolated $X$-ray pulsar is a blackbody
with typical temperature $T \sim 0.1 KeV$. Now Eq.~(\ref{A.19}) shows that for typical pulsar magnetic field $B_{12} \sim
1$  the electrons trapped at $\sim \bar{z}$ are able to absorb thermal photons in the observed range. Note that for
magnetars the typical blackbody temperature is  $T \sim 1.0 \; KeV$ (the relevant temperature is the polar cup blackbody
temperature), and $B_{12} \sim 10^2$. So that Eq.~(\ref{A.19}) gives  $\omega_B \; \sim \; 0.85 \; KeV$, indicating that
also for magnetars the trapped electrons may efficiently absorb the thermal photons from the polar cups.
\\
Let us discuss, now, the radiative transitions from trapped electrons. To evaluate the rate of transitions we need to
evaluate:
\begin{equation}
\label{A.22}
 \left | \frac{i e}{m_e} \; < n^{'}, j^{'} | \; \exp(- i \vec{k} \cdot \vec{r} ) \; \vec{l} \cdot \vec{\nabla} \; | n, j >
 \right |^2 \;  \; \; ,
\end{equation}
where:
\begin{equation}
\label{A.23}
| n, j > \; = \;  \zeta_j(z) \; \phi_n(x,y)  \; \; \; , \; \; \vec{l} \cdot \vec{k} \; = \; 0 \; \; \; ,
\end{equation}
$\vec{k}$ and $\vec{l}$ being the photon momentum and polarization respectively. Without loss in generality, we may write:
\begin{equation}
\label{A.24}
 \vec{k} \; = \; ( k_\bot , 0,  k_3)  \; \; \; , \; \; \; \vec{l} \; = \; ( 0, 1, 0 ) \; \; \; ,
\end{equation}
so that:
\begin{equation}
\label{A.25}
 \left | \frac{i e}{m_e} \; < n^{'}, j^{'} | \; \exp(- i \vec{k} \cdot \vec{r} ) \vec{l} \cdot \vec{\nabla} \; | n, j >
 \right |^2 \; =  \; \frac{e^2}{m_e^2} \;I_{n^{'},n} \; J_{j^{'},j} \; \; \; ,
\end{equation}
where:
\begin{equation}
\label{A.26}
I_{n^{'},n} \;  = \; \left |  \; < n^{'} | \; \exp(- i k_\bot x )  \frac{\partial}{\partial y}  \; | n >
                  \right |^2 \;  \; \; \; ,
\end{equation}
\begin{equation}
\label{A.27}
J_{j^{'},j} \;  = \;  \left |  \; <  j^{'} | \; \exp(- i k_3 z )  \; |  j >
                       \right |^2 \;  \;  \; \; .
\end{equation}
Obviously we have also:
\begin{equation}
\label{A.28}
 | \vec{k} | \; = \;  E_{n^{'},j^{'}} \; - \;  E_{n,j} \; = \;   \omega_B \; (n^{'} \; - \; n ) \; + \;
                      \omega_m \; (j^{'} \; - \; j )  \; \; \; .
\end{equation}
We are interested in  processes where trapped electrons absorb thermal photons with energy $ \sim  \omega_B$ (red lines in
Fig~\ref{fig_18}) and emits photons with $ | \vec{k} | \ll \omega_B$ (blue lines in Fig.~\ref{fig_18}). Since  $\omega_m
\sim 10^{-12} \; eV$, from Eq.~(\ref{A.28}) it follows that $\Delta j \gg 1$. Then, the transitions $ j \rightarrow j^{'}$
cannot be induced by the electromagnetic field. Indeed, these transitions are induced by thermal collisions. To see this,
we note that for $j \gg 1$ the wave function  $\zeta(z)$ is quasi classical:
\begin{equation}
\label{A.29}
 \zeta(z) \; \sim  \; \exp [ - i p(z) z ] \; \; \; ,
\end{equation}
where $p(z)$ is the quasi classical momentum. Now, Eq.~(\ref{A.27}) means that $J_{j^{'},j}$ is different from zero if the
momentum is conserved. However, for vastly different quasi momenta the overlap of the quasi classical wave function is
very small. So that we are lead to the conclusion that $| k_3 | \ll |  k_\bot |$. As a consequence we see that the
transitions must be induced by collisions, so that we may safely assume $|J_{j^{'},j}| \simeq 1$. To evaluate
$I_{n^{'},n}$ we need the wave functions $\phi_n(x,y)$:
\begin{equation}
\label{A.30}
 \phi_n(x,y) \; =  \; \frac{\exp( - i p_x x )}{\sqrt{2 \pi}}  \; \sqrt{\frac{m_e \omega_B}{n! 2^n}} \; \frac{1}{\pi^{\frac{1}{4}}}
 \; \exp[ - \frac{m_e \omega_B}{2} (y - \frac{p_x}{m_e \omega_B})^2 ] \; \; H_n[ \sqrt{m_e \omega_B}(y - \frac{p_x}{m_e \omega_B})]
 \; \; .
\end{equation}
The most important transition is for $n^{'}=1 , n=0$. A standard calculations gives:
\begin{equation}
\label{A.31}
I_{1,0} \;  = \; \frac{m_e \omega_B}{2} \; \left ( 1 \; - \;  \frac{k_\bot^2}{2 m_e \omega_B}  \right )^2 \;
               \exp \left [ - \frac{k_\bot^2}{2 m_e \omega_B} \right ] \; \; \; ,
\end{equation}
so that:
\begin{equation}
\label{A.32}
 \left | \frac{i e}{m_e} \; < n^{'}, j^{'} | \; \exp(- i \vec{k} \cdot \vec{r} ) \; \vec{l} \cdot \vec{\nabla} \; | n, j >
 \right |^2 \; \simeq \; \frac{e^2 \omega_B}{2 m_e} \; \left ( 1 \; - \;  \frac{k_\bot^2}{2 m_e \omega_B}  \right )^2
  \; \exp \left [ - \frac{k_\bot^2}{2 m_e \omega_B} \right ] \; .
\end{equation}
Note that in our approximation  $|\vec{k}| = k \simeq |k_\bot| \ll \omega_B$. Thus, we have:
\begin{equation}
\label{A.33}
 \left | \mathcal{A}_{if} \right |^2 \; \equiv \; \left | \frac{i e}{m_e} \; < n^{'}, j^{'} | \; \exp(- i \vec{k}
   \cdot \vec{r} ) \; \vec{l} \cdot \vec{\nabla} \; | n, j >
   \right |^2 \; \simeq \; \frac{e^2 \omega_B}{2 m_e} \; \; \; .
\end{equation}
It is clear that Eq.~(\ref{A.33}) leads to a power law emission flux with the cutoff $k \ll \omega_B$. Moreover, to get
the flux we need to assume some energy distribution for the trapped electrons, which, however, goes beyond the aim of the
present paper. Nevertheless, there is a general aspect which is worthwhile to stress. We have seen that the soft spectrum
in isolated pulsar can be understood as thermal photons reprocessed by electrons trapped in the magnetosphere above the
polar cups. These electrons absorb thermal photons with frequency $\sim \; \omega_B $ and emit photons with frequencies
$\ll \; \omega_B$. So that the number  of these electrons is proportional to the number of thermal photons with energy
$\sim \; \omega_B $.  We know that $\omega_B \sim T$, where $T$ is the blackbody temperature. As a consequence, recalling
that the number density of thermal photons scale as $\sim T^3$, we have that the soft fluxes in magnetars should be about
a factor $(T_{magnetar}/T_{pulsar})^3 \sim 10^3$ greater than the one in isolated $X$-ray pulsars, in fair agreement with
observations. \\
Let us conclude this Appendix by roughly estimating the soft flux $F_\omega$. The probability for emission of a photon in
the frequency range $\omega$, $\omega + d \omega$ is:
\begin{equation}
\label{A.34}
 \frac{d P}{dt \; d\omega} \; = \; \frac{\omega}{2 \pi} \;  \left | \mathcal{A}_{if} \right |^2 \; \delta [\omega \; -
\; (E_{1,j^{'}} \; - \;  E_{0,j}) ] \; \; \; ,
\end{equation}
where we used $ \omega = |\vec{k}| = k \simeq |k_\bot|$, and the delta function ensures the energy conservation
Eq.~(\ref{A.28}). Summing  over the  degenerate final states of the almost continuum spectrum, and using $d\omega =
\omega_m \; dj^{'}$, we get:
\begin{equation}
\label{A.35}
 \frac{d P}{dt \; d\omega} \; \simeq \; \frac{1}{2 \pi} \;  \frac{\omega}{\omega_m} \; \frac{e^2 \omega_B}{2 m_e}  \; \; \; .
\end{equation}
Finally, the spectral flux is obtained  multiplying by the photon energy $\omega$ and by the number of active electrons.
Let $n_{act}$ be the number density of active electrons, we have:
\begin{equation}
\label{A.36}
 F_{\omega} \; \simeq \; \frac{1}{2 \pi} \;  \frac{\omega^2}{\omega_m} \; \frac{e^2 \omega_B}{2 m_e}  \; n_{act} \; V  \; \; ,
\; \; \; \omega \; \ll \; \omega_B \; \; \; ,
\end{equation}
where $V$ is the volume of the emitting region. Note that $ F_{\omega}$ has a Rayleigh-Jeans power law form with an upper
cutoff which, however, cannot be easily estimated without a precise knowledge of electron energy distribution. Using
Eqs.~(\ref{A.17}) and (\ref{A.19}), we rewrite Eq.~(\ref{A.36}) as:
\begin{equation}
\label{A.37}
 F_{\omega} \; \simeq \; 8.2 \; 10^{-6} \; \frac{ergs}{sec \; Hz} \; (\frac{\omega}{1 eV})^2 \; B_{12}^{\frac{3}{8}}
 \; n_{13}^{\frac{1}{8}}  \; \; n_{act} \; V  \; \; \; \; \; .
\end{equation}
The number density of active electrons is not easily estimated without the knowledge of the energy distribution of
electrons trapped into the magnetosphere. In general, $n_{act}$ depends on the number of thermal photons with energy
$\omega \sim \omega_B$. Nevertheless, we may estimate  the needed number density $n_{act}$ by comparing with observed soft
fluxes. For typical isolated $X$-ray pulsar with $B_{12} \sim 1$ and $n_{13} \sim 1$ we infer:
\begin{equation}
\label{A.38}
 F_{\omega} \; \simeq \;  \; 10^{13} \; \frac{ergs}{sec \; Hz} \; \; , \; \; \omega \; \simeq \; 1 eV \; \; .
\end{equation}
So that, assuming for the emitting volume the reasonable value $V \, \sim \, 10^{10} \; cm^3$, from Eq.~(\ref{A.37}) we
get:
\begin{equation}
\label{A.39}
 n_{act} \; \sim  \;  \; 10^{8} \; cm^{-3} \; \sim \; 10^{-5} \; n \; \; \; .
\end{equation}
It is interesting to compare Eq.~(\ref{A.39}) with the number density of active electrons obtained assuming that trapped
electrons have an uniform distribution. In this case, observing that the probability for transitions from the $n=0$ to
$n=1$ Landau levels is given by Eq.~(\ref{A.32}) with $k_\bot \simeq \omega_B$, we get:
\begin{equation}
\label{A.40}
 n_{act} \; \simeq  \;  \frac{e^2 \omega_B}{2 m_e}  \; n  \;  \; \; .
\end{equation}
So that, for typical isolated $X$-ray pulsars we have:
\begin{equation}
\label{A.41}
 n_{act} \; \simeq \; 2.4 \; 10^{-5} \; n \; \; \; ,
\end{equation}
that, indeed, is in reasonable agreement with our estimate Eq.~(\ref{A.39}).

%

%
\end{document}